\documentclass[conference]{IEEEtran}
\usepackage{cite}
\usepackage{amsmath,amssymb,amsfonts}
\usepackage{algorithmic}
\usepackage{graphicx}
\usepackage{textcomp}
\usepackage{xcolor}
\usepackage[hyphens]{url}
\usepackage{fancyhdr}
\usepackage[bookmarks=true,breaklinks=true,letterpaper=true,colorlinks,citecolor=blue,linkcolor=blue,urlcolor=blue]{hyperref}

\usepackage{dirtree} 

\usepackage{booktabs} 

\usepackage{pifont} 

\usepackage[tight]{subfigure} 

\usepackage{multirow} 

\usepackage{tikz} 

\usepackage{caption}

\usepackage[normalem]{ulem}

\usepackage{listings} 

\usepackage[letterspace=-0,tracking=true,final]{microtype}

\newcommand{\circled}[1]{\tikz[baseline=(myanchor.base)] \node[circle,fill=.,inner sep=.5pt] (myanchor) {\color{-.}\footnotesize #1};}

\def\fixme#1{\bgroup \color{red}{[{#1}]}\egroup}

\newcommand{\rev}[1]{\textcolor{black}{#1}}

\fancypagestyle{firstpage}{
  \fancyhf{}
  
  \fancyhead[C]{\scriptsize{This article has been accepted for publication in 57th IEEE/ACM International Symposium on Microarchitecture (MICRO), 2024. Best paper runner-up (author version).}\\~\\~\\~\\~\vspace{0pt}} 
  \fancyfoot[C]{\thepage}
}

\title{A Mess of Memory System Benchmarking, Simulation and Application Profiling \vspace{-.0cm}} 

\author{\IEEEauthorblockN{Pouya~Esmaili-Dokht\IEEEauthorrefmark{1}\IEEEauthorrefmark{2}, Francesco~Sgherzi\IEEEauthorrefmark{1}\IEEEauthorrefmark{2}, Val\'eria~Soldera~Girelli\IEEEauthorrefmark{1}\IEEEauthorrefmark{2}, Isaac~Boixaderas\IEEEauthorrefmark{1}, Mariana~Carmin\IEEEauthorrefmark{1},\\ Alireza~Monemi\IEEEauthorrefmark{1}, Adrià~Armejach\IEEEauthorrefmark{1}\IEEEauthorrefmark{2}, Estanislao~Mercadal\IEEEauthorrefmark{1}, Germ\'an~Llort\IEEEauthorrefmark{1}, Petar~Radojkovi\'c\IEEEauthorrefmark{1}, Miquel~Moreto\IEEEauthorrefmark{1}\IEEEauthorrefmark{2},\\Judit Gim\'enez\IEEEauthorrefmark{1}\IEEEauthorrefmark{2}, Xavier~Martorell\IEEEauthorrefmark{1}\IEEEauthorrefmark{2}, Eduard~Ayguad\'e\IEEEauthorrefmark{1}\IEEEauthorrefmark{2}, Jesus~Labarta\IEEEauthorrefmark{1}\IEEEauthorrefmark{2}, Emanuele Confalonieri\IEEEauthorrefmark{3},\\Rishabh Dubey\IEEEauthorrefmark{3}, Jason Adlard\IEEEauthorrefmark{3}}
\IEEEauthorblockA{Barcelona Supercomputing Center\IEEEauthorrefmark{1}, Unversitat Politecnica de Catalunya\IEEEauthorrefmark{2}, Micron Technology\IEEEauthorrefmark{3}}
\IEEEauthorblockA{
\{pouya.esmaili, francesco.sgherzi, valeria.soldera, isaac.boixaderas, mariana.carmin, alireza.monemi, adria.armejach\}@bsc.es\\
\{lau.mercadal, german.llort, petar.radojkovic, miquel.moreto, judit, xavier.martorell, eduard.ayguade, jesus.labarta\}@bsc.es\\
\{econfalo, rdubeya, jadlard\}@micron.com
}}

\begin{document}
\maketitle
\thispagestyle{firstpage}
\pagestyle{plain}


\begin{abstract}
The Memory stress~(Mess) framework provides a unified view of the  
memory system benchmarking, simulation and application profiling.   

The \emph{Mess benchmark} provides a holistic and detailed memory system characterization. 
It is based on hundreds of measurements that are represented as a family of bandwidth--latency curves. 
The benchmark increases the coverage of all the previous tools and leads to new findings in the behavior of the actual and simulated memory systems.  
We deploy the Mess benchmark to characterize Intel, AMD, IBM, Fujitsu, Amazon and NVIDIA servers  with DDR4, DDR5, 
HBM2 and HBM2E memory.    
The \emph{Mess memory simulator} uses bandwidth--latency concept for the memory performance simulation. 
We integrate Mess with widely-used CPUs simulators enabling modeling of all high-end memory technologies. 
The Mess simulator is fast, easy to integrate and it closely matches the actual system performance. 
By design, it enables a quick adoption of new memory technologies in hardware simulators. 
Finally, the \emph{Mess application profiling} positions the application in the bandwidth--latency space of the target memory system. 
This information can be correlated with other application runtime activities and the source code, leading to a better overall understanding of the application's behavior. 

The current Mess benchmark release covers all major CPU and GPU ISAs, \textbf{x86}, \textbf{ARM}, \textbf{Power}, \textbf{RISC-V}, and \textbf{NVIDIA's PTX}. 
We also release as open source the ZSim, gem5 and OpenPiton Metro-MPI integrated with the Mess simulator for DDR4, DDR5, Optane, HBM2, HBM2E  and CXL memory expanders. 
The Mess application profiling is already integrated into a suite of production HPC performance analysis tools.

\end{abstract}

\section{Introduction}
\label{sec:Introduction}



The importance of the main memory in the overall system’s design~\cite{Wulf:memory-wall,Sites:MemoryStupid, Saulsbury:memwall-solution} drives significant effort for memory system benchmarking, simulation, and memory-related application profiling. 
Although these three memory performance aspects are inherently interrelated, they are analyzed with distinct and decoupled tools.
\textbf{Memory benchmarks} typically report
the maximum sustainable memory bandwidth~\cite{intel:advisor, mccalpin:streamBenchmark,Peng:camp} 
or performance of the bandwidth-limited application kernels~\cite{Dongarra:HPCG}. 
This is sometimes complemented with latency measurements in unloaded memory systems~\cite{lmbench,Google:multichase,Sanchez:SandyBridge} or 
for a small number of memory-usage scenarios~\cite{Intel:MLC,Gottscho:X-Mem}.  
\textbf{Memory simulators} determine the memory system response time for a given traffic. 
Simple simulators model memory with a fixed latency, 
or calculate its service time based on queueing theory or simplified DDR protocols~\cite{carlson:sniperUpdated,carlson:sniperOriginal,sanchez:zsim,Jason:gem5,Gulur:memModel3queue,Miller:graphite,Ganesh:queueModel}. 
Dedicated cycle-accurate memory simulators consider detailed memory device sequences and timings~\cite{chatterjee:usimm,Wang:DRAMSIM,Shangli:dramsim3,kim:ramulator,Haocong:Ramulator2,steiner:dramsys4}. 
\textbf{Application profiling tools} determine whether applications are memory bound 
based on the memory access latency~\cite{Dean:profileMe, Helm:perfmemplus, Intel:ProgrammingGuide}, 
the position in the Roofline model~\cite{williams:roofline, ilic:roofline-cache-aware} 
or the the memory-related portion of the overall CPI stack~\cite{Yasin:topdown}. 

\begin{figure}[!t]
 \centering
 \includegraphics[width=\linewidth]{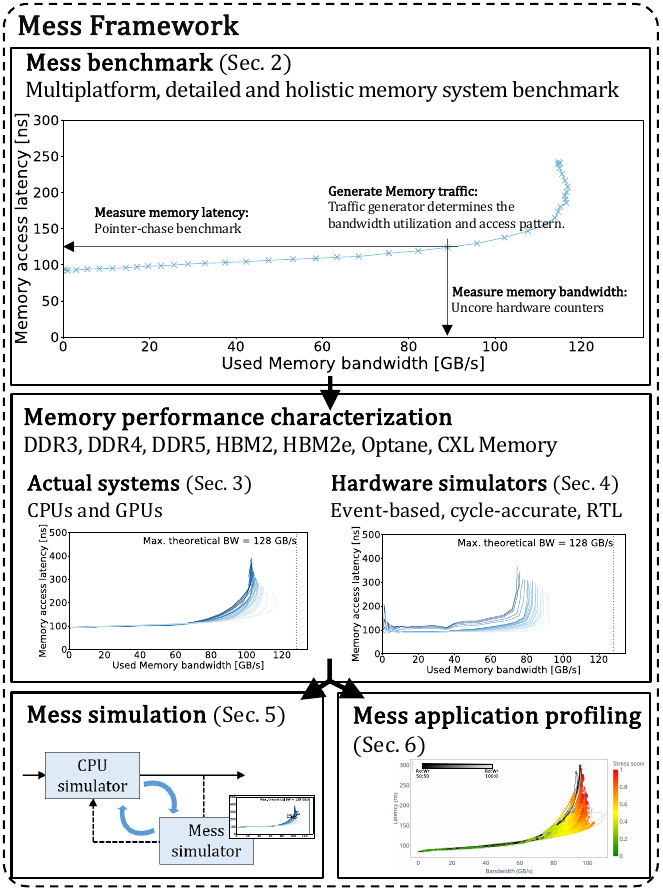}
 \caption{Mess framework: The Mess benchmark describes the memory system performance with a family of bandwidth--latency curves.
We deploy the benchmark to characterize memory systems of actual servers and hardware simulators. 
The Mess bandwidth–latency curves can be also used for the memory performance simulation and memory-related application profiling.}
  \label{fig:bandwidth-latency-curve-build1}
\end{figure}

Our study argues that the memory system benchmarking, simulation and application profiling 
can and should be based on a \textbf{unified view of memory system performance}. 
We provide this view with the \textbf{Memory stress~(Mess) framework}
comprised of the Mess benchmark, \rev{analytical memory system simulator} and application profiling tool (Figure~\ref{fig:bandwidth-latency-curve-build1}).   
\textbf{Mess benchmark} (Section~\ref{sec:memory-stress-benchmark}) 
describes the memory system performance with a \textbf{family of bandwidth--latency curves}. 
The benchmark covers the full range of the memory traffic intensity, from the unloaded to fully-saturated memory system.  
It also considers numerous compositions of read and write operations, plotted with different shades of blue in 
Figure~\ref{fig:bandwidth-latency-curve-build1} (middle).   
The Mess benchmark is designed for holistic and detailed memory system characterization,  
and it is easily adaptive to different target platforms. 
%
The current benchmark release covers all major CPU and GPU ISAs: \textbf{x86}, \textbf{ARM}, \textbf{Power}, \textbf{RISC-V}, and NVIDIA's \textbf{Parallel Thread Execution~(PTX)}~\cite{mess:urlLink}. 


\looseness -1 We deploy the Mess benchmark to 
characterize \textbf{Intel, AMD, IBM, Fujitsu and Amazon servers} as well as \textbf{NVIDIA GPUs} with \textbf{DDR4, DDR5, HBM2 and HBM2E memory} (Section~\ref{sec:mess-characterization-actual-systems}).   
We report and discuss a wide range of memory system behavior
even for the hardware platforms with the same main memory configuration. 
These differences are 
especially pronounced in the high-bandwidth areas
which have the greatest impact on memory-intensive applications. 
\rev{We also detect and analyze scenarios is which increase of the memory request rate leads to the degradation of the measured bandwidth.}


We use the Mess benchmark to evaluate memory system simulation of event-based \textbf{ZSim}~\cite{sanchez:zsim}, 
cycle-accurate \textbf{gem5}~\cite{Jason:gem5} and RTL simulator \textbf{OpenPiton Metro-MPI}~\cite{Lopez-Paradis:Metro-Mpi} (Section~\ref{sec:mess-characterization-simulator})
We evaluate different internal memory models and widely-used external memory simulators, \textbf{DRAMsim3}~\cite{Shangli:dramsim3},
\textbf{Ramulator}~\cite{Shangli:dramsim3} and \textbf{Ramulator~2}~\cite{Haocong:Ramulator2}. 
Unfortunately, all tested memory simulators, 
including well-established and trusted gem5 DDR models and cycle-accurate memory simulators, poorly resemble the actual system performance. 
The simulators show an unrealistically low load-to-use latency (starting at 4\,ns), 
high memory bandwidth (exceeding 1.8$\times$ the maximum theoretical one), and a simulation error of tens of percents for memory-intensive benchmarks, STREAM~\cite{mccalpin:streamBenchmark}, LMbench~\cite{lmbench} and Google multichase~\cite{Google:multichase}. 
\rev{We detect two sources of these errors: the Zsim interface towards the external memory simulators, 
and the imprecise DRAMsim3 and Ramulator model of the row-buffer utilization.}    
Finally, although it was not the initial design target, we also detected a case 
in which holistic and detailed Mess benchmarking diagnosed a bug in the coherency protocol generated by the OpenPiton framework.
%




Apart from the memory system characterization, 
the Mess bandwidth--latency curves can be also used for the memory performance simulation (Section~\ref{sec:detailed-hardware-simulation}). 
We develop and integrate the \textbf{Mess simulator} in the \textbf{ZSim, gem5}, and \textbf{OpenPiton Metro-MPI simulators}, enabling simulation 
of high-end memory systems based on \textbf{DDR4}, \textbf{DDR5}, \textbf{Optane}, \textbf{HBM2}, and \textbf{HBM2E} technologies, and \textbf{Compute Express Link~(CXL)}~\cite{CXL2020}. 
The Mess integration is easy, based on the standard interfaces between the CPU and external memory simulators~\cite{sanchez:zsim,Jason:gem5,wenisch:simflex}.  
The Mess simulator closely matches the actual memory systems performance.     
The simulation error for STREAM, LMbench and Google multichase benchmarks is between  
0.4\% and 6\%, which is significantly better than any other memory models we tested. 
Mess is also fast. For example, Mess integrated with ZSim introduces only 26\% simulation time increment over the fixed-latency memory model, 
while the speed-up over the Ramulator and DRAMsim3 ranges between 13$\times$ and 15$\times$.
%
%
Mess also removes the current lag between the emergence of memory technologies and development of the reliable simulation models. 
For example, Mess is the first memory simulator that models \textbf{CXL memory expanders}, 
enabling further research on these novel memory devices.  
The simulation is based on the bandwidth--latency curves
obtained from the memory manufacturer’s SystemC hardware model (Section~\ref{sec:novel-memory-systems}).
%

%

%
%

\looseness -1 Finally, the Mess memory bandwidth--latency curves enhance application profiling and performance analysis (Section~\ref{sec:paraver-profet}). 
\textbf{Mess application profiling} determines positions of application execution time segments on the corresponding memory bandwidth--latency curves. The application memory stress can be combined with the overall application timeline analysis and can be linked to the source code.  
The Mess application profiling is already integrated into a suite of \textbf{production HPC performance analysis tools}~\cite{BSCTools}.  
%

The inherent dependency between the used memory bandwidth and the access latency is by no means a new phenomenon, 
and the community has known about it at least for a couple of decades~\cite{Jacob:bw-lat-initial}. 
The Mess framework extends the previous work in three important aspects. 
Previous studies typically use a single bandwidth--latency memory curve to illustrate a \textbf{general memory system behavior}~\cite{Jacob:bw-lat-initial,Helm:perfmemplus,Gottscho:X-Mem,helm:bw-latCurve,Yang:optaneCurve,Joseph:optaneReportLong}.
The Mess benchmark is designed for \textbf{holistic, detailed and close-to-the-hardware} memory system performance characterization 
of the \textbf{particular system under study}.
To reach this objective, the Mess benchmark is developed directly in the assembly, and the experiments are tailored to minimize and mitigate the impact of the system software. 
%
It detects and quantifies some aspects of the memory systems behavior not discussed in previous studies,   
such as the impact of the read and write memory traffic on performance, 
or discrepancies between different memory systems, both in actual platforms and hardware simulators. 
%
%
%
%
Second, the Mess framework \textbf{tightly integrates the memory performance characterization into the memory simulation}. 
The Mess simulator avoids complex memory system simulation 
and \rev{analytically} adjusts the rate of memory instructions (provided by the CPU simulator) to the actual memory performance.
The accuracy of this simulation approach relies on the input memory performance characterization. 
This characterization therefore has to be holistic, detailed and specific to the system under study, 
closely matching the Mess benchmark design.  
%
%
Third, the framework \textbf{closely couples the memory-related profiling of hardware platforms and applications}.  
Similar to the Mess simulator, the Mess application profiling itself is uncomplicated, 
and its real value comes from the application analysis in the context of the memory system characteristic. 
The quality of the memory characterization therefore directly impacts the the quality of the overall analysis. 
For this reason, the analysis has to be performed with detailed Mess-like memory performance description. 
%


%

\looseness -1  The Mess benchmark is released as open source and it is ready to be used in \textbf{x86}, \textbf{Power}, \textbf{ARM}, and \textbf{RISC-V} CPUs and \textbf{NVIDIA} GPUs~\cite{mess:urlLink}. 
The release also contains all bandwidth--latency measurements shown in the paper, including the \textbf{CXL expander curves} provided by the memory manufacturer. 
We also release as open source the 
\textbf{ZSim}, \textbf{gem5} and \textbf{OpenPiton Metro-MPI} integrated with the Mess simulator supporting \textbf{DDR4}, \textbf{DDR5}, \textbf{Optane}, \textbf{HBM2}, \textbf{HBM2E} and \textbf{CXL expanders}~\cite{messim:urlLink}.
Also, the public releases of the \textbf{production HPC performance analysis tools} already include the Mess application profiling extension~\cite{messparaver:urlLink}.
The released tools are ready to be used by the community for better understanding of the current and exploration of future memory systems.


\section{Mess benchmark}
\label{sec:memory-stress-benchmark}


%
This section describes the Mess benchmark use, 
and analyzes its characterization of actual platforms and hardware simulators. 

\subsection{Mess benchmark: Memory bandwidth--latency curves} 
\label{sec:BW-lat_curves}

\looseness -1 The Mess memory characterization comprises tens of bandwidth--latency curves, 
each corresponding to a specific ratio between the read and write memory traffic. 
%
%
%
%
The Mess benchmark kernels cover the whole range of memory operations, from 100\%-\texttt{loads} to 100\%-\texttt{stores}. 
The 100\%-\texttt{load} kernel generates a 100\%-read memory traffic, 
while the 100\%-\texttt{store} kernel creates  50\%-read/50\%-write traffic. 
This is because the contemporary CPUs deploy the write-allocate cache policy~\cite{Jouppi:write-allocate}. 
With this policy, each store instruction
first reads data from the main memory to the cache, then modifies
it, and finally writes it to the main memory once the cache line is
evicted. Each store instruction, therefore, does not correspond to
a single memory write, but to one read and one write.\footnote{Memory traffic with more than 50\% of writes can be generated 
with streaming stores that directly write data to the main memory. 
\rev{The public Mess repository already includes the benchmark with x86 streaming (non-temporal) stores. 
We are working on the equivalent benchmark version for ARM and Power CPUs (zero cache line fill instructions), and NVIDIA GPUs.}} 


The Mess bandwidth--latency curves are illustrated in Figure~\ref{fig:bandwidth-latency-curve-build1} (middle). 
The curves with different composition of read and write traffic are plotted with different shades of blue.  
%
%
%
Each curve is constructed based on tens of measurement points that cover the whole range of memory-traffic intensity.  
Figure~\ref{fig:bandwidth-latency-curve-build1}~(top) illustrates the construction process for one of the curves. 
The $x$-axis of the chart corresponds to memory bandwidth, monitored with hardware counters~\cite{Linux:perf,Terpstra:PAPI,Treibig:likwid, cuda:cupti}.   
The memory access latency is measured with a pointer-chase benchmark~\cite{Verdejo:microbenchmarks, sanchez:zsimEnhance} executed on one CPU core or one GPU Stream Multiprocessor~(SM). 
This determines the $y$-axis position of the measurement.
\rev{The pointer-chase is selected because it is simple and portable to different actual hardware platforms and simulators.  
Optionally, memory latency could be measured with instruction-based sampling approach ~\cite{Westcott:instructionSampling} 
available in some current architectures~\cite{Drongowski:AMDIBS,Srinivas:PowerIBS,Intel:ProgrammingGuide}.} 
 
Running pointer-chase alone measures the unloaded memory access latency. 
To measure the latency of the loaded memory system,  
concurrently with the pointer-chase, on the remaining CPU cores or GPU SMs, 
we run a memory traffic generator.
\rev{The memory access latency is still measured only for the pointer-chase benchmark. The purpose of the traffic generator is to create memory accesses that will collide with the pointer-chase in the loaded memory system. }
The generator is designed to create memory traffic with configurable memory bandwidth utilization and read/write ratio. \rev{Each CPU core and GPU SM traverses two separate arrays, one with load and one with store operations. Therefore, the overall memory traffic is complex, determined by  a sequential accesses within each array, but also by the interleaving between memory request from distinct arrays. 
The Mess benchmark covers a large range of row-buffer hit/empty/miss rates, e.g. between
35/43/22\% and 84/13/3\% in state-of-the-art Intel architectures (Section~\ref{sec:mem-simulation-error}).  }
 
Both, pointer-chase and traffic generator are implemented in assembly to minimize any compiler intervention. The detailed implementations of pointer- chase and traffic generator is presented in Appendix~\ref{app:Mess-implementation}. 
To minimize the latency penalties introduced by the 
TLB misses and the page walk, the Mess data-structures are allocated in huge memory pages. Additionally, at runtime, these overheads are monitored with hardware counters 
and subtracted from the memory latency measurements. 
The Mess benchmark release~\cite{mess:urlLink} includes the source code and its detailed description.

\subsection{Validation}
\label{sec:mess-validation}

The Mess benchmark exceeds the coverage of 
all existing memory benchmarks and tools. Still, these tools can validate some of the Mess measurements.
 
The unloaded memory system latency can be measured with  LMbench and Google multichase in CPU platforms, and P-chase~\cite{Mei:P-chaseGPU} in GPUs. 
We used these benchmarks to validate the Mess unloaded latency measurements in all hardware platforms under study.
%
In all experiments, Mess closely matches the LMbench, Google multichase and P-chase results. 

\looseness -1 
In Intel systems, the maximum sustained memory bandwidth can be measured with the Intel Advisor~\cite{intel:advisor}. 
In the Skylake, Cascade Lake and Sapphire Rapids servers under study, the Mess benchmark matches the Advisor measurements, with a difference below 1\%.


\looseness -1 The Intel Memory Latency Checker~(MLC)~\cite{Intel:MLC}, 
can measure the memory latency for a selected memory traffic intensity, i.e. memory bandwidth. 
The memory bandwidth can be fine-tuned, 
but the tool provides a sparse analysis of different traffic compositions, i.e. read and write memory operations. 
%
We compare the Intel MLC results and the corresponding subset of the Mess measurements 
for all Intel platforms under study. 
The MLC and Mess results show the same trend, with slightly lower ($<$5\%) latencies reported by the Mess benchmark. 
This is because the Mess is designed for close-to-the-hardware memory characterization, and, unlike the MLC, it excludes the latency penalties introduced by the OS overheads, the TLB misses and the page walk.

\subsection{Performance analysis}
\label{sec:Mess-metrics}
\begin{figure}[!t]
 \centering
 \includegraphics[width=\linewidth]{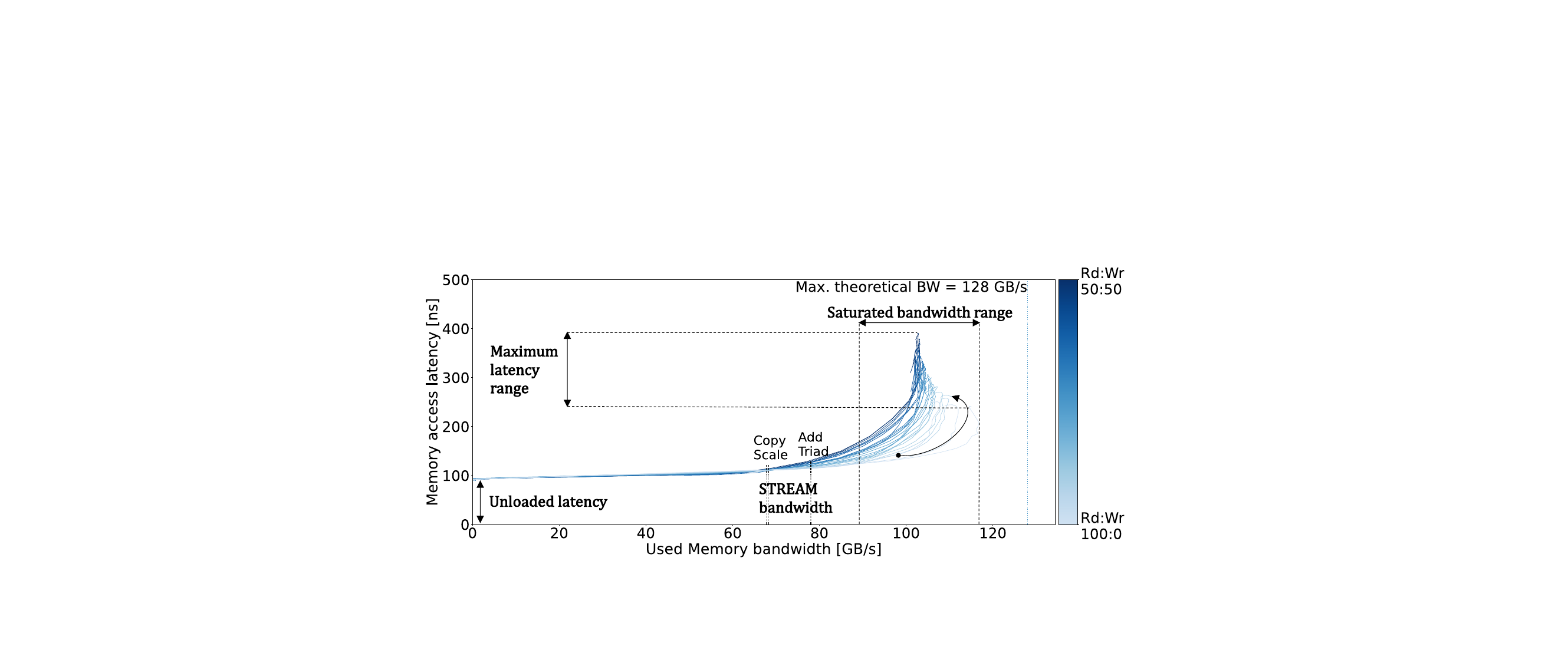} 
 \caption{The Mess benchmark models the memory system performance with a family of bandwidth--latency curves.}
 \label{fig:bandwidth-latency-curve-metric}
\end{figure}

\looseness -1  Figure~\ref{fig:bandwidth-latency-curve-metric} illustrates the Mess benchmark performance analysis with an example of the 
24-core Intel Skylake server with six DDR4-2666 memory channels~\cite{Intel:non-temporal}.  
The figure confirms a general trend of the memory access latency, which is initially roughly constant 
and then increases with higher memory pressure (i.e. bandwidth) due to resource contention among parallel accesses~\cite{Jacob:bw-lat-initial,Radulovic:PROFET,Clapp:initial-bw-lat-with-rw-wrong}. 

Detailed and close-to-the-hardware Mess characterization reveals some memory system aspects not discussed by previous studies.  
The most important one is the impact of the read and write memory traffic.
The best performance, the lowest latency and the highest achieved bandwidth, are obtained for 100\%-read traffic. 
Memory writes reduce the memory performance and reach the saturation point sooner. 
This is due to the extra timing constraints such as $t_{WR}$ and $t_{WTR}$, which come with memory write operations~\cite{jedec:ddr4}. 
We detect this behavior for all the Intel, AMD, IBM, Fujitsu, and Amazon servers as well as NVIDIA GPUs used in the study with DDR4, DDR5, HBM2 and HBM2E memory (Section~\ref{sec:mess-characterization-actual-systems}).
However, we see a very different write traffic impact on CXL memory expanders~\cite{CXL2020}. 
This behavior of CXL memory expanders is analyzed in Section~\ref{sec:novel-memory-systems}.

\looseness -1 Apart from the memory bandwidth--latency curves, we use the Mess benchmark to derive memory system performance metrics, also depicted in Figure~\ref{fig:bandwidth-latency-curve-metric}, for quantitative comparison of different memory systems.  
In addition to the commonly-used unloaded memory latency, the detailed Mess characterization quantifies the \textbf{maximum latency range} of memory access latencies for all read/write ratios and the \textbf{saturated bandwidth range}.  
The memory system is saturated when any further increase in the memory system pressure, i.e. bandwidth, leads to a high increase in the memory access latency. We consider that the saturated bandwidth area starts at the point in which the memory access latency doubles the unloaded latency. 
\rev{In some of the platforms under study, we also detect that the increase in the Mess memory access rate
causes a memory bandwidth decline, while the access latency continues to increase. This behavior is observed in the bandwidth–latency curves as a ``wave form'' seen in Figure~\ref{fig:bandwidth-latency-curve-metric}.} 
The causes for this memory behavior are analyzed in Section~\ref{sec:mess-characterization-actual-systems}. 
To the best of our knowledge, our study is the first one to detect and analyze this suboptimal memory system behavior. 


%
%
\looseness -1  \rev{Figure~\ref{fig:bandwidth-latency-curve-metric} also depicts the bandwidth reported by the \textbf{STREAM benchmark} (vertical dashed lines).
 STREAM and Mess provide complementary memory bandwidth analysis. 
While STREAM is the \emph{de facto} standard for measuring  application-level 
sustained memory bandwidth~\cite{mccalpin:streamBenchmark,Deakin:gpu-stream},   
the Mess benchmark provides the microarchitecture view and considers all memory traffic visible by hardware counters. 
Differences between these two approaches are analyzed in the next section.}

\begin{table*}[t!]
\caption{CPU and GPU platforms under study: Quantitative memory performance comparison.}
\vspace{-.29cm}
\label{tab:hpc-server-description}
\centering
\resizebox{1\textwidth}{!}{%
\begin{tabular}{@{}l|llllllll@{}}
\hline
\toprule
Platform			& \begin{tabular}[l]{@{}l@{}}Intel Skylake \\ Xeon\,Platinum\cite{Intel:non-temporal}\end{tabular}          & \begin{tabular}[l]{@{}l@{}}Intel Cascade Lake \\Xeon Gold\cite{Intel:non-temporal}\end{tabular}  & \begin{tabular}[l]{@{}l@{}}AMD Zen\,2 \\ EPYC 7742\cite{AMD:Zen2}\end{tabular} & \begin{tabular}[l]{@{}l@{}}IBM Power\,9 \\ 02CY415\cite{Sadasivam:Power9}\end{tabular}   &   \begin{tabular}[l]{@{}l@{}}Amazon \\ Graviton\,3\cite{Amazon:Graviton3}\end{tabular} &  \begin{tabular}[l]{@{}l@{}}Intel Sapphire\,Rapids\\  Xeon\,Platinum\cite{Intel:Sapphire-Rapids}\end{tabular} &  \begin{tabular}[l]{@{}l@{}}Fujitsu \\ A64FX\cite{Fujitsu:A64FX}\end{tabular} & \begin{tabular}[l]{@{}l@{}}NVIDIA Hopper\\  H100\cite{Choquette:H100-GPU}\end{tabular}  \vspace{-.05cm} \\
\midrule
Released					& 2015    				& 2019  				& 2019     				& 2017 				& 2022				& 2023					& 2019				& 2023				\\
Cores @ frequency 			& 24 @2.1\,GHz  		& 16	@2.3\,GHz  		& 64 @2.25\,GHz    		& 20 @2.4\,GHz  	  	& 64	@2.6\,GHz  		& 56 @2GHz				& 48	@2.2\,GHz		& 132 SMs@1.1GHz		\\
Main memory 				& 6$\times$DDR4-2666	& 6$\times$DDR4-2666	& 8$\times$DDR4-3200	& 8$\times$DDR4-2666	& 8$\times$DDR5-4800 	& 8$\times$DDR5-4800		& 4$\times$HBM2		& 4$\times$HBM2E		\\
Theoretical\,bandwidth		& 128\,GB/s			& 128\,GB/s			& 204\,GB/s			& 170\,GB/s 			& 307\,GB/s 			& 307\,GB/s 				& 1024\,GB/s			& 1631\,GB/s	\vspace{-.08cm}	\\
\bottomrule
\hline
\multicolumn{9}{l}{\textit{\% of the max theoretical bandwidth}} \\
Saturated\,bandwidth range	& 72--91\%			& 68--87\%			& 57--71\%			&  67--91\%			& 63--95\% 			& 60--86\% 				& 72--92\%			& 51--95\% \\
\rev{STREAM\,kernels:\,Application view}				& 53--61\%			& 51--57\%			& 46--51\%			&  32--36\%			& 78--82\%			& 63--66\%				& 49–55\%			& 64--69\% \\
\hline
Unloaded latency			& 89\,ns				& 85\,ns				& 113\,ns				&  96\,ns				& 122\,ns 	 			& 109\,ns					& 129\,ns				& 363\,ns   \\
Maximum\,latency range		& 242--391\,ns			& 182--303\,ns			& 257--657\,ns			& 238--546\,ns 			& 332--527\,ns			& 238--406\,ns 				& 338--428\,ns			& 699--1433\,ns	\vspace{-.08cm}\\

\bottomrule
\hline
\end{tabular}
}
\vspace{-.4cm}
\end{table*}


\section{Performance characterization: Actual systems}
\label{sec:mess-characterization-actual-systems}

\begin{figure}[!t]%
\centering
\includegraphics[width=.8\linewidth]{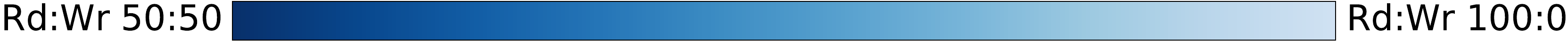}\\
\vspace{-.1cm}
\hspace{-0.1cm}\subfigure[][\fontsize{6.9}{0}\selectfont Intel\,Skylake\,with\,6$\times$\,DDR4-2666.]{%
\label{fig:char-bw-lat-result-skylake}%
\includegraphics[width=.5\linewidth]{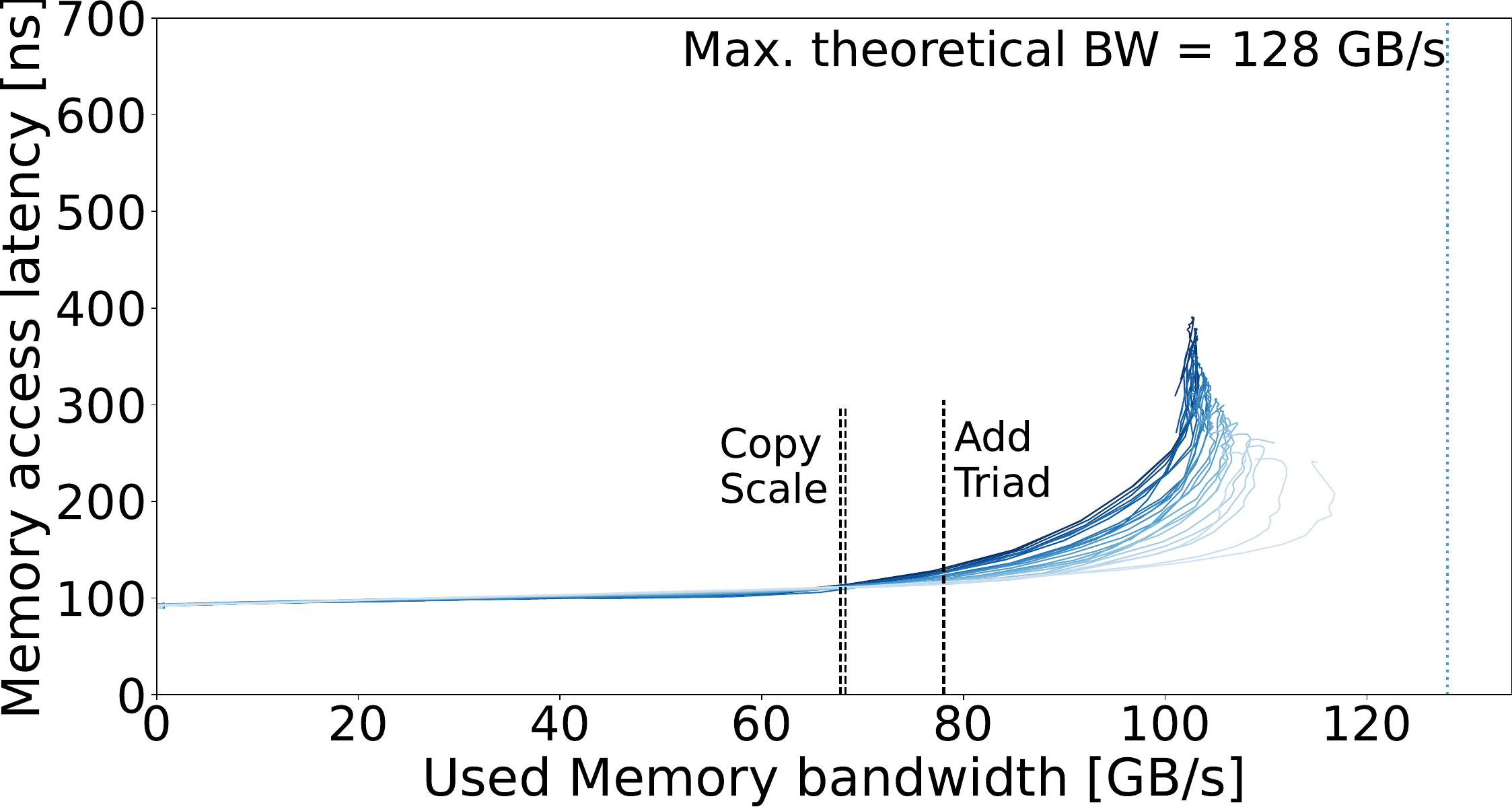}}~%
\subfigure[][\fontsize{6.9}{0}\selectfont Intel\,Cascade\,Lake\,with\,6$\times$\,DDR4-2666.]{%
\label{fig:}%
\includegraphics[width=.5\linewidth]{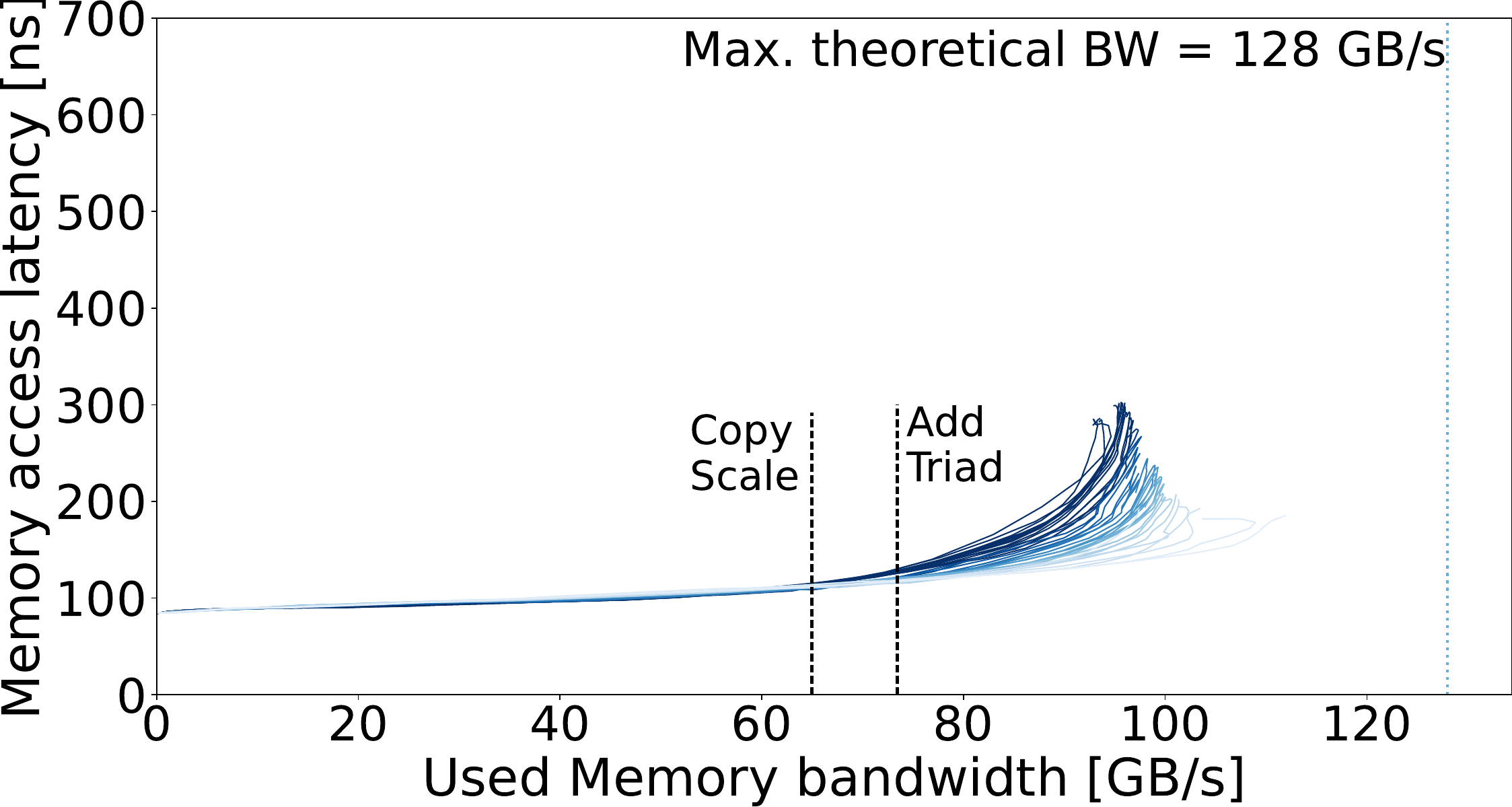}}\\%
\hspace{-0.1cm}\subfigure[][\fontsize{6.9}{0}\selectfont AMD\,Zen2\,with\,8$\times$\,DDR4-3200]{%
\label{fig:}%
\includegraphics[width=.5\linewidth]{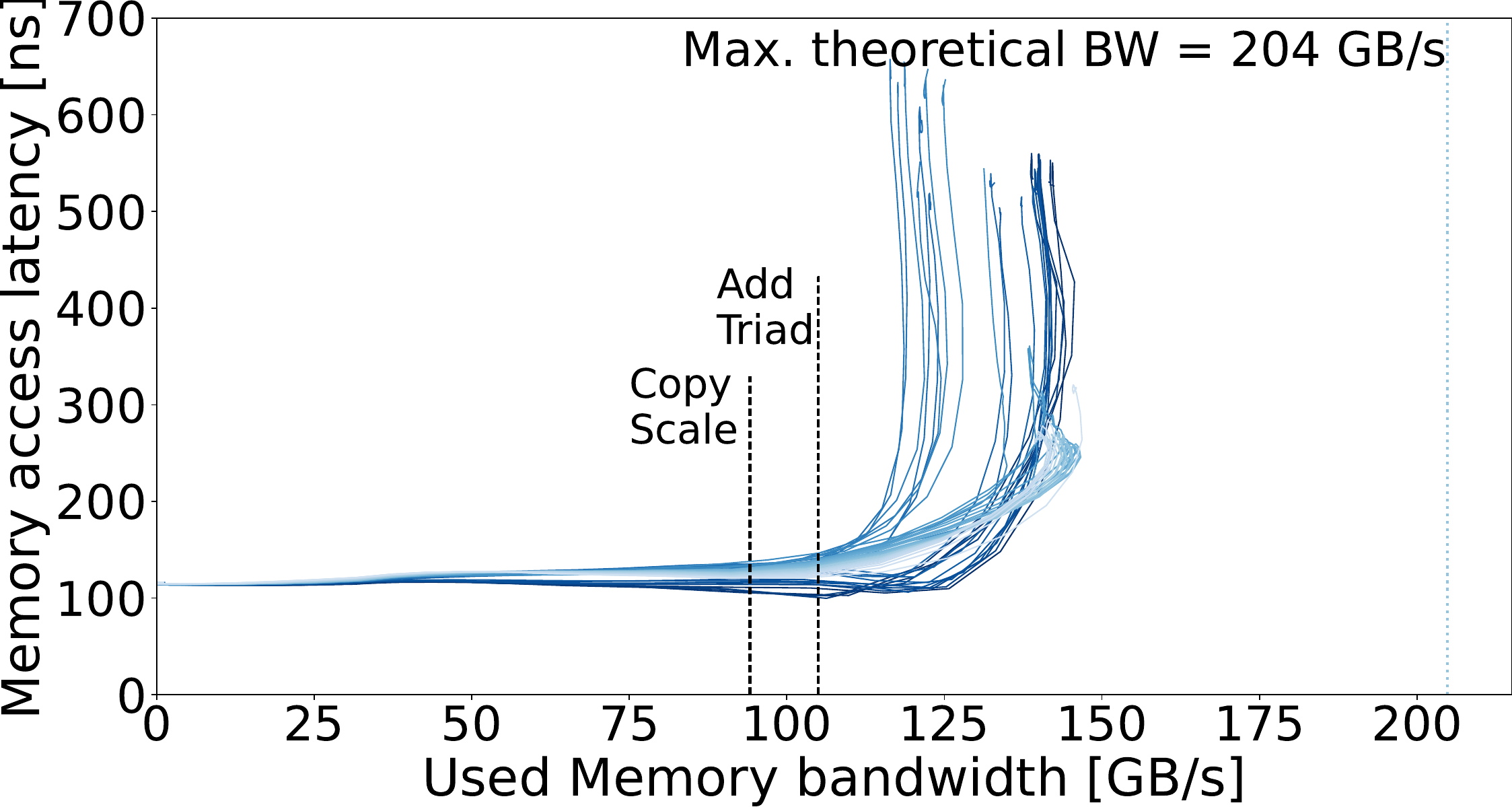}}~%
\subfigure[][\fontsize{6.9}{0}\selectfont IBM\,Power\,9\,with\,8$\times$\,DDR4-2666.]{%
\label{fig:}%
\includegraphics[width=.5\linewidth]{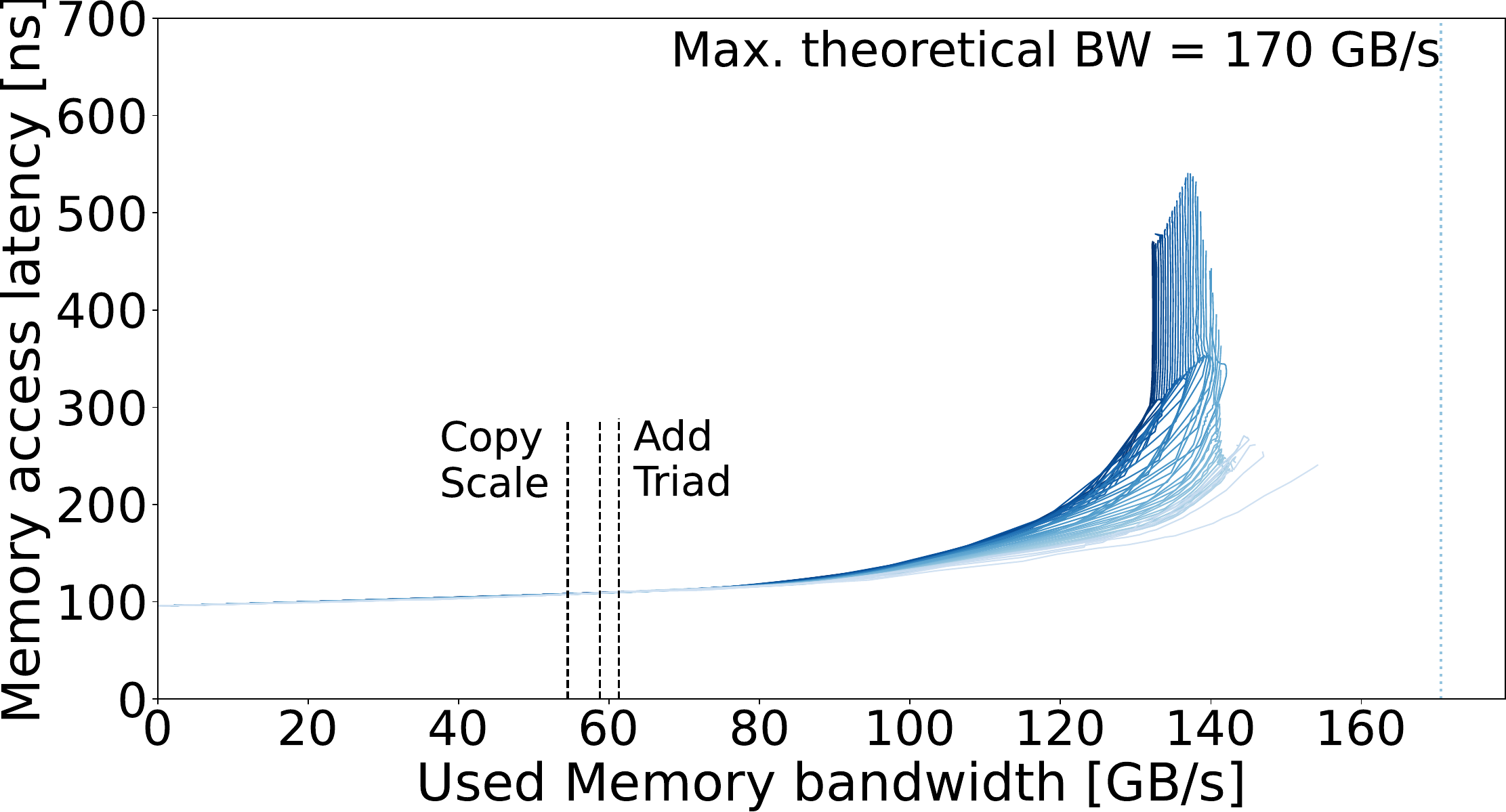}}\\%
\hspace{-0.1cm}\subfigure[][\fontsize{6.9}{0}\selectfont Amazon\,Graviton3\,with\,8$\times$\,DDR5-4800.]{%
\label{fig:char-bw-lat-result-graviton}%
\includegraphics[width=.5\linewidth]{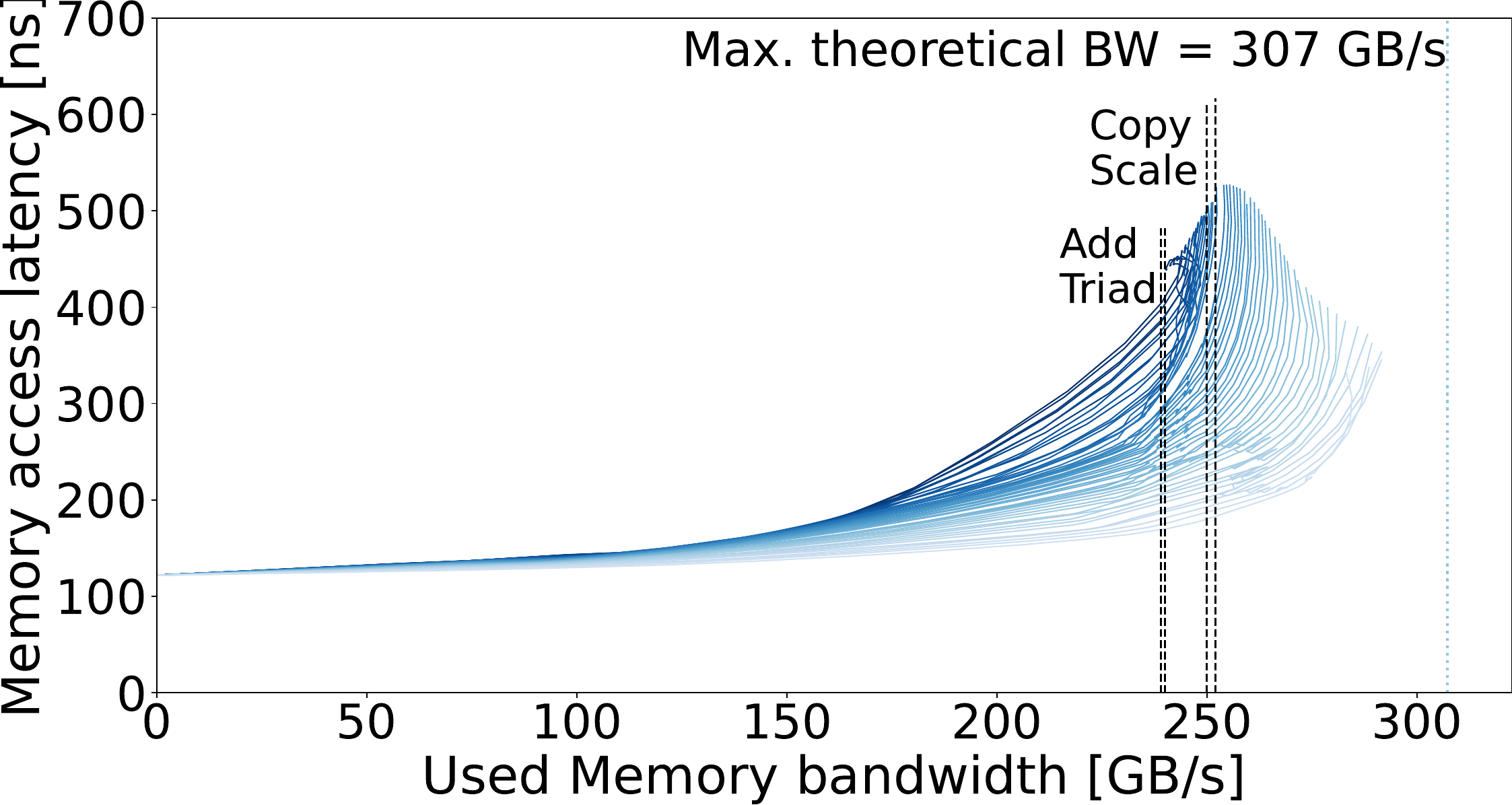}}~%
\subfigure[][\fontsize{6.6}{0}\selectfont Intel\,Sapphire\,Rapids\,with\,8$\times$\,DDR5-4800.]{%
\label{fig:char-bw-lat-result-sapphirerapids}%
\includegraphics[width=.5\linewidth]{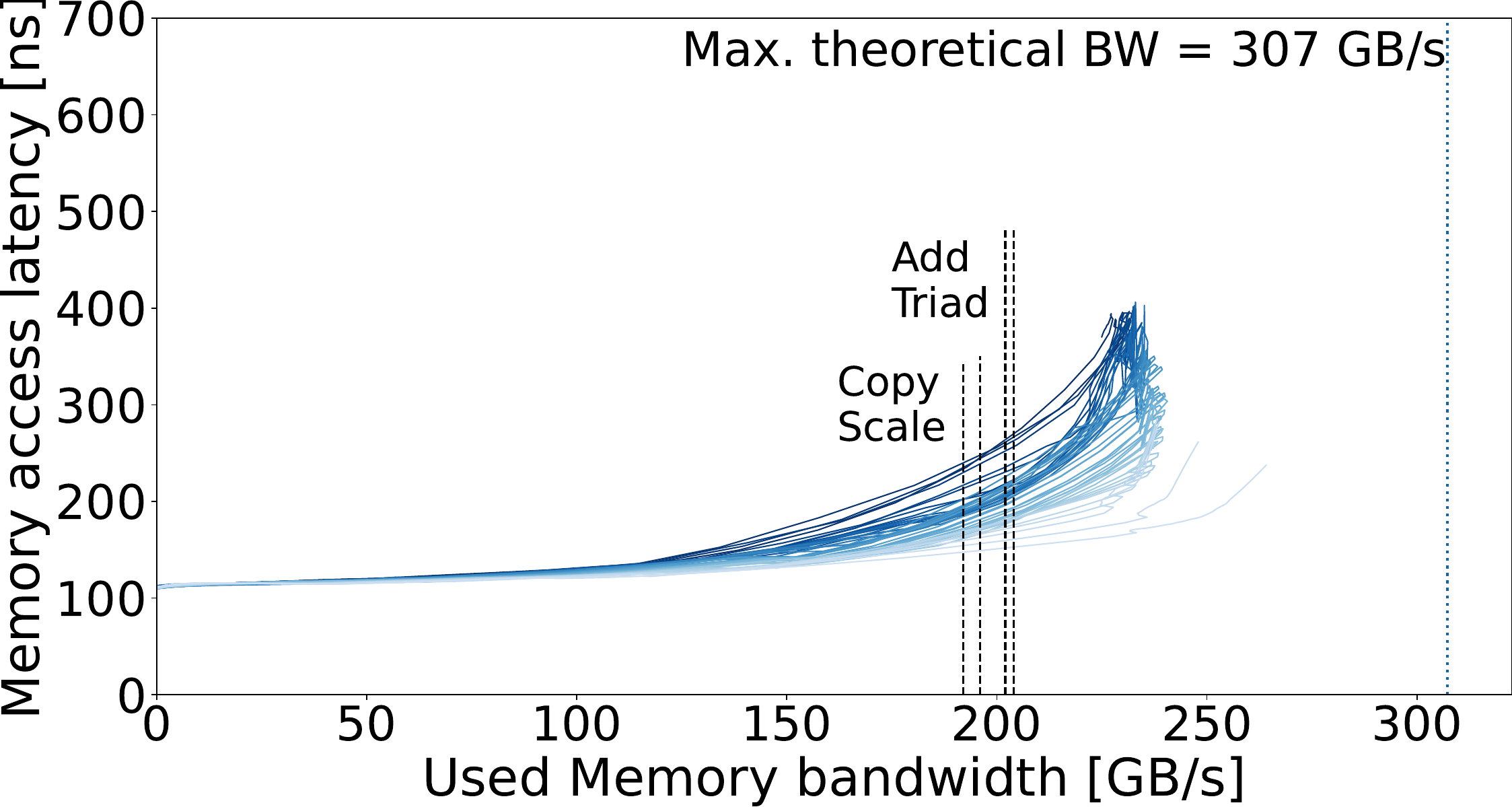}}\\%
\hspace{-0.1cm}\subfigure[][\fontsize{6.9}{0}\selectfont Fujitsu\,A64FX\,with\,4$\times$\,HBM2.]{%
\label{fig:char-bw-lat-result-a64fx}%
\includegraphics[width=.5\linewidth]{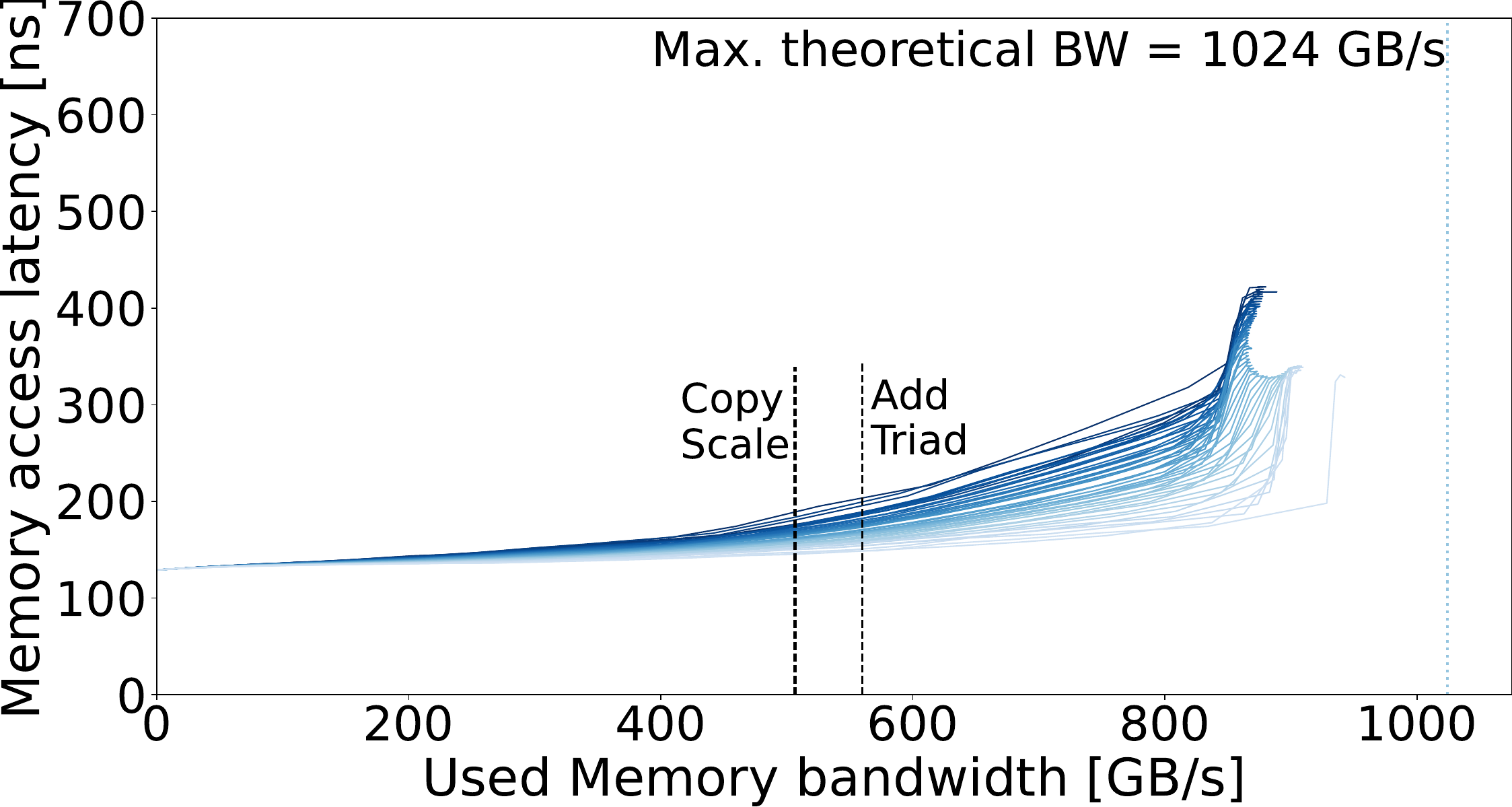}}~%
\subfigure[][\fontsize{6.9}{0}\selectfont NVIDIA\,H100\,with\,4$\times$\,HBM2E.]{%
\label{fig:char-bw-lat-result-h100}%
\includegraphics[width=.5\linewidth]{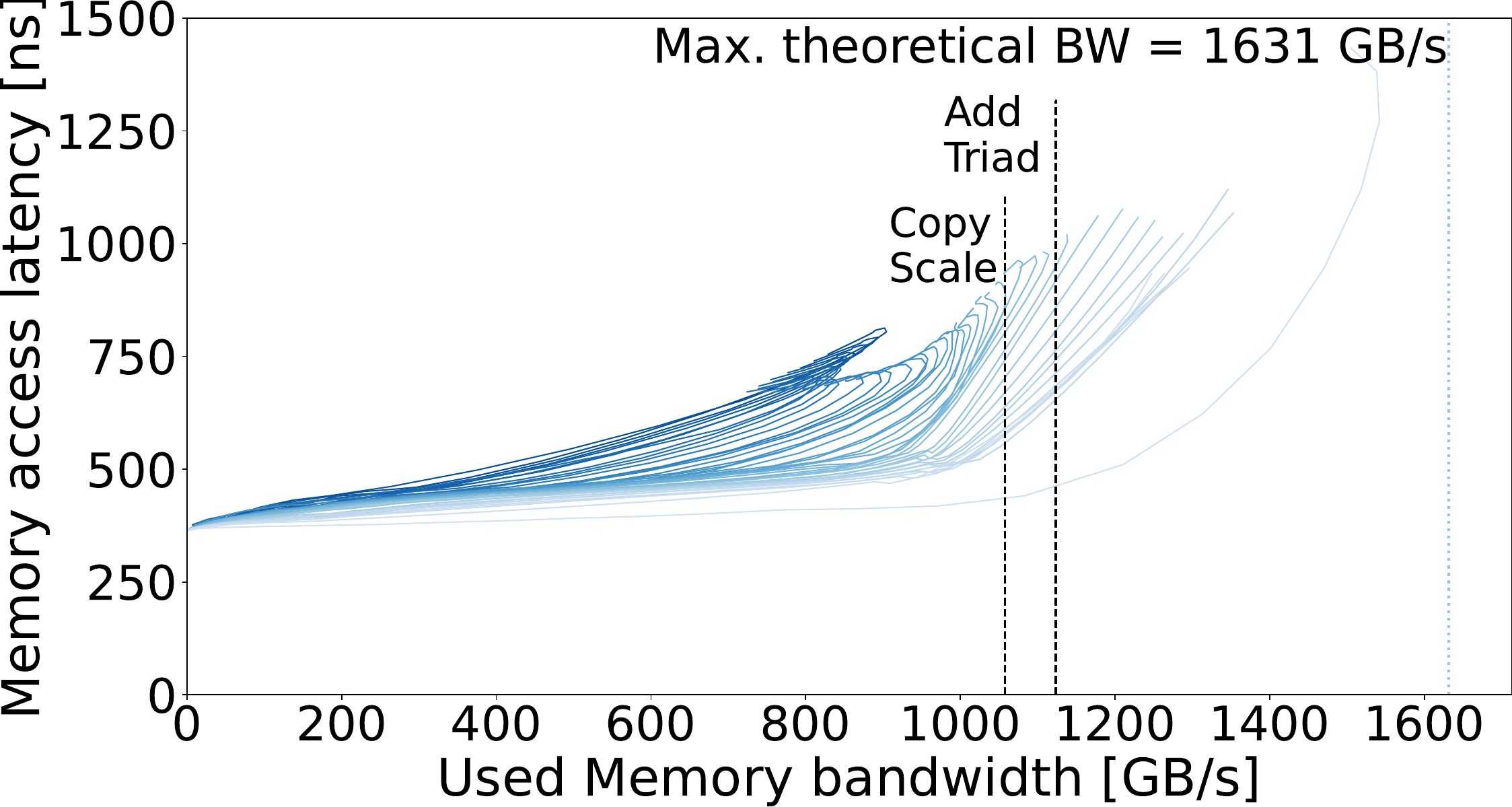}}%
\caption[A set of four subfigures.]{We detect a wide range of memory system behavior, even for the hardware platforms with the same memory standard.}
\label{fig:char-bw-lat-result}%
\end{figure}

We use the Mess measurements to compare the memory system performance of Intel, AMD, IBM, Fujitsu and Amazon servers as well as NVIDIA GPUs with DDR4, DDR5, HBM2 and HBM2E. The platform and memory system characteristics are listed in Table~\ref{tab:hpc-server-description} 
while Figure~\ref{fig:char-bw-lat-result} shows their bandwidth--latency curves. 
The \textbf{unloaded memory latency} varies significantly between different platforms. 
It ranges from 85\,ns in the Cascade~Lake server with DDR4, 
to 122\,ns in the A64FX servers with HBM2 and 
129\,ns in the Graviton~3 with DDR5 main memory. 
This difference should not be directly associated to the main memory. 
For example, the AMD~Zen2 comprises DDR4 technology with practically the same command latencies (in nanoseconds) as the
Intel Cascade~Lake server, and still it shows almost 30\,ns higher unloaded memory latency. 
This is because the load-to-use latency considers the time memory requests spend within the CPU chip, including the cache hierarchy and network on chip. These timings can differ significantly between different CPU architectures. 
We detect the highest unloaded memory latencies in the chips with the largest number of cores (Zen2, A64FX, Graviton~3) indicating that at least a portion of this latency is likely to be attributed to the network on chip which is larger and more complex in these architectures. NVIDIA H100 GPU shows higher unloaded latency due to its massive number of arithmetic processing units, lower on-chip frequency, and complex memory hierarchy~\cite{Mei:P-chaseGPU,Hestness:gpuVsCPU}.

We also detect a wide \textbf{maximum latency range} between different platforms, and even within a single platform for different read and write memory traffic. Maximum memory latency is a primary concern in real-time systems, and it is less critical in high-performance computing~(HPC). Still it has to be bound to guarantee quality of service of some HPC applications.  
We believe that our study may open a discussion about the sources of different maximum latencies in different systems, e.g. due to the different lengths of the memory queues, and about its desirable limits. 

\looseness -1 All platforms except AMD~Zen2 CPU and NVIDIA~H100 show a similar \textbf{saturated bandwidth range}, between approximately 70\% and 90\% of the maximum theoretical bandwidth. 
%
The maximum achieved bandwidth cannot reach the theoretical one because part of it is 
``lost'' due to factors such as DRAM refresh cycles blocking the entire chip, page misses causing precharge and activate cycles, and timing restrictions at the bank, rank, and channel levels~\cite{Eyerman:bwLatStack}. 
The efficiency of the memory bandwidth utilization also depends on the memory controller design. 
As expected (see Section~\ref{sec:Mess-metrics}), the best utilization is achieved for 100\%-read memory traffic, and it reduces with the increment of the memory writes. 
AMD~Zen2 is an exception in two ways. 
First, its saturated bandwidth range is significantly lower, 57--71\% of the maximum theoretical one. 
Second, it does not follow the expected impact of the write traffic on the bandwidth utilization. 
The traffic with the maximum rate of memory writes shows a very good performance, very close to the 100\%-read traffic, 
while the main drop is detected for a mixed, e.g. 60\%-read/40\%-write, traffic.

\rev{In some AMD Zen\,2, Intel Skylake and Intel Cascade Lake bandwidth--latency curves, 
increase in the Mess memory access rate, after some point, leads to decline in the measured bandwidth.\footnote{These findings are consistent with two recent studies which report that running high-bandwidth benchmarks on all CPU cores may lead to lower memory bandwidths w.r.t. the experiment in which some cores are not used~\cite{Peng:camp,Velten:AMDPerformanceUnderPopulation}.} 
In Amazon Graviton\,3, Intel Sapphire Rapids and NVIDIA H100 this behavior is frequent 
for the memory traffic with high percent of writes.
We explored these findings in the Cascade Lake servers in which we had access to the row-buffer hardware counters.  
In the experiments that experienced the bandwidth decline, we detect a significant increase in row-buffer misses. 
In case of a row-buffer miss, the current content of the row is stored in the memory array 
and the correct row is loaded into the row-buffer. 
These additional operations increase the memory access time and reduce the effective device bandwidth. 
To confirm these findings,
we increased the memory system pressure by removing four out of six DIMMs in each socket.  
In these experiments, we detected a large ``wave-form'' segments in all bandwidth--latency curves. 
And indeed, the measurement confirmed that the measured bandwidth drop is highly correlated with the row-buffer miss rate.}

\looseness -1 
\rev{The results also show the \textbf{difference between the STREAM and Mess measurements}. 
STREAM benchmark reports four data-points: the maximum sustained bandwidth of Copy, Scale, Add and Triad kernels.
Mess provides a wide range of measurements for different memory traffic intensity and read/write ratios. 
STREAM measures the application-level data traffic estimated based on the application execution time, size of the data structures, and number of load and store instruction in each STREAM kernel.  
%
Mess measures the architecture-level memory bandwidth measured with hardware counters. 
This includes all the application data traffic, but also all the microarchitecture data transfers.  
For example, STREAM bandwidth calculation assumes one memory read for each load instruction and one memory write for each store.
In state-of-the-art HPC servers with write-allocate cache policy, each store instruction does not correspond to a single memory write, but to one read and one write (see Section~\ref{sec:BW-lat_curves}). 
For this reason, Mess benchmark reports higher maximum measured bandwidths. 
This difference is clearly visible in IBM~Power\,9, A64FX and all Intel platforms. 
In Graviton\,3 and NVIDIA H100, the results reported by STREAM are very close to the maximum Mess measurements for the corresponding read/write ratio. This would correspond to a architecture with a write-through cache policy.
The second cause of the difference between the maximum Mess and STREAM bandwidths is the memory traffic composition. 
The Mess benchmark achieves the maximum bandwidth for a 100\%-read traffic. 
The write memory traffic, present in all STREAM kernels, adds timing constrains and reaches sooner the bandwidth saturation point.
Our results show that different platforms show distinct relation between the application-level STREAM 
and architecture-level Mess measurements.   
Therefore, we believe that the memory bandwidth analysis should include both approaches. 
As a part of future work, we plan to compare the STREAM and Mess measurements in 
high-end architectures that dynamically adjust the level of write-allocate traffic~\cite{laukemann:cloverleaf,papazian:IceLake}.}

\section{Performance characterization: Memory simulators}
\label{sec:mess-characterization-simulator}

Mess benchmark can also be used to characterize memory simulators and compare them with the actual systems they intent to model. 
We illustrate this Mess capability with the gem5, ZSim, and OpenPiton Metro-MPI simulators with different internal memory models and widely-used external memory simulators, DRAMsim3, Ramulator and Ramulator~2. 

\subsection{gem5}
\label{sec:gem5-characterization}
\begin{figure}[!t]%
\centering
\includegraphics[width=.8\linewidth]{graphics/horizontal-legend}\\
\vspace{-.15cm}
\subfigure[][\fontsize{7.7}{0}\selectfont Amazon\,Graviton3:\,8$\times$DDR5-4800]{%
\label{fig:gem5-bw-lat-result-characterization-hardware}%
\hspace{-0.1cm}\includegraphics[width=.5\linewidth]{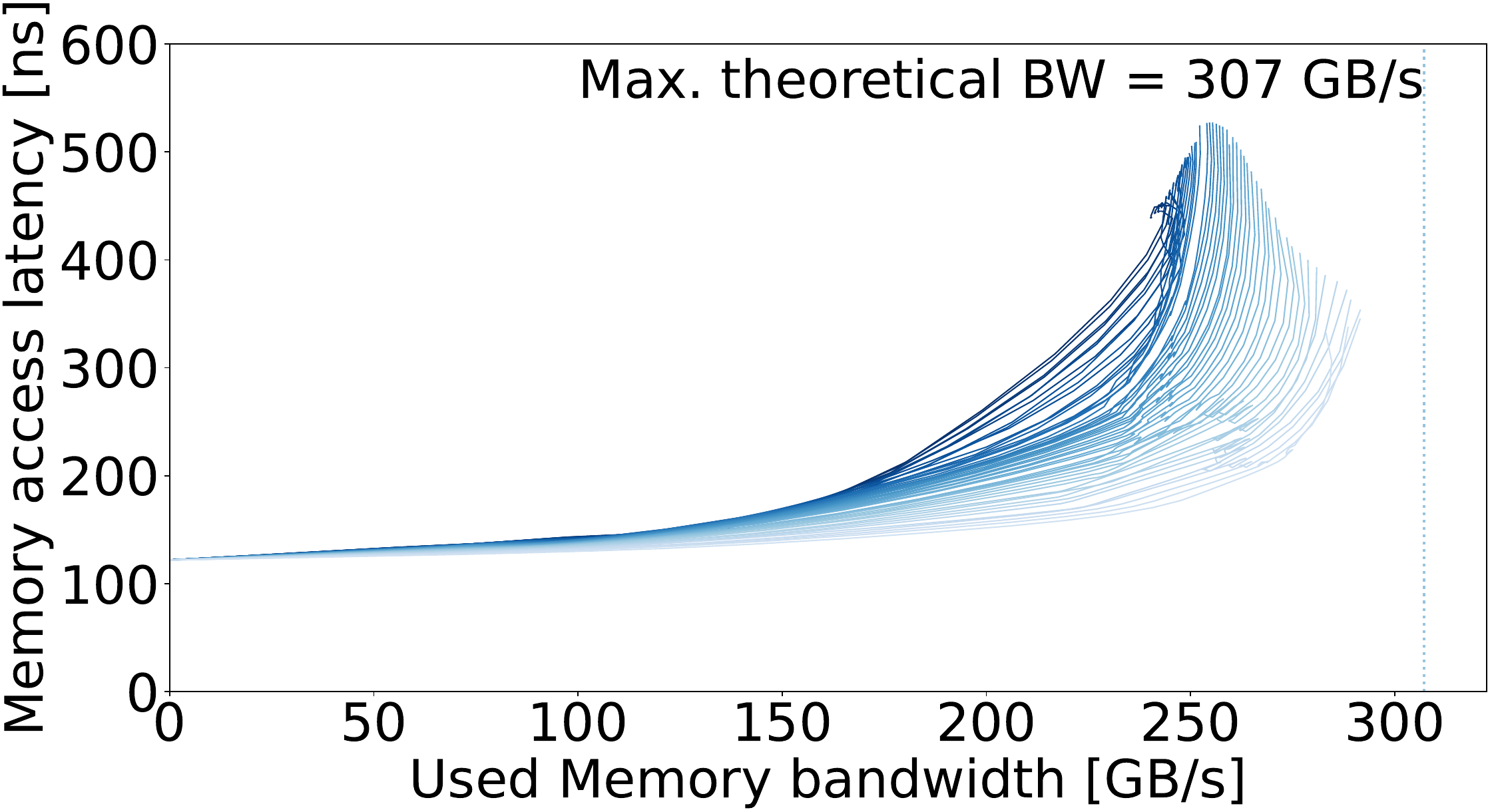}}~%
\subfigure[][gem5: Simple memory model]{%
\label{fig:}%
\includegraphics[width=.5\linewidth]{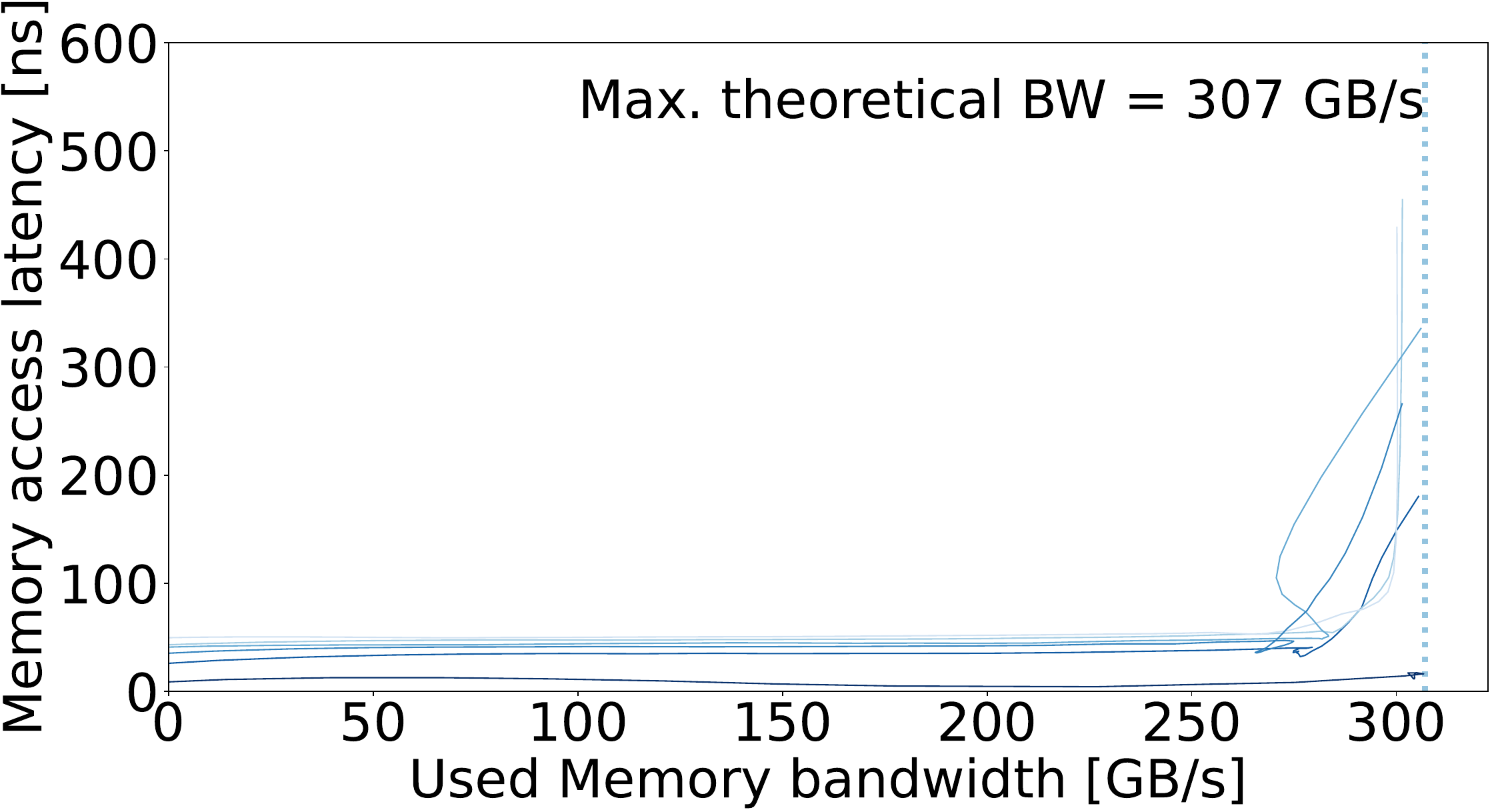}}\\%
\subfigure[][gem5: Internal DDR model]{%
\label{fig:}%
\includegraphics[width=.5\linewidth]{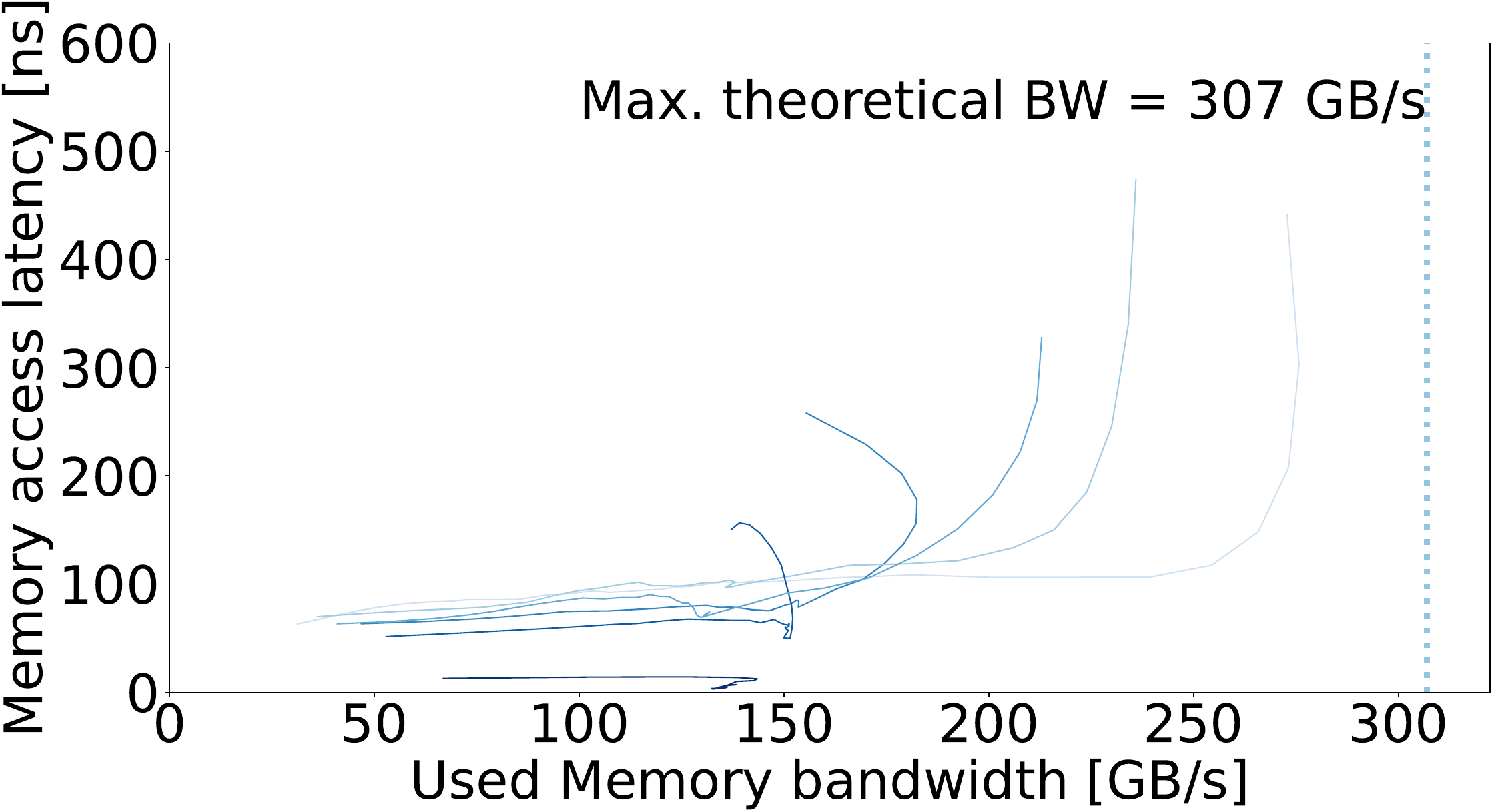}}~%
\subfigure[][gem5+Ramulator2]{%
\label{fig:gem5-bw-lat-result-characterization-ramulator2}%
\includegraphics[width=.5\linewidth]{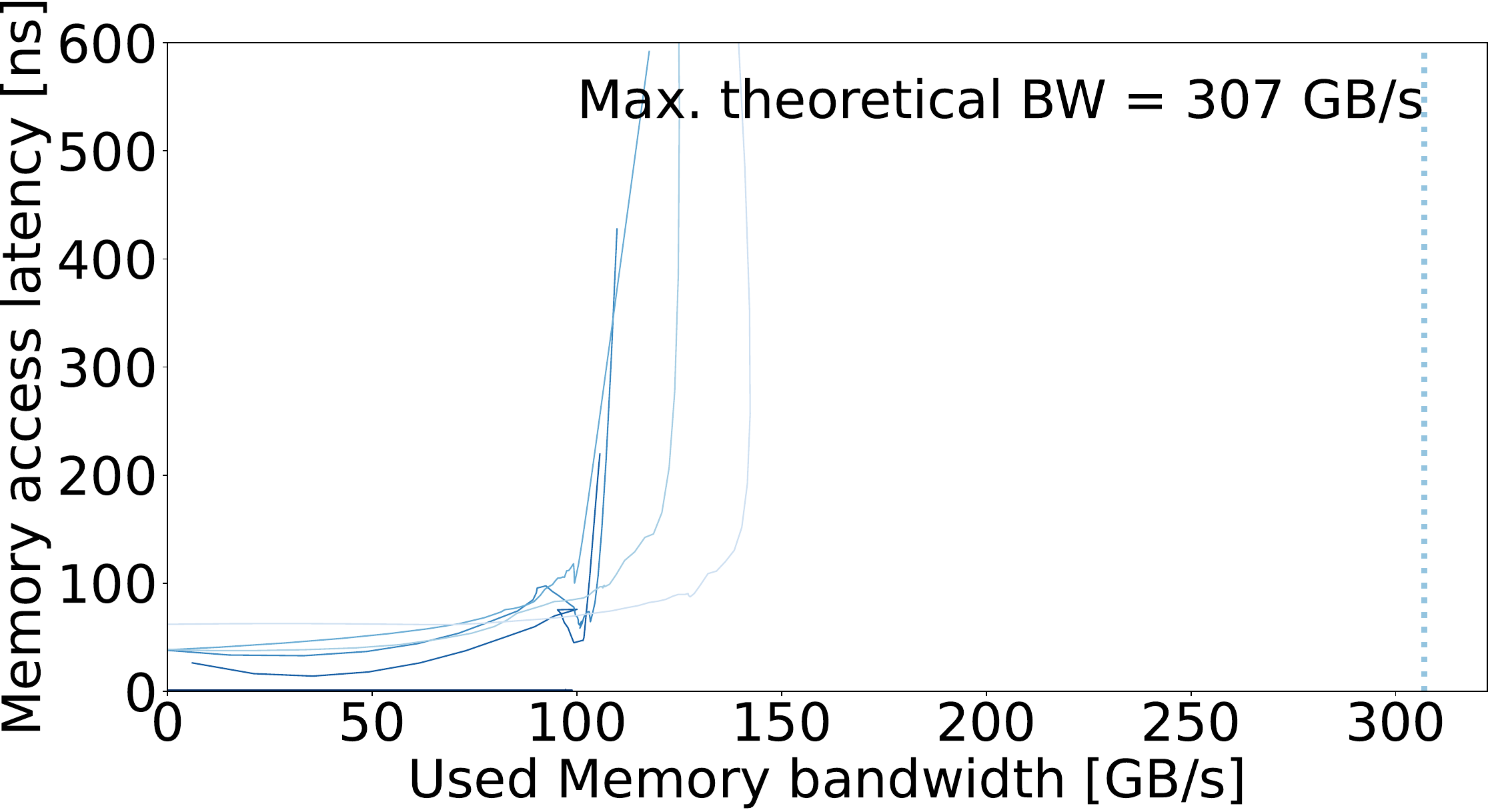}}\\%
\caption[A set of four subfigures.]{Memory performance: Amazon Graviton3 server vs. gem5 memory models.}
\label{fig:gem5-bw-lat-result-characterization}%
\end{figure}

The gem5~\cite{Jason:gem5} is a cycle-accurate full-system simulator.
In our experiments, the simulator is configured to model the Graviton~3 server with 64 Neoverse~N1 cores~\cite{Amazon:Graviton3}.
The cache hierarchy includes 64\,KB of 4-way L1 instruction and data cache, 1\,MB of 8-way private L2 cache and 64\,MB of 16-way shared L3.
The main memory system has eight DDR5-4800 memory channels. 
%
Figure~\ref{fig:gem5-bw-lat-result-characterization} compares Mess bandwidth--latency curves of the actual server 
with \textbf{gem5 simple memory model}, more complex \textbf{gem5 internal DDR model} and gem5 connected to \textbf{Ramulator~2}. 
To maintain a reasonable simulation time, we model each system with a family of six curves, from 50\% to 100\% read memory traffic with a 10\% step.

\looseness -1 Practically in the whole bandwidth range, 
the gem5 \textbf{simple memory model} delivers a fixed latency of 4--49\,ns. 
The latency increases only when the bandwidth asymptotically approaches its theoretical maximum. 
Contrary to the Graviton~3 server, the highest latencies are measured for a 100\%-read traffic, and the latency drops with the percent of memory writes. 
Also, unlike in the actual platform, for some memory traffic, increasing the bandwidth reduces the memory access latency. 
For example, the 50\%-read/50\%-write traffic reaches the lowest simulated latency of only 4\,ns at the 200\,GB/s bandwidth. 
The same traffic in the actual system has the memory access latency of 261\,ns. 
 
\looseness -1  The more detailed \textbf{internal DDR model} shows small improvements over the simple memory model, 
but still poorly resembles the actual system performance. 
The simulated latencies are unrealistically low, most of them in the range of 14--100\,ns. 
Similarly to the gem5 simple memory model, the latencies drops with the percent of memory writes. 
For all the curves except 100\%-read, the saturated bandwidth is significantly lower from the one measured on the actual system. Again, the error increases with the percent of memory writes. 

\looseness -1  \rev{As the internal gem5 memory models, \textbf{gem5+Ramulator~2} simulates unrealistically low memory latencies and the error increases with the ratio of memory writes. 
In addition to this, the curves experience a sharp, nearly vertical rise between 100\,GB/s and 130\,GB/s, 
which is less than a half of the actual measured bandwidth. 
Surprisingly, the most complex and trusted memory model shows the highest simulation error. Some sources of this error will be analyzed in 
Section~\ref{sec:mem-simulation-error}.}

\subsection{ZSim}
\label{sec:ZSim-characterization}

\looseness -1 We select ZSim~\cite{sanchez:zsim} as a representative of event-based hardware simulators. 
We use publicly-available ZSim modeling 24-core Intel Skylake processor connected to six DDR4-2666 channels~\cite{Simulator:infrastructureLink}. The cache hierarchy of the modeled CPU includes 64\,KB of 8-way L1 instruction and data cache, 1\,MB of 16-way private L2 cache and 33\,MB of 11-way shared L3.
The simulator is extensively evaluated against the actual hardware platform~\cite{esmaili:acm}. 
The ZSim comprises three internal memory models: fixed-latency, M/D/1 queue model and the internal DDR model. 
Also it is already connected to Ramulator~\cite{kim:ramulator} and DRAMsim3~\cite{Shangli:dramsim3}. 
To avoid any simulator integration error, we use the ZSim+DRAMSim3 released by the University of Maryland (DRAMSim3 developers)~\cite{DRAMsim3:repo}, and ZSim+Ramulator from The SAFARI Research Group at the ETH University~(Ramulator developers)~\cite{Oliveira:amov-ZSim-ramulator}.
%

\begin{figure}[!t]%
\centering
\includegraphics[width=.8\linewidth]{graphics/horizontal-legend}\\
\vspace{-.15cm}
\hspace{-0.1cm}\subfigure[][Intel Skylake with 6$\times$DDR4-2666.]{%
\label{fig:simulators-intel-skylake-curves}%
\includegraphics[width=.5\linewidth]{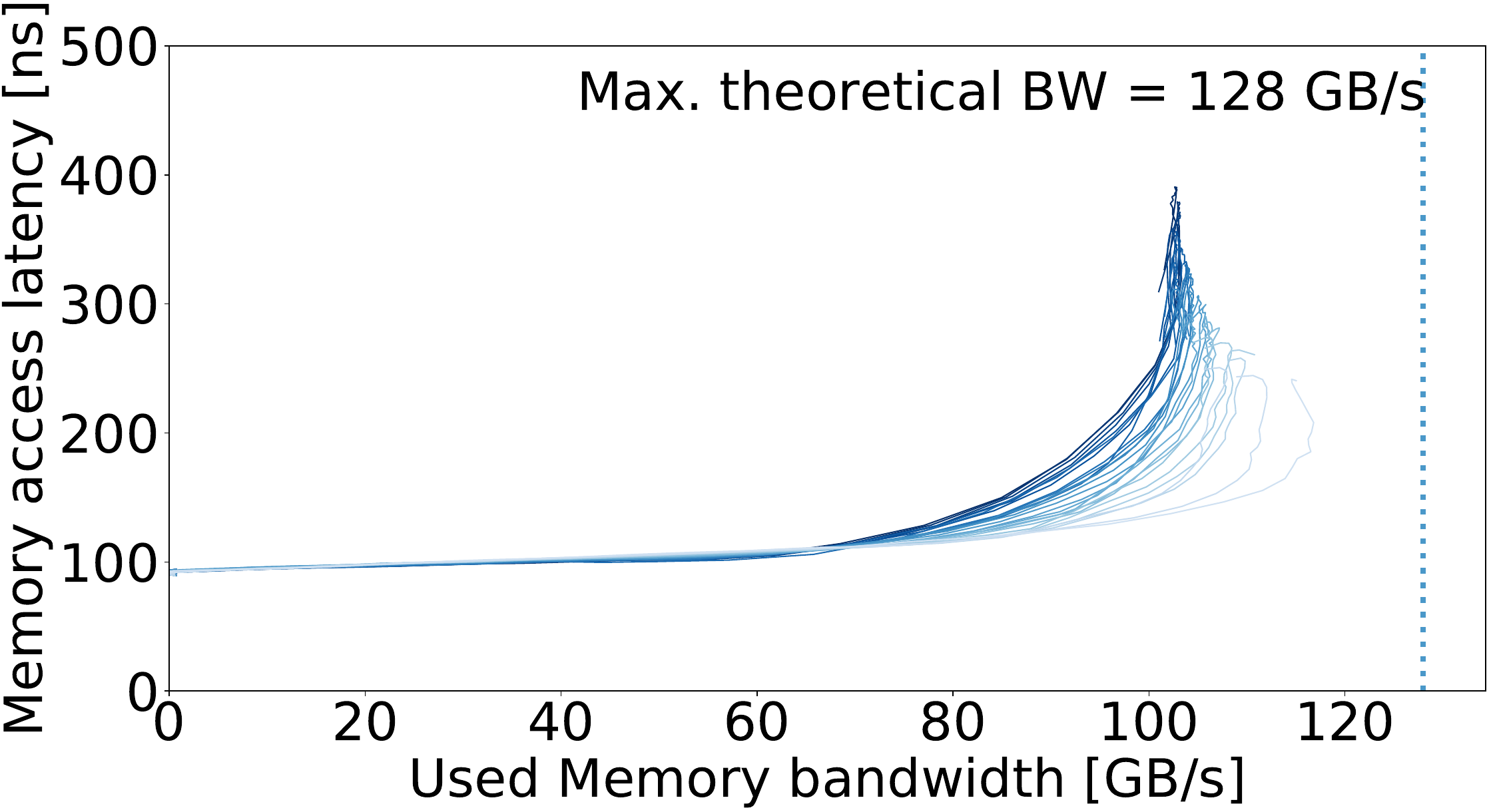}}~%
\subfigure[][ZSim: Fixed-latency model]{%
\label{fig:}%
\includegraphics[width=.5\linewidth]{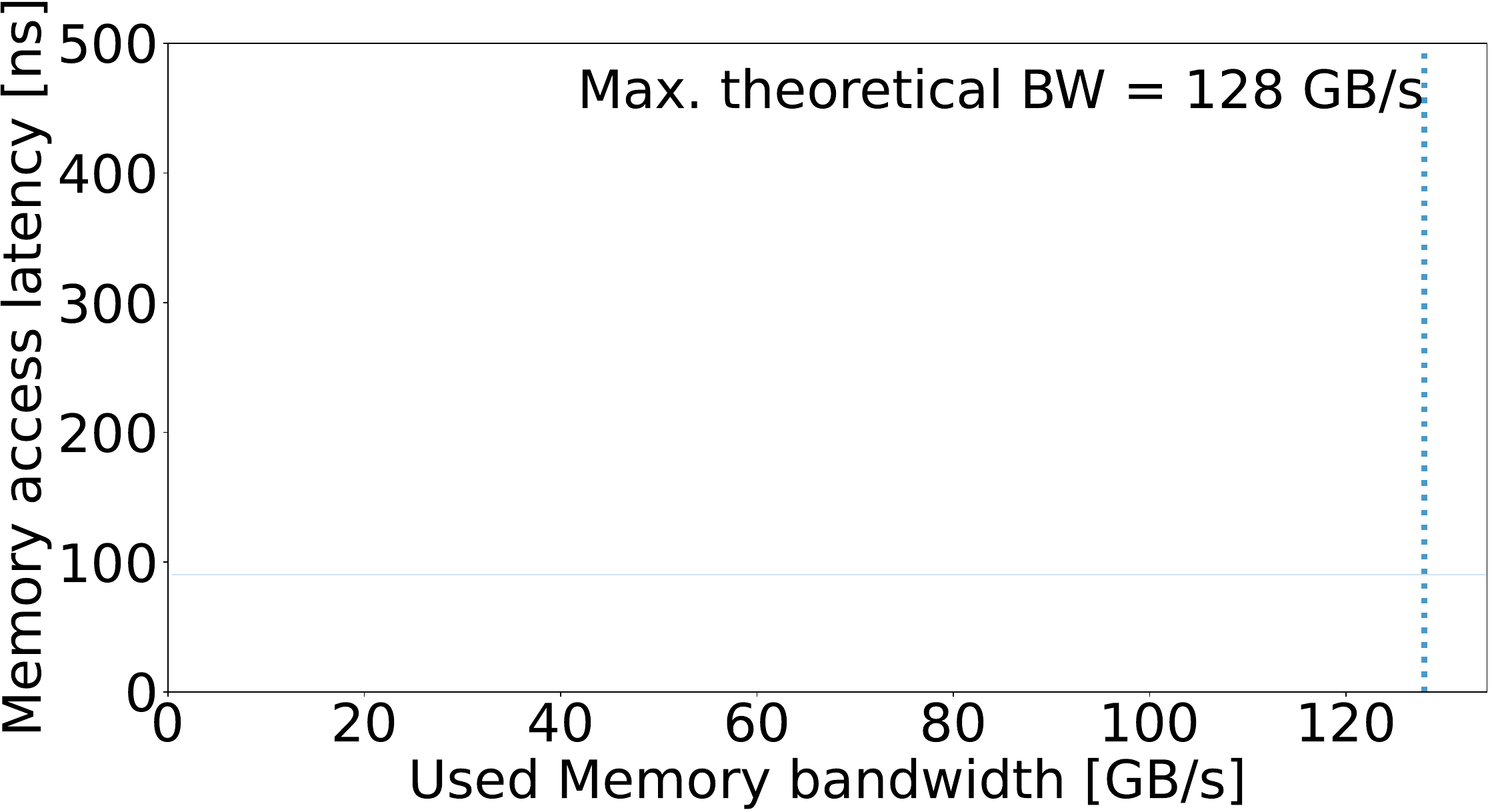}}\\
\hspace{-0.1cm}\subfigure[][ZSim: M/D/1 Queue model]{%
\label{fig:}%
\includegraphics[width=.5\linewidth]{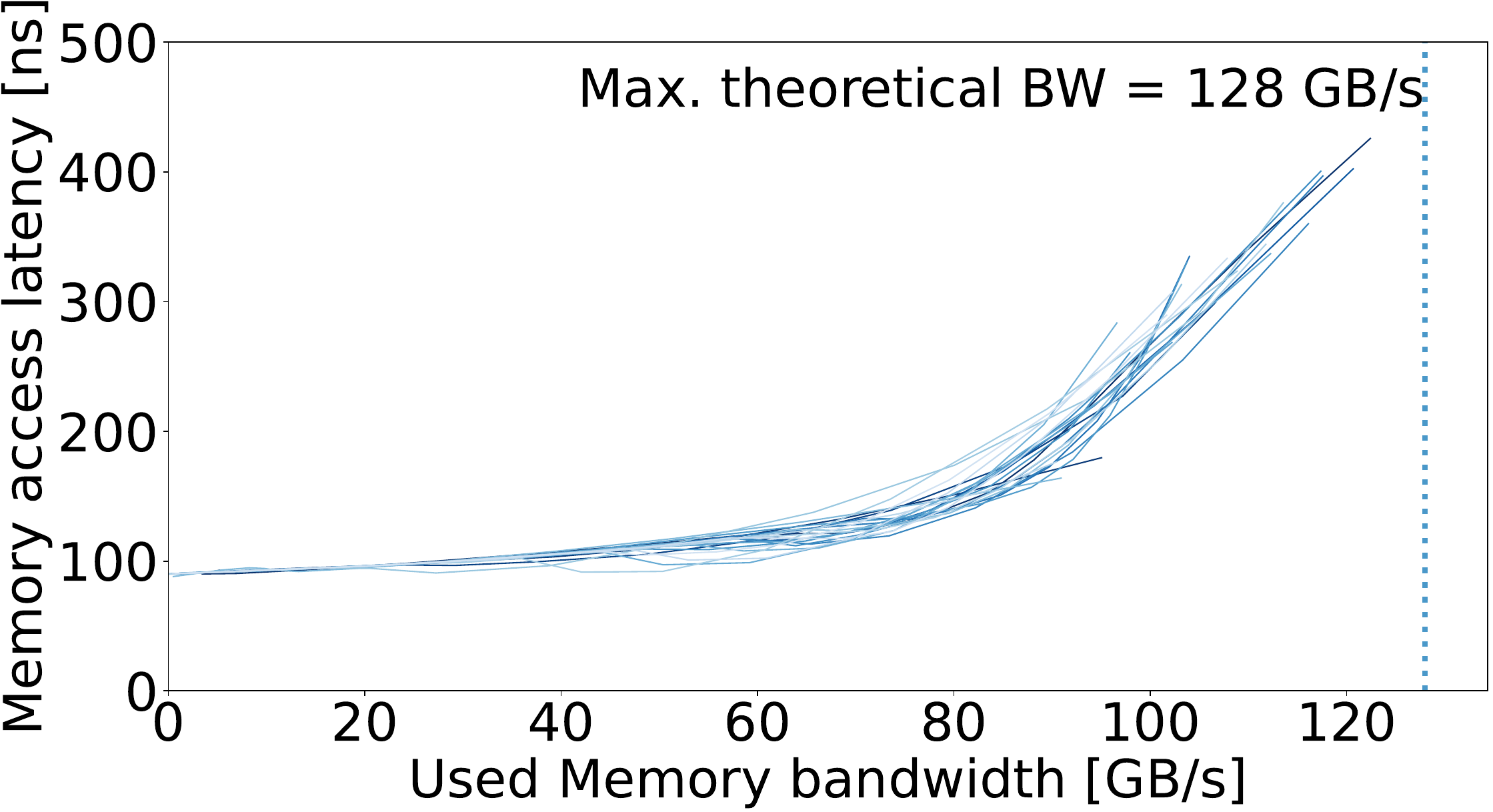}}~%
\subfigure[][ZSim: Internal DDR model]{%
\label{fig:}%
\includegraphics[width=.5\linewidth]{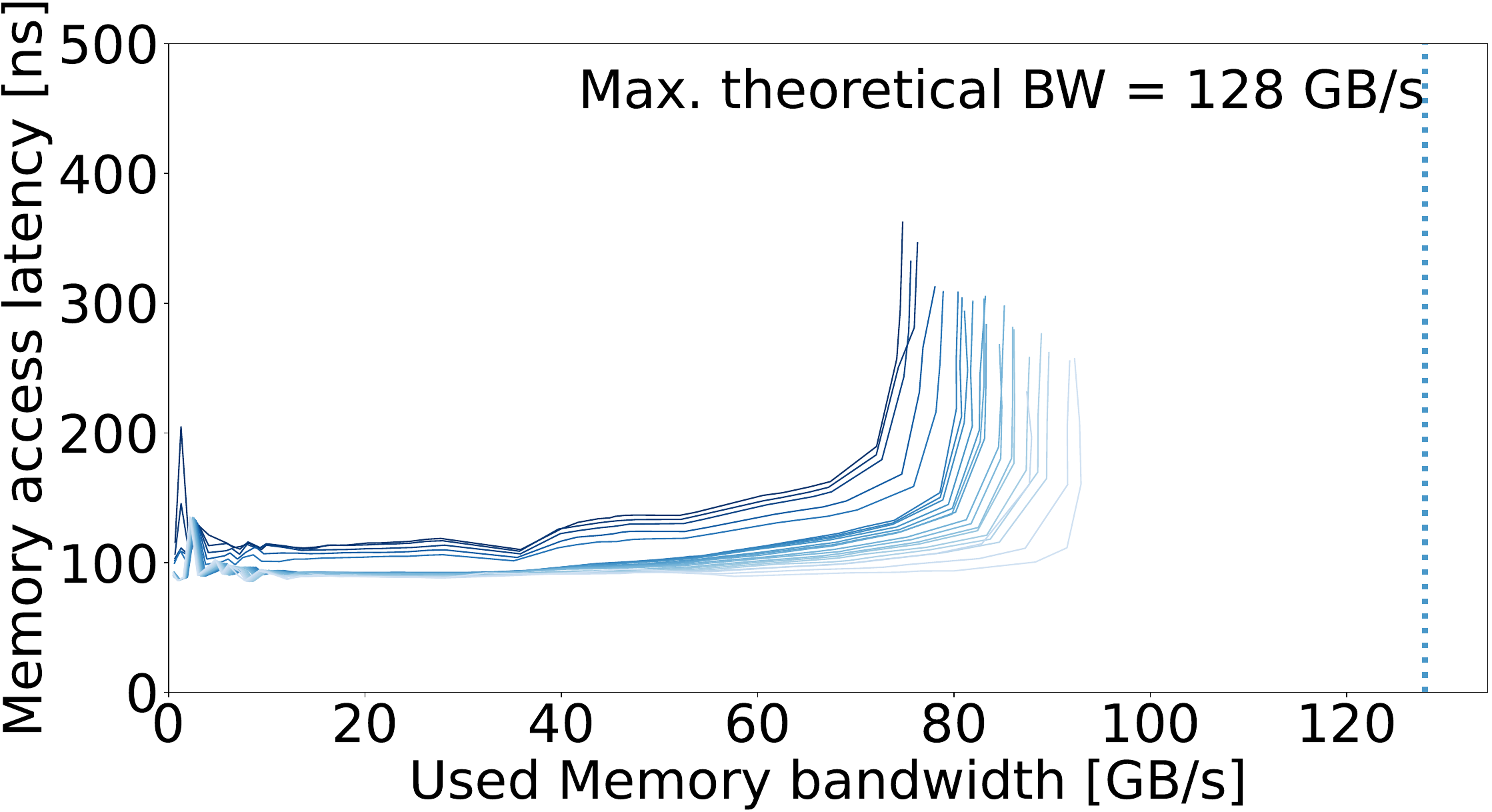}}\\
\hspace{-0.1cm}\subfigure[][ZSim+DRAMsim3 simulator.]{%
\label{fig:}%
\includegraphics[width=.5\linewidth]{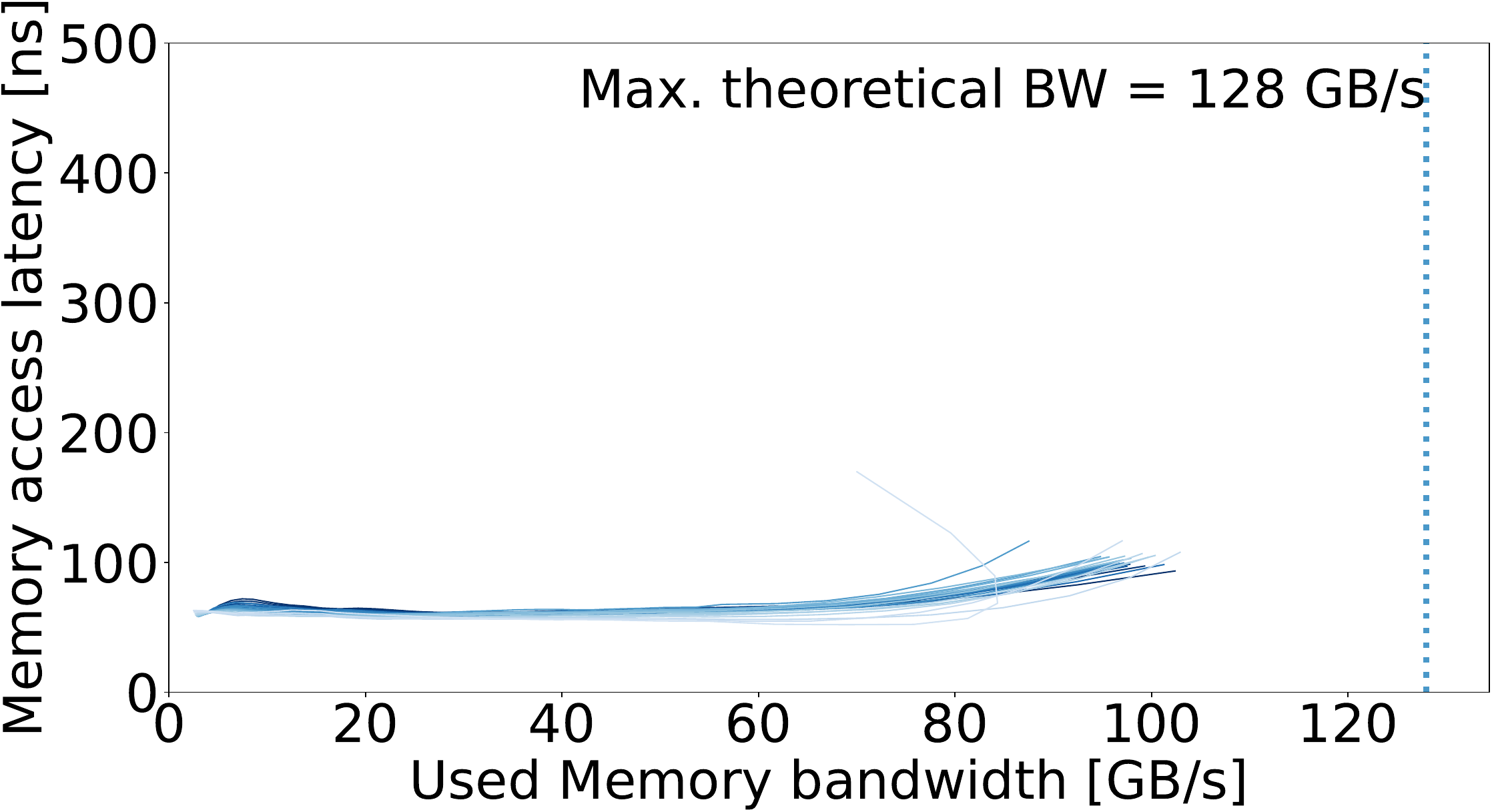}}~%
\subfigure[][ZSim+Ramulator simulator.]{%
\label{fig:}%
\includegraphics[width=.5\linewidth]{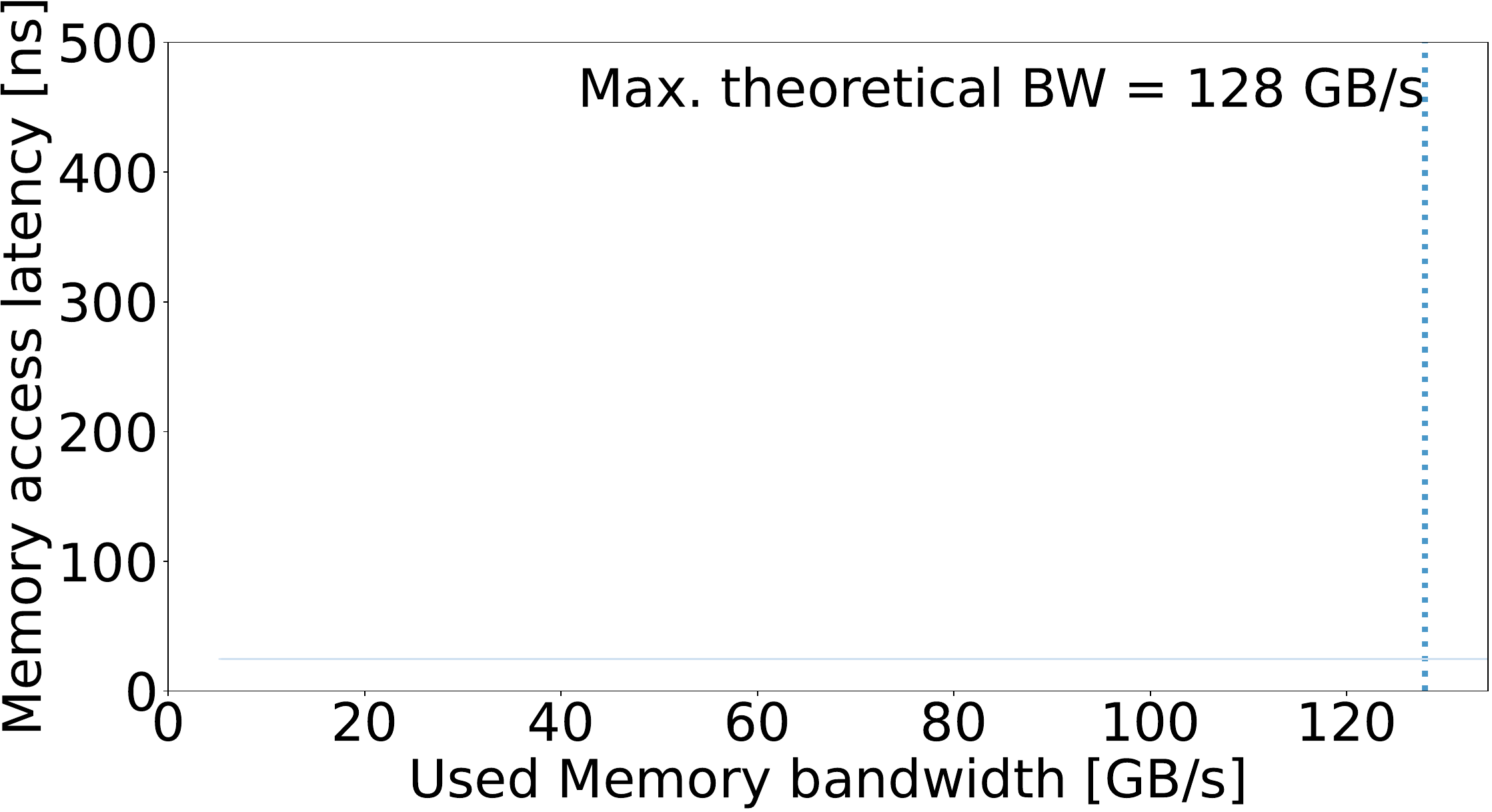}}\\%
\caption[A set of four subfigures.]{Memory performance: Intel Skylake server vs. ZSim memory models.}
\label{fig:zsim-bw-lat-result-characterization}%
\end{figure}

Figure~\ref{fig:zsim-bw-lat-result-characterization} compares the Mess bandwidth--latency curves of the actual server with \textbf{all five ZSim memory simulation approaches}. 
%
%
As expected, the \textbf{fixed-latency memory model} provides a constant latency in the whole bandwidth domain. 
Given that this latency is configured by a user, it can be set to match the unloaded memory latency in the actual system.  
On the down side, the memory bandwidth provided by this model is unrealistic:  
the maximum simulated bandwidth is 342\,GB/s, which exceeds the maximum theoretical one by 2.7$\times$. 
The \textbf{M/D/1 queues} correctly model the memory system behavior in the linear part of the curves. 
The modeling of the system saturation is less accurate. 
The queue model does show some difference between read and write memory traffic, 
but the reported performance does not correspond to the actual system trend in which increasing the write traffic lowers the performance. 
The \textbf{internal DDR model} 
correctly emulates the linear and saturated segments of the curves and the impact of the memory writes. 
However, the simulator underestimates the saturated bandwidth area to 69--93\,GB/s, 
significantly below the 92--116\,GB/s measured in the actual system. 
Also, the simulator excessively penalizes the memory writes which is seen as a wider spread of the curves with a higher write memory traffic. 
Finally, we detect some unrealistic memory-latency peaks in the low-bandwidth 1--4\,GB/s curve segments. 
The \textbf{DRAMsim3} shows a similar trend as the M/D/1 queue model in the linear segments of the memory curves,  
with some latency error, 52--63\,ns in the DRAMsim3 versus 89--109\,ns in the actual system.   
The simulator does not model the saturated bandwidth area. 
Finally, the \textbf{Ramulator} provides a fixed 25\,ns latency in the whole bandwidth area and for all memory traffic configurations. 
Also, similar to the fixed-latency model, the simulated bandwidth is unrealistic, exceeding by 1.8$\times$ the maximum theoretical one. 

Our evaluation of the memory models and detailed hardware simulators detected major discrepancies w.r.t. the actual memory systems performance. 
DRAMsim3, Ramulator and Ramulator~2 are considered \emph{de facto} standard for the memory system simulation. 
All of them are evaluated against the manufacturer's Verilog model and they show no violation of the JEDEC timings~\cite{jedec:ddr3,jedec:ddr4}. 
\rev{DRAMsim3 is evaluated for DDR3 and DDR4~\cite{Shangli:dramsim3}, Ramulator for DDR3~\cite{kim:ramulator}, and Ramulator~2 for DDR4~\cite{Haocong:Ramulator2}.}  
However, as our results demonstrate, this does not guarantee that the simulators properly model the memory system performance. 
\rev{In Section~\ref{sec:mem-simulation-error}, we will use Mess benchmark to analyze some causes of these discrepancies.}

\subsection{RTL simulators: OpenPiton and Metro-MPI}
\label{sec:characterization-openpiton}
OpenPiton framework~\cite{Balkind:OpenPiton} provides an open-source RTL implementation of a tiled architecture based on Ariane RISC-V cores~\cite{Balkind:OpenPitonandAriane,Leyva:OpenPitonupdateBSC,Balkind:OpenPitonandAriane2}. 
Developed in the Verilog RTL, the OpenPiton simulation is slow, especially for large number of cores. 
We use the OpenPiton simulation accelerated by Metro-MPI~\cite{Lopez-Paradis:Metro-Mpi}. 
This approach uses Verilator~\cite{Snyder:Verilator} to convert the RTL code of each tile into a cycle-accurate \texttt{C++} simulation model. 
Then, all the tiles are simulated in parallel and their interconnect communication is done with the MPI programming interface. 

In our experiments, the OpenPiton framework is configured to generate 64-core Ariane architecture which includes 16\,KB of 4-way L1 instruction and data cache, and 4\,MB of 4-way shared L2 cache.
The main memory is originally modeled with a single-cycle latency, and it is recently extended with a fixed-latency model~\cite{Leyva:OpenPitonupdateBSC}. 
Our Mess measurements 
confirm that both models deliver the expected load-to-use latency.  
Also, as expected, we see no difference between read and write memory traffic, leading to a perfect overlap of the curves. 
The only difference is in the maximum observed memory bandwidth. 
For a single-cycle memory latency, 100\%-read memory traffic achieves 32\,GB/s, limited by the memory concurrency of the 64 in-order Ariane cores. 
Memory writes do not stall the cores, so the achieved memory bandwidth increases with the write memory traffic ratio. 
Still, a small 2-entry miss status holding registers~(MSHR) limits the memory bandwidth to 47\,GB/s for 50\%-read/50\%-write traffic.  
We detect the same trend for the fixed memory model. 
 
\looseness -1 The Mess evaluation of the OpenPiton Metro-MPI resulted in an unexpected discovery: in some experiments we detected significantly higher memory write traffic than anticipated. 
By analyzing the system behavior for various Mess configurations,  
we connected the extra memory traffic to the unnecessary eviction of the data from the last-level cache. 
Instead of evicting only the dirty cache lines, the system was evicting all of them.  
The source of the error is the coherency protocol generated by the OpenPiton framework. 
The error was reported to the OpenPiton developers and they confirmed its existence. 


\subsection{Sources of memory simulation errors}
\label{sec:mem-simulation-error}

\begin{figure}[!t]%
\centering
\includegraphics[width=.8\linewidth]{graphics/horizontal-legend}\\
\vspace{-.15cm}
\subfigure[][Ramulator2: 8$\times$DDR5-4800]{%
\label{fig:trace-driven-ramulator2}%
\includegraphics[width=.5\linewidth]{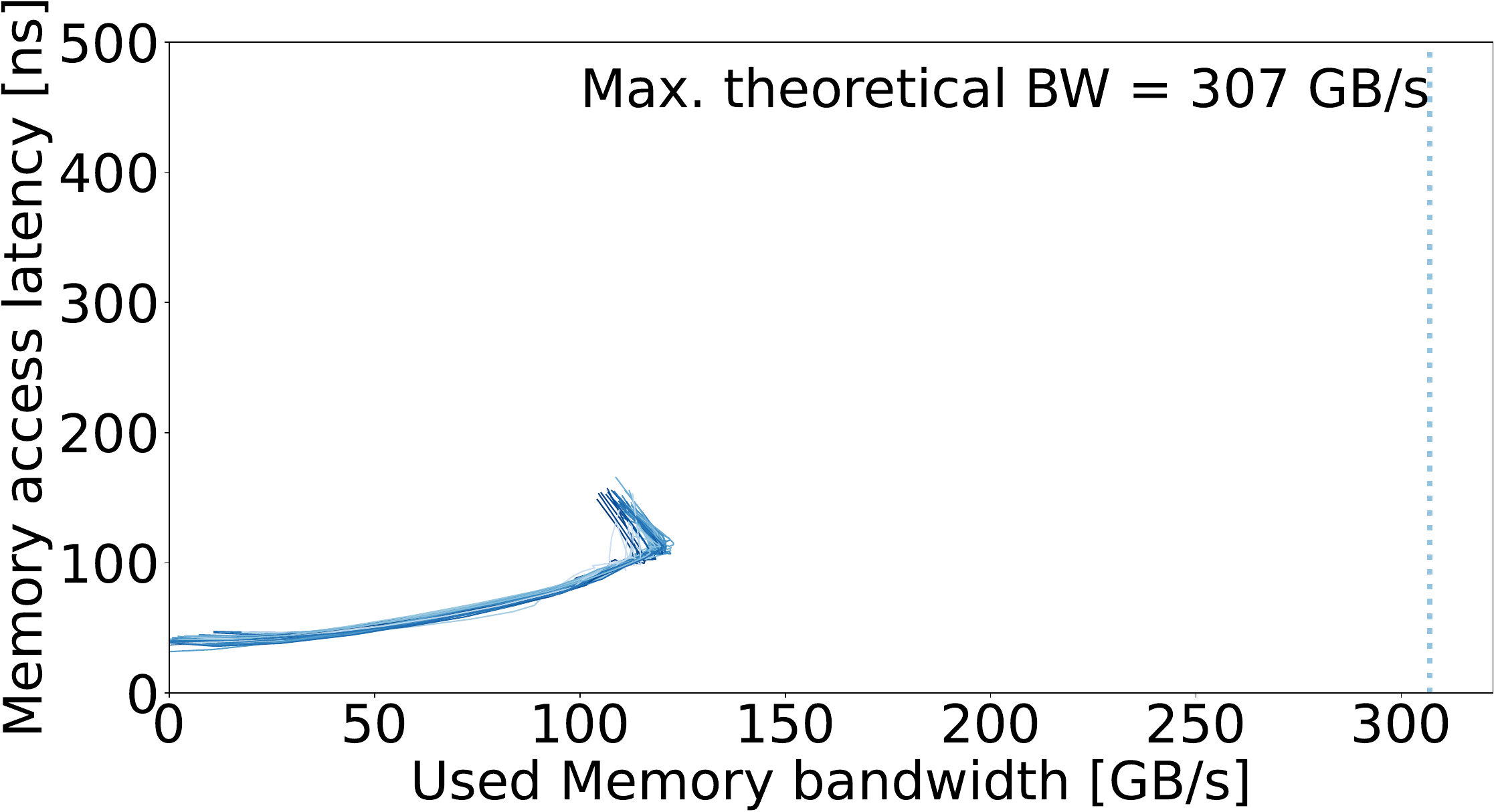}}\\
\hspace{-0.1cm}\subfigure[][DRAMsim3: 6$\times$DDR4-2666]{%
\label{fig:trace-driven-dramsim3}%
\includegraphics[width=.5\linewidth]{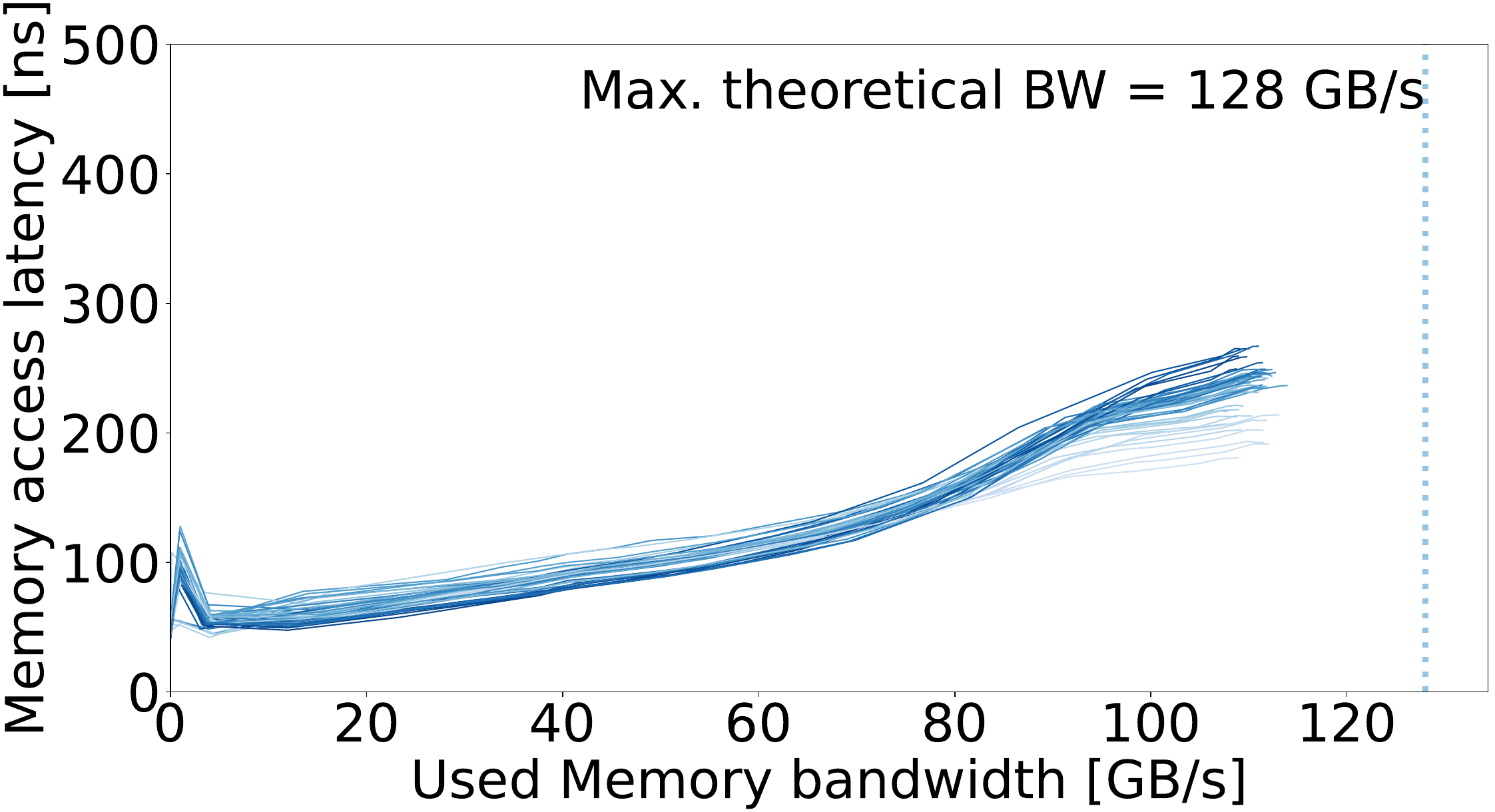}}~%
\subfigure[][Ramulator: 6$\times$DDR4-2666]{%
\label{fig:trace-driven-ramulator}%
\includegraphics[width=.5\linewidth]{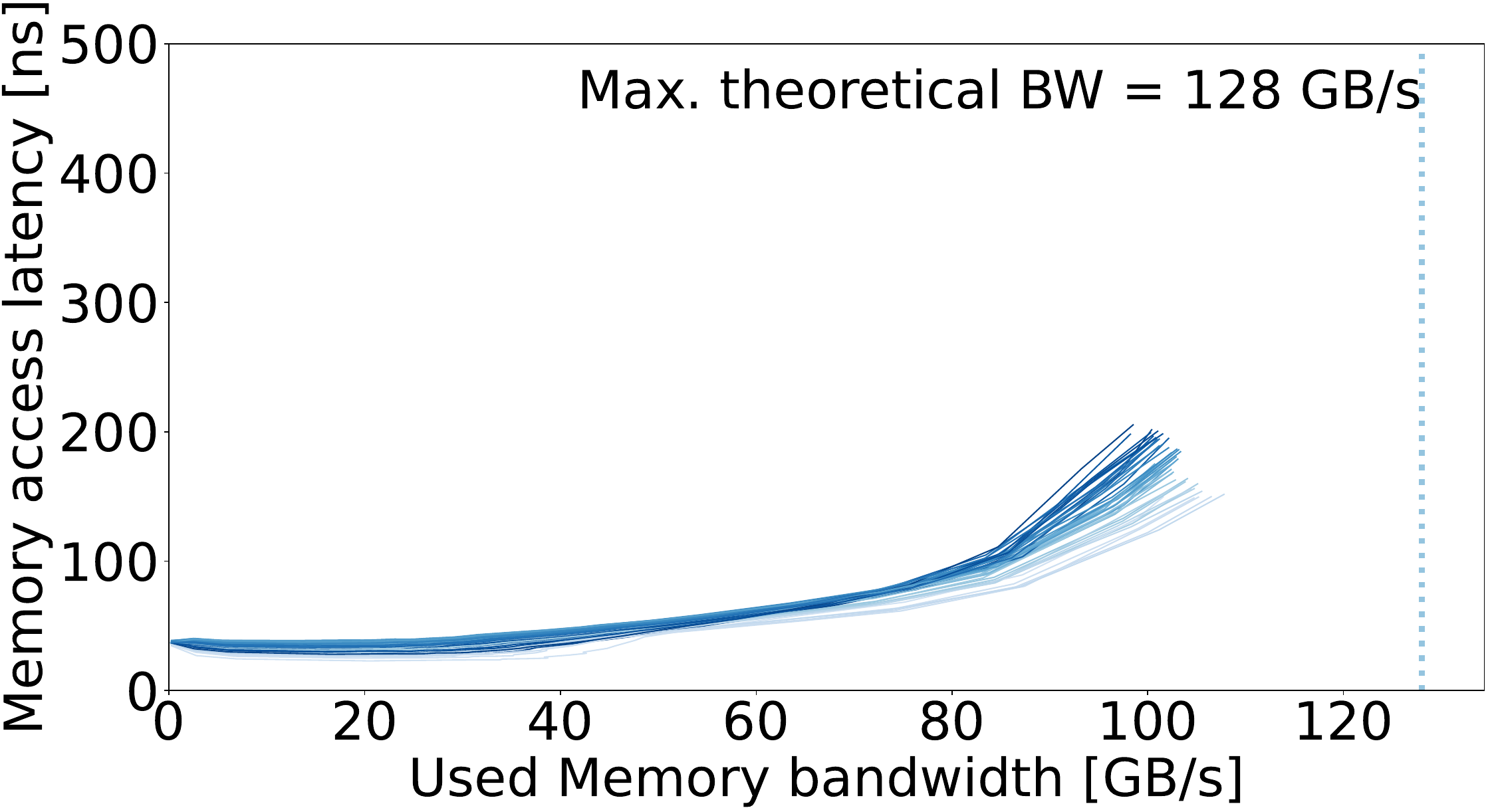}}\\
\caption[A set of four subfigures.]{Memory performance: Trace-driven cycle-accurate simulators.}
\label{fig:cycle-accurate-bw-lat-result-characterization}%
\end{figure}

%
\rev{To exclude any simulation error caused by the CPU simulators or their memory interfaces,   
we perform a \textbf{trace-driven DRAMsim3, Ramulator and Ramulator~2 simulation}. 
The detailed Mess memory traces are collected from its ZSim simulation, 
and they include the addresses of all memory read and write operations. 
To account for the timings of non-memory operations, 
DRAMsim3 traces contain simulation cycles in which the memory requests reach the memory controller, 
The Ramulator and Ramulator~2 traces include the number of non-memory instructions between the consecutive memory operations. 
The Mess traces may contain some timings errors w.r.t. actual execution, 
so the DRAMsim3, Ramulator, and Ramulator~2 simulations may not match the exact bandwidth--latency point.  
Still, the correct memory simulation should provide data-points that are located on the actual bandwidth--latency curves.
Figure~\ref{fig:cycle-accurate-bw-lat-result-characterization} shows the trace-driven Mess evaluation of  DRAMsim3, Ramulator and Ramulator~2. 
The charts report the round-trip memory access latency from the memory controller, 
so it is expected that the simulated curves are somewhat below the actual load-to-use measurements.}  

\rev{The trace-driven \textbf{Ramulator~2} show the same bandwidth--latency trends as the gem5+Ramulator~2 simulations 
(see Figure~\ref{fig:gem5-bw-lat-result-characterization-ramulator2}) . The simulated memory latency is unrealistically low and the maximum simulated memory bandwidth is only 126\,GB/s which is less than a half of the 292\,GB/s measured in the actual system. This indicates that the main source of the large simulation error is indeed Ramulator~2.}   


\rev{The conclusions are somewhat different for \textbf{DRAMsim3 and Ramulator} 
as their trace-driven memory bandwidth--latency curves show better general trends than the corresponding Zsim-driven simulations. 
This indicates that a part of the ZSim+DRAsim3 and ZSim+Ramulator simulation errors reported in Figure~\ref{fig:zsim-bw-lat-result-characterization} is 
caused by the simulators' interfaces.  
Our finding is aligned with the previous studies that report issues in the integration of the event-based CPU simulators with cycle-accurate memory models~\cite{sanchez:zsimEnhance, Shang:misscycle, Eyerman:IRMM}.} 

\rev{However, trace-driven DRAMsim3 and Ramulator also show important discrepancies w.r.t. actual bandwidth--latency curves.
\textbf{DRAMsim3 }simulated latency starts at 68\,ns. 
Apart from the peak at 5\,GB/s,
the latency increases linearly with the bandwidth.  
The curves for different read/write ratios are spread and 
intertwined in whole bandwidth range. 
Below 70\,GB/s the curves with the highest write traffic ratio have the lowest latency.   
We detect no bandwidth saturation, and all the curves 
linearly reach the maximum bandwidth of 113\,GB/s.    
\textbf{Ramulator} shows a better general trend: 
roughly-constant latencies below 40\,GB/s, 
a light latency increase until approximately 85\,GB/s, 
and a higher inclination in the final segments of the curves. 
Still, 
the latency is unrealistically low, starting at only 25\,ns (100\%-read traffic, 20--40\,GB/s), 
different read/write curves are spread in all bandwidths, 
and the saturated behavior differs significantly from the actual one. 
}
 
\begin{figure}[!t]%
\centering
\includegraphics[width=.85\linewidth]{graphics/row-buffer-legend}\\
\vspace{-.18cm}%
\subfigure[][100\%-read]{%
\label{fig:row-buffer-statistics-read}%
\hspace{-0.1cm}\includegraphics[width=.5\linewidth]{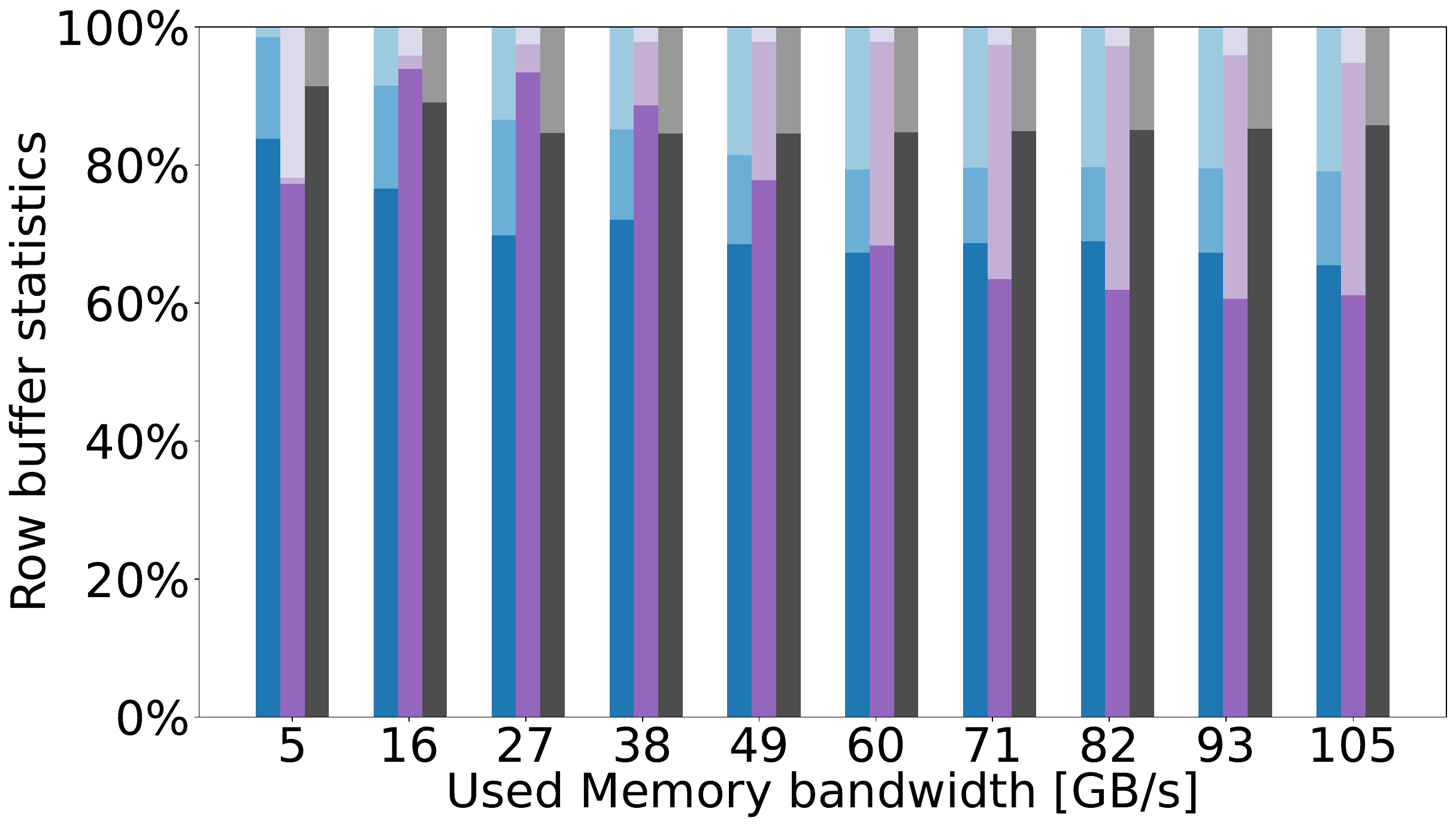}}~%
\subfigure[][50\%-read/50\%-write]{%
\label{fig:row-buffer-statistics-write}%
\includegraphics[width=.5\linewidth]{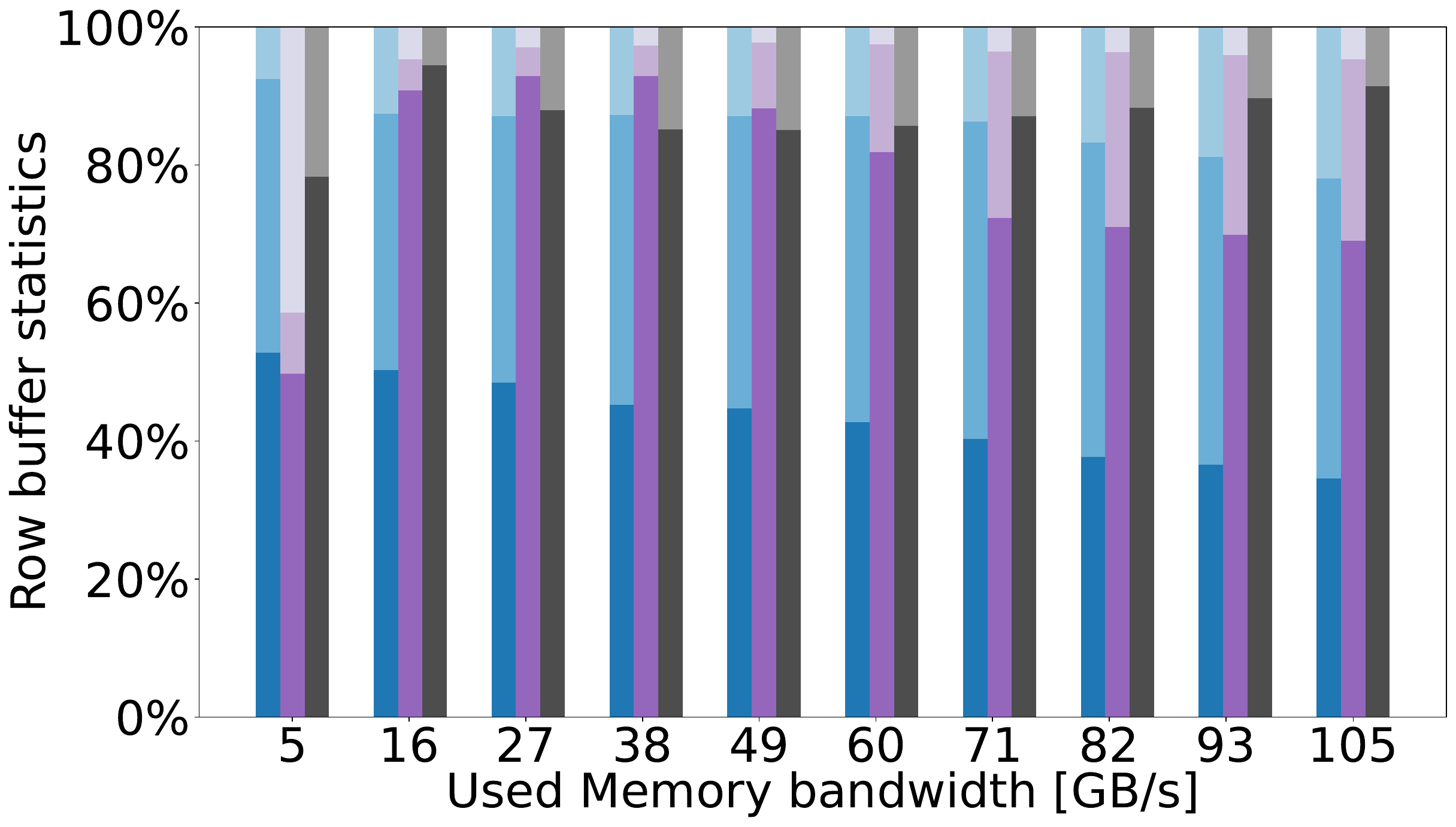}}\vspace{-.0cm}\\%
\caption[A set of four subfigures.]{Row buffer statistics: Actual hardware \emph{vs.} DRAMsim3 \emph{vs.} Ramulator.}
\label{fig:row-buffer-statistics}%
\end{figure}

\rev{To understand better the underlying causes of the trace-driven DRAMsim3 and Ramulator simulation errors, 
we compare their \textbf{row-buffer hit, empty and miss statistics} with to the measurements from the actual Intel platform.
We could not perform the same analysis for the Ramulator~2
because its baseline architecture, Amazon Graviton3 with 8$\times$DDR5-4800, does not support the row-buffer measurements. 
Figure~\ref{fig:row-buffer-statistics} shows a subset of the results for the 100\%-read and 50\%-read/50\%-write memory traffic, 
which is sufficient to show a general trend. 
For a 100\%-read traffic and low memory bandwidth utilization, the actual system has 84\% row-buffer hits, 
13\% empty buffers and 3\% misses. 
As expected, higher memory bandwidth utilization decreases the hit ratio, and increases the empty pages and misses.  
Also, as we increase the write traffic, the row-buffer utilization degrades, compare Figure~\ref{fig:row-buffer-statistics-read}~and~\ref{fig:row-buffer-statistics-write}.} 
 
\rev{\textbf{DRAMsim3} shows a very different behavior. 
In most of the experiments, we measure 84--93\% row-buffer hit rate with 7--16\% of the misses. 
We detect the highest hit-rates for the dominantly-read and dominantly-write traffic, 
while intermediate read/write ratios have lower values. 
These hit-rates match the vertical spread we see in the DRAMsim3 bandwidth-latency curves for low and medium bandwidths (80\,GB/s).   
The curves with high hit-rates have the lowest latencies. 
For the curves with lower hit-rates, the latency increase by up to 20\,ns.
In 2\,GB/s DRAMsim3 experiments some of the read/write ratios have a surprisingly low row-buffer hit-rates ($<35\%$), 
perfectly matching the Figure~\ref{fig:trace-driven-dramsim3} memory latency peak at this bandwidth data-point.} 
%
%
%
\rev{The \textbf{Ramulator} row-buffer statistics resemble better the actual measurements, see Figure~\ref{fig:row-buffer-statistics}. 
Still we detect some discrepancies that, interestingly, have similar trends as the DRAMsim3 results. 
Again, we detect the highest hit-rates for the dominantly-read and dominantly-write traffic, 
while intermediate read/write ratios have lower values.  
For $>40\%$ write traffic, Ramulator hit-rates greatly exceed the actual ones in the whole bandwidth range (Figure~\ref{fig:row-buffer-statistics-write}). 
As in DRAMsim3, these hit-rates closely match the vertical spread of the Ramulator latency simulations. }
%

\rev{Comparison of the \textbf{actual platform} row-buffer statistics and the latency measurements resulted in an interesting finding. 
For low and moderate bandwidth utilization ($<$70\,GB/s), 
the write traffic increase leads to a notably worse row-buffer utilization, 
but this does not translate into higher memory access latencies. 
It seems that the actual system is capable to mask the row-buffer contentions and delays. 
We do not see this behavior in the DRAMsim3 and Ramulator simulations.  
Overall, our analysis of the row-buffer statistics and its correlation with the bandwidth--latency curves
provides some first steps in the analysis of the memory simulation errors. 
We believe that our work and publicly-released Mess benchmark will motivate the community to
continue this exploration. } 



 
\rev{The memory access pattern has a significant impact on the row-buffer utilization and the overall memory system performance. 
The Mess benchmark 
concurrently executes 
random-access pointer-chase and the multi-process memory traffic generator. 
Each process traverses two arrays, one with load and one with store operations.
Each array is accessed sequentially, 
but, the overall memory access pattern is complex due to the 
concurrent traversal of tens of distinct arrays located in main memory. 
Mess traffic generator can be easily extended to cover different array access patterns. 
Some of these patterns are strided access, e.g. targeting a new row-buffer in each operation,  
or the random access, e.g. the RandomAccess test in HPC Challenge benchmark developed to measure Giga Updates Per Second~(GUPS) in a system~\cite{luszczek:HPCC}.}  
\section{Mess simulator}
\label{sec:detailed-hardware-simulation}

In this section we will present the Mess \rev{analytical memory system simulator} and show how it significantly improves the memory simulation accuracy and enables a quick adoption of new memory technologies in hardware simulators.

\subsection{Design} 
\label{sec:bw-lat-memory-simulator}

\looseness -1 The CPU and memory simulators are typically connected in the following way:   
the CPU simulator issues the memory operations 
and the memory simulator determines their latencies. 
%
%
The Mess simulator does this \rev{analytically} based on the application's position in the memory bandwidth--latency curves. 
This process is complex
due to the inherent dependency between the memory system latency, 
timings of the memory operations and all dependent instructions, and the generated memory bandwidth. 
%
We simplify the problem by designing the Mess simulator not to compute the exact memory latency for a given memory traffic,
but to detect and correct discrepancies between the 
memory access latency 
and the simulated bandwidth. 
This approach, together with the fundamental principle of application's position in the memory bandwidth--latency curves, 
enables the Mess simulator to surpass the accuracy of all other memory simulators, while remaining simple and fast. 
\rev{The only memory system parameter required by the Mess model is the family of the bandwidth--latency curves. 
The curves can be measured on the actual hardware (Section~\ref{sec:mess-characterization-actual-systems}) 
or can be provided by the manufacturers, e.g. based on their detailed hardware model, as we will discuss in Section~\ref{sec:novel-memory-systems}.} 

\begin{figure}[t]%
\centering
\includegraphics[width=.7\linewidth]{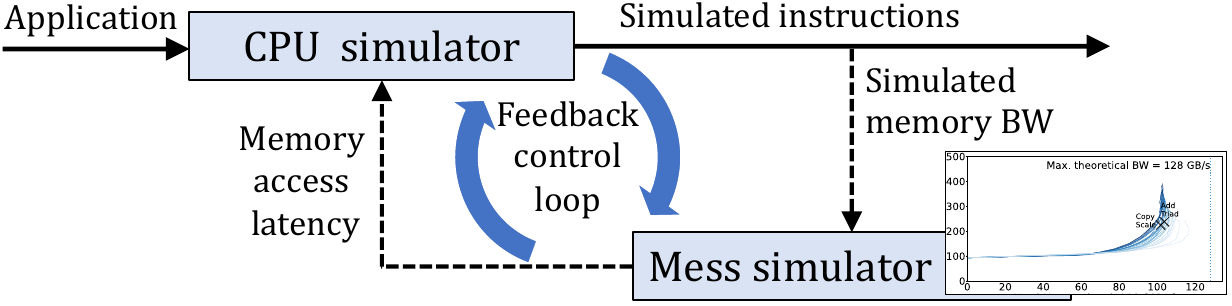}\\
\caption{Mess feedback control loop adjusts the simulated memory access latency.}
\label{fig:simulation-platform-total}%
\end{figure}

%
%
The Mess simulator acts as a feedback controller~\cite{Franklin:PIDController} from the classical control theory~\cite{Goodwin:ControlSystemDesign}, illustrated in Figure~\ref{fig:simulation-platform-total}. 
%
\rev{The simulation can start from any memory access latency, e.g. the unloaded one. 
This latency is used by the CPU simulator which generates memory memory reads and writes.
The Mess simulator observers this simulated memory bandwidth,  
positions it at the corresponding memory bandwidth--latency curve,   
and controls whether it coincides with the memory latency used in the CPU simulation. 
If this is not the case, the memory latency is being adjusted with an iterative process we describe later. 
}

The control process is performed at the end of each simulation window, which, in our experiments, comprises  1000 memory operations. 
This is much smaller than the length of the application phases~\cite{Eyerman:bwLatStack}, 
so the transition error between different application phases 
has a negligible impact on the simulator's accuracy.

\begin{figure}[t]%
\centering
\includegraphics[width=1\linewidth]{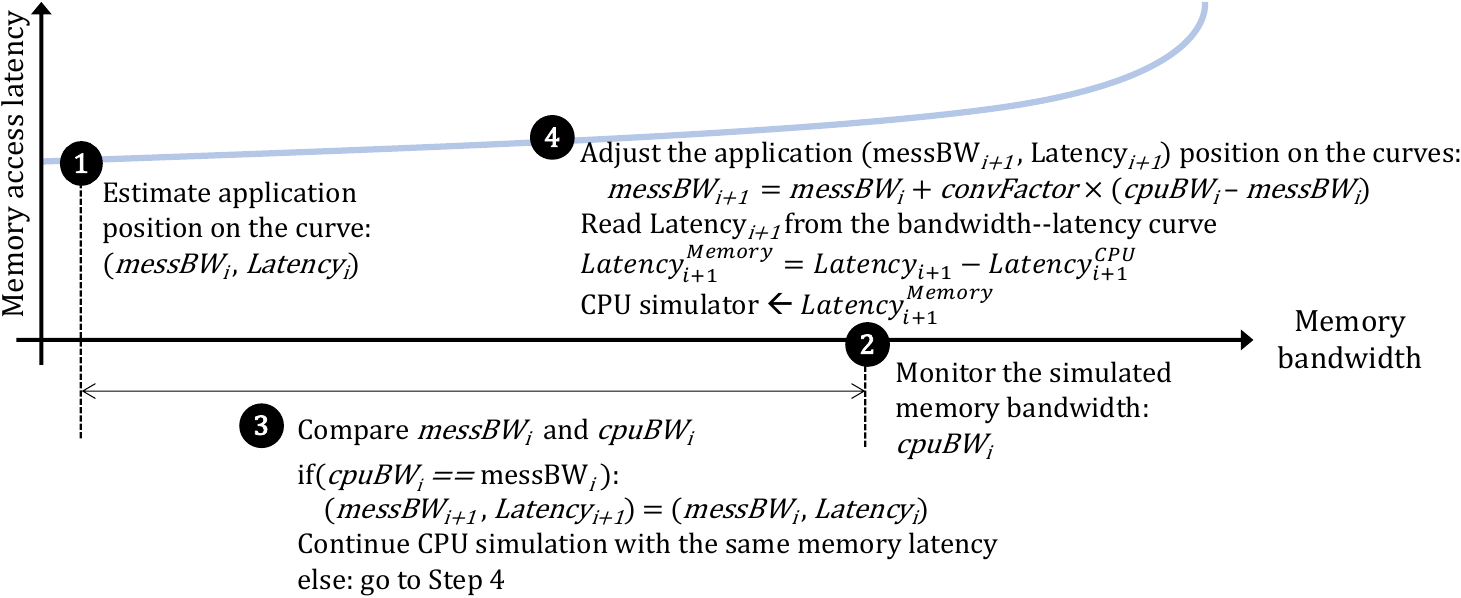}\\
\caption{One control loop iteration: Mess simulator monitors the simulated memory bandwidth, $cpuBW_{i}$, and compares it with the $messBW_{i}$ estimated at the beginning of the simulation window. If a major difference is detected, the Mess simulator adjusts the application position in the bandwidth--latency curves.}
\label{fig:simulation-platform-total2}%
\end{figure}

Figure~\ref{fig:simulation-platform-total2} describes one iteration of the Mess simulator control loop.  
For simplicity, the figure shows a single bandwidth–latency curve.
We start with the Mess estimate of the application's bandwidth--latency position 
in the $i^{th}$ simulation window, $(messBW_{i},~Latency_i)$\,\circled{1}. 
\rev{The curve is selected based on the 
read/write ratio of the simulated memory traffic, and the $x$-axis position 
is determined by $messBW_{i}$.}
From that point on, all the issued memory requests are simulated with $Latency_i$.
At the end of the simulation  window, 
the Mess simulator monitors the simulated memory bandwidth, $cpuBW_{i}$\,\circled{2},  
and compares it with the $messBW_{i}$ estimated at the beginning of the window\,\circled{3}. 
 
If the simulation is in a steady-state and the application did not change its behavior, there will be no major difference between $cpuBW_{i}$ and $messBW_{i}$. This confirms the consistency in the simulated memory access latency, the CPU timings and the achieved memory bandwidth. 
Therefore, the CPU simulation in the next window will continue with the same memory latency. 

Otherwise, a difference between $cpuBW_{i}$ and $messBW_{i}$ suggests inconsistent simulated memory latency and bandwidth. 
This can happen, for example, if the application changes its behavior.   
Figure~\ref{fig:simulation-platform-total2} illustrates the case in which the application increases the frequency of memory request 
leading to the higher bandwidth: $cpuBW_{i} > messBW_{i}$\,\circled{3}. 
In this case, the simulated memory bandwidth $cpuBW_{i}$ does not correspond to the memory $Latency_i$ used in the CPU simulation. 
To address this inconsistency, the Mess simulator adjusts the predicted application position in the bandwidth--latency curves. 
The objective of this adjustment is not to reach the correct $(BW, Latency)$ position in a single iteration. 
%
The next Mess estimate, $(messBW_{i+1},~Latency_{i+1})$, will be positioned in-between $messBW_{i}$ and $cpuBW_{i}$\,\circled{4}.
The exact position is determined based on 
the user-defined convergence factor: $messBW_{i+1} = messBW_i + convFactor \times (cpuBW_{i} - messBW_{i})$. 
The approach is based on the proportional--integral controller mechanism from the control theory~\cite{Franklin:PIDController,Goodwin:ControlSystemDesign}. 
Finally, the Mess uses $messBW_{i+1}$ to read the $Latency_{i+1}$ from the corresponding bandwidth--latency curve. 
\rev{The $Latency_{i+1}$ is a load-to-use latency.
This includes the time spent in the CPU cores, cache hierarchy and network on chip, already considered in the CPU simulation. 
In the final step, the $Latency_{i+1}$ is adjusted by these CPU timings 
\mbox{$Latency_{i+1}^{Memory} = Latency_{i+1} - Latency_{i+1}^{CPU}$}. 
The next simulation window starts with the Mess providing the updated $Latency_{i+1}^{Memory}$ to the CPU simulator.}

\subsection{Evaluation}
\label{sec:Mess-model-validation-sim-all}

The Mess simulator is integrated with ZSim and gem5, and evaluated against the actual hardware. 
We compare the simulated and actual bandwidth--latency curves as well as the performance of memory-bound benchmarks: STREAM~\cite{mccalpin:streamBenchmark}, LMbench~\cite{lmbench}, and Google multichase~\cite{Google:multichase}.

\subsubsection{ZSim}
\label{sec:ZSim-validation-sim-total}

\begin{figure}[!t]%
\centering
\includegraphics[width=.8\linewidth]{graphics/horizontal-legend}\\
\vspace{-.15cm}
\subfigure[][\fontsize{7.4}{0}\selectfont ZSim: 24 cores, 6$\times$ DDR4-2666]{%
\label{fig:ZSim-curves-result-mess-model-skylake}%
\includegraphics[width=.5\linewidth]{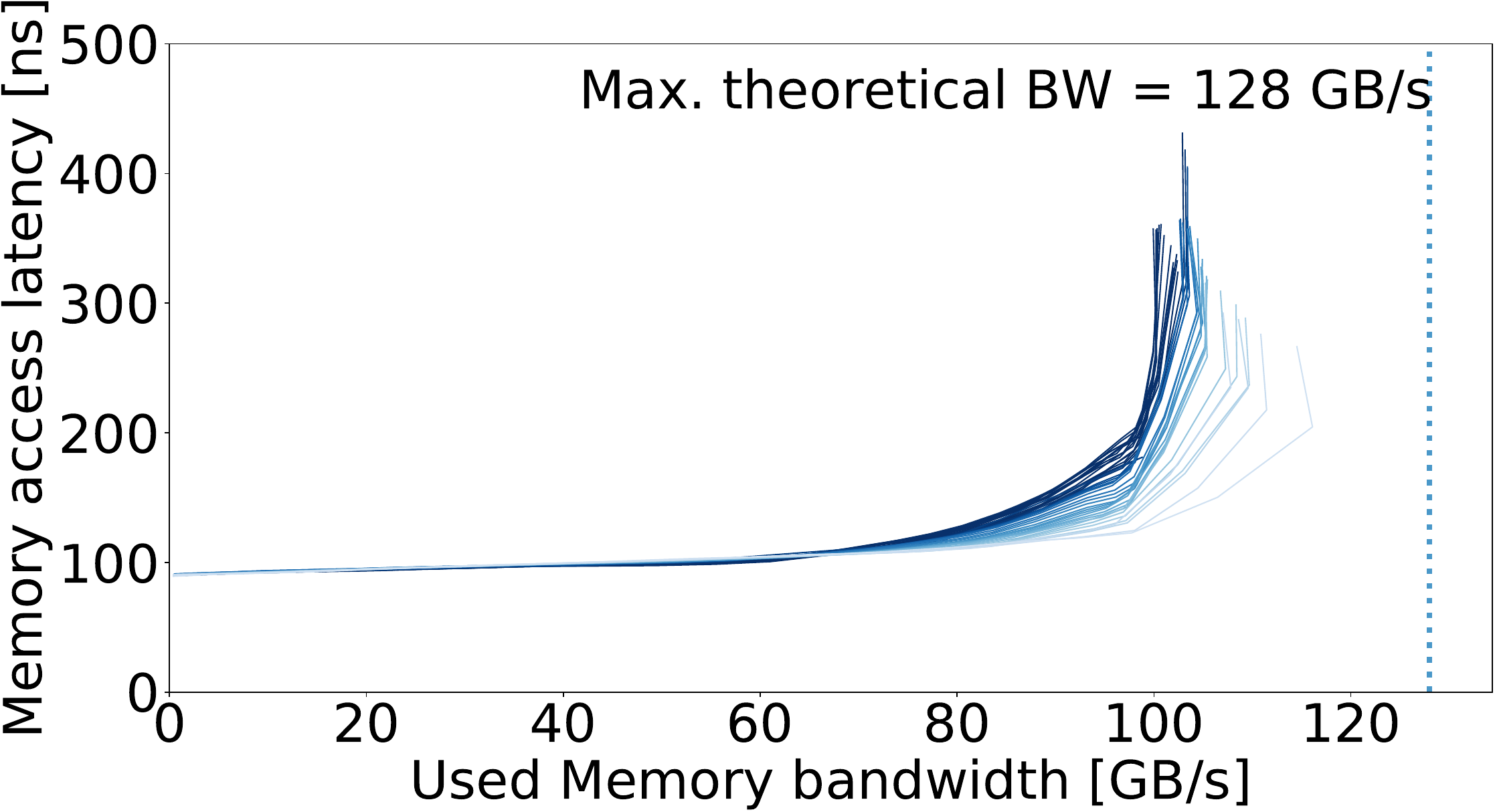}}\\%
\subfigure[][\fontsize{7.4}{0}\selectfont ZSim: 58 cores, 8$\times$ DDR5-4800]{%
\label{fig:ZSim-curves-result-mess-model-graviton}%
\includegraphics[width=.5\linewidth]{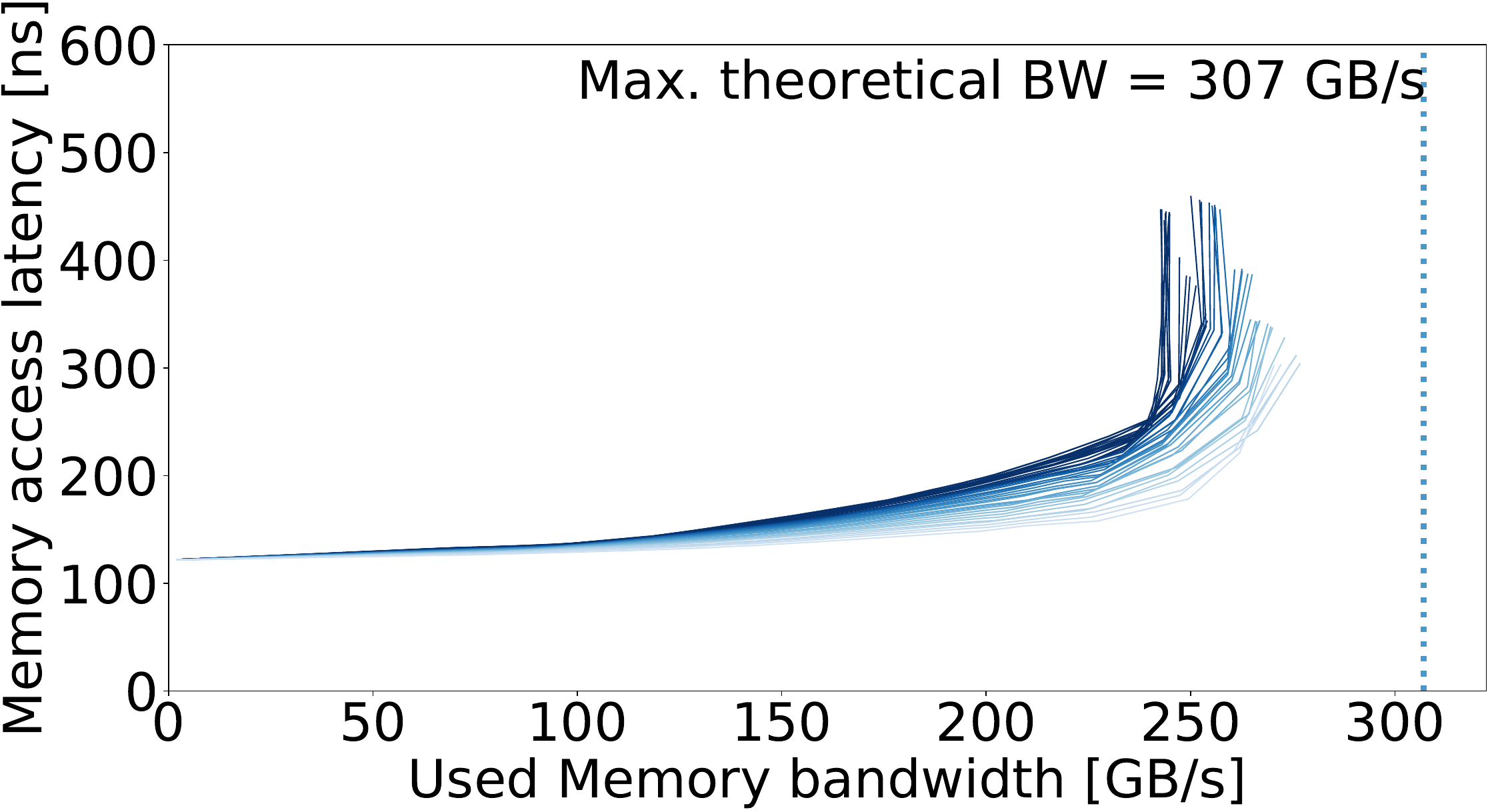}}~%
\subfigure[][\fontsize{7.4}{0}\selectfont ZSim:\,192\,cores,\,32$\times$\,HBM2\,channels]{%
\label{fig:ZSim-curves-result-mess-model-fujitsu}%
\includegraphics[width=.5\linewidth]{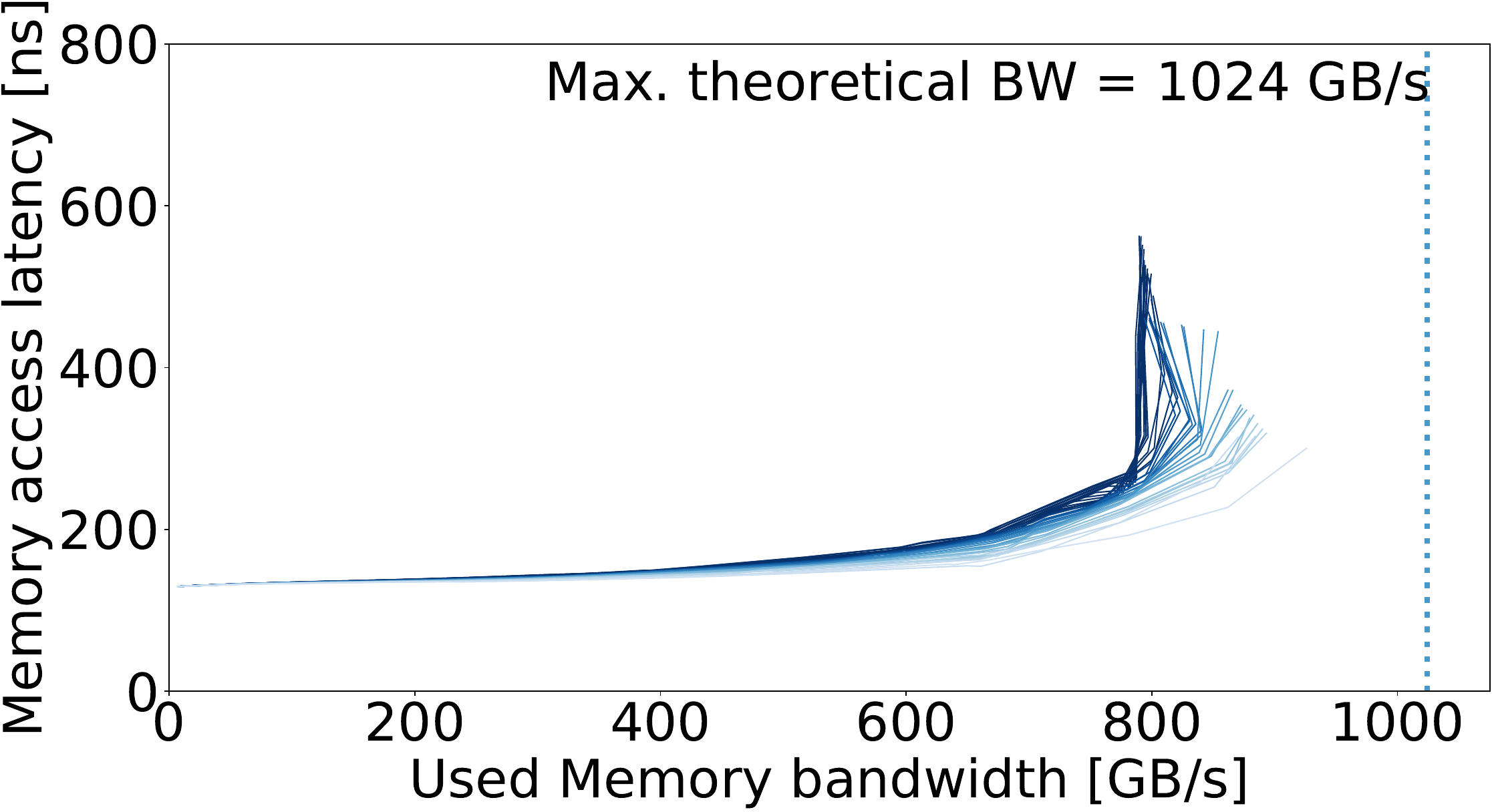}}%
\caption[A set of four subfigures.]{ZSim with the Mess simulator closely matches the actual memory systems.}
\label{fig:ZSim-bw-lat-curves-validation}%
\end{figure}

Figure~\ref{fig:ZSim-bw-lat-curves-validation} shows the DDR4, DDR5 and HBM2 bandwidth--latency curves measured with the ZSim connected to the Mess simulator.\footnote{The Mess simulator also supports the Intel Optane technology. Optane's bandwidth--latency curves are measured on a 16-core Cascade Lake server with 6$\times$DDR4-2666 16\,GB and 2$\times$Intel Optane 128\,GB memory in App Direct mode. Intel Optane technology is discontinued since 2023, so we do not analyze its performance characteristics and simulation.}
The configurations of the simulators match the actual Intel Skylake with 24-core and six DDR4-2666 memory channels. 
The simulated Mess curves, depicted in Figure~\ref{fig:ZSim-curves-result-mess-model-skylake}, 
closely resemble the actual memory systems performance (Figure~\ref{fig:char-bw-lat-result-skylake}).  
The simulation error of the unloaded memory latency is below 1\%, and it is around 3\% for the maximum latencies.  
The difference between the simulated and the actual saturated bandwidth range is only 2\%. 
Figures~\ref{fig:ZSim-curves-result-mess-model-graviton}~and~\ref{fig:ZSim-curves-result-mess-model-fujitsu} show the ZSim+Mess simulation results for the high-end DDR5 and HBM memories.   
To saturate the 8-channel DDR5-4800 and 32-channel HBM2, we increase the number of simulated cores to 58 and 192, respectively. 
Again, the simulated bandwidth--latency curves closely resemble the performance of the corresponding actual memory systems (Figures~\ref{fig:char-bw-lat-result-a64fx}~and~\ref{fig:char-bw-lat-result-graviton}). 

\begin{figure}[t]%
\centering
\includegraphics[width=1.02\linewidth]{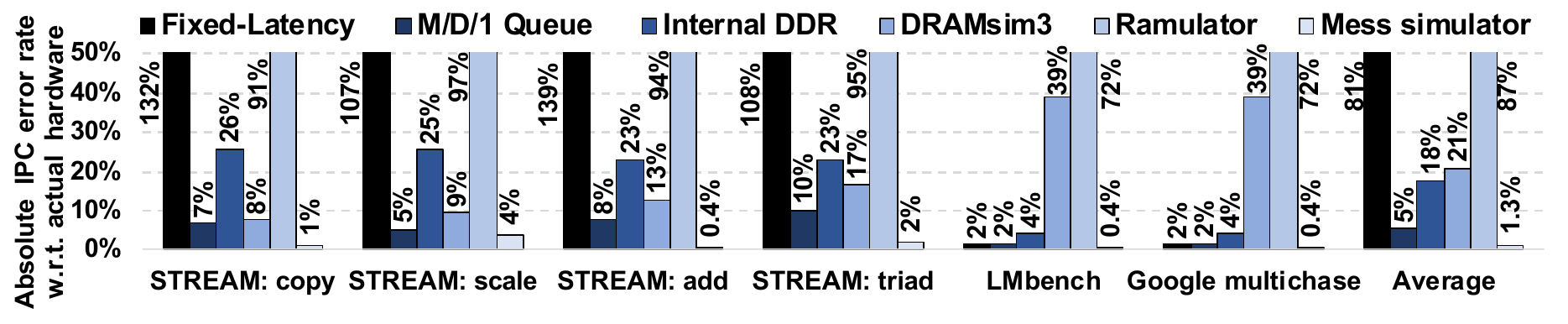}\\
\caption{The ZSim+Mess simulation error for STREAM, LMbench and Google multichase is only 1.3\%.}
\label{fig:simulation-platform-benchmark-zsim}%
\end{figure}

\looseness -1 Figure~\ref{fig:simulation-platform-benchmark-zsim} shows the evaluation results, w.r.t. to the actual Intel Skylake server, of all six ZSim memory models when running memory intensive STREAM~\cite{mccalpin:streamBenchmark}, LMbench~\cite{lmbench} and Google multichase~\cite{Google:multichase}. 
The simulation errors are closely correlated with the similarity between the simulated and actual bandwidth--latency curves. 
The Mess shows the best accuracy with only 1.3\% average error, followed by the M/D/1 and internal DDR model. 
The fixed-latency simulation and Ramulator show the highest errors of more than 80\%. 
The Mess simulator is also fast. 
It increases the simulation time by only 26\% higher w.r.t. a simple fixed-latency memory, 
and it is 2\% and 15\% faster than the M/D/1 and internal DDR model. 
The ZSim+Mess simulation speed-up over the ZSim+Ramulator and ZSim+DRAMsim3 is 13$\times$ and 15$\times$.

\subsubsection{gem5}
\label{sec:gem5-validation-sim-total}

\begin{figure}[!t]%
\centering
\includegraphics[width=.8\linewidth]{graphics/horizontal-legend}\\
\vspace{-.15cm}
\hspace{-0.1cm}\subfigure[][\fontsize{7.69}{0}\selectfont gem5: 16 cores \& 1$\times$\,DDR5-4800]{%
\label{fig:bw-lat-result-gem5}%
\includegraphics[width=.5\linewidth]{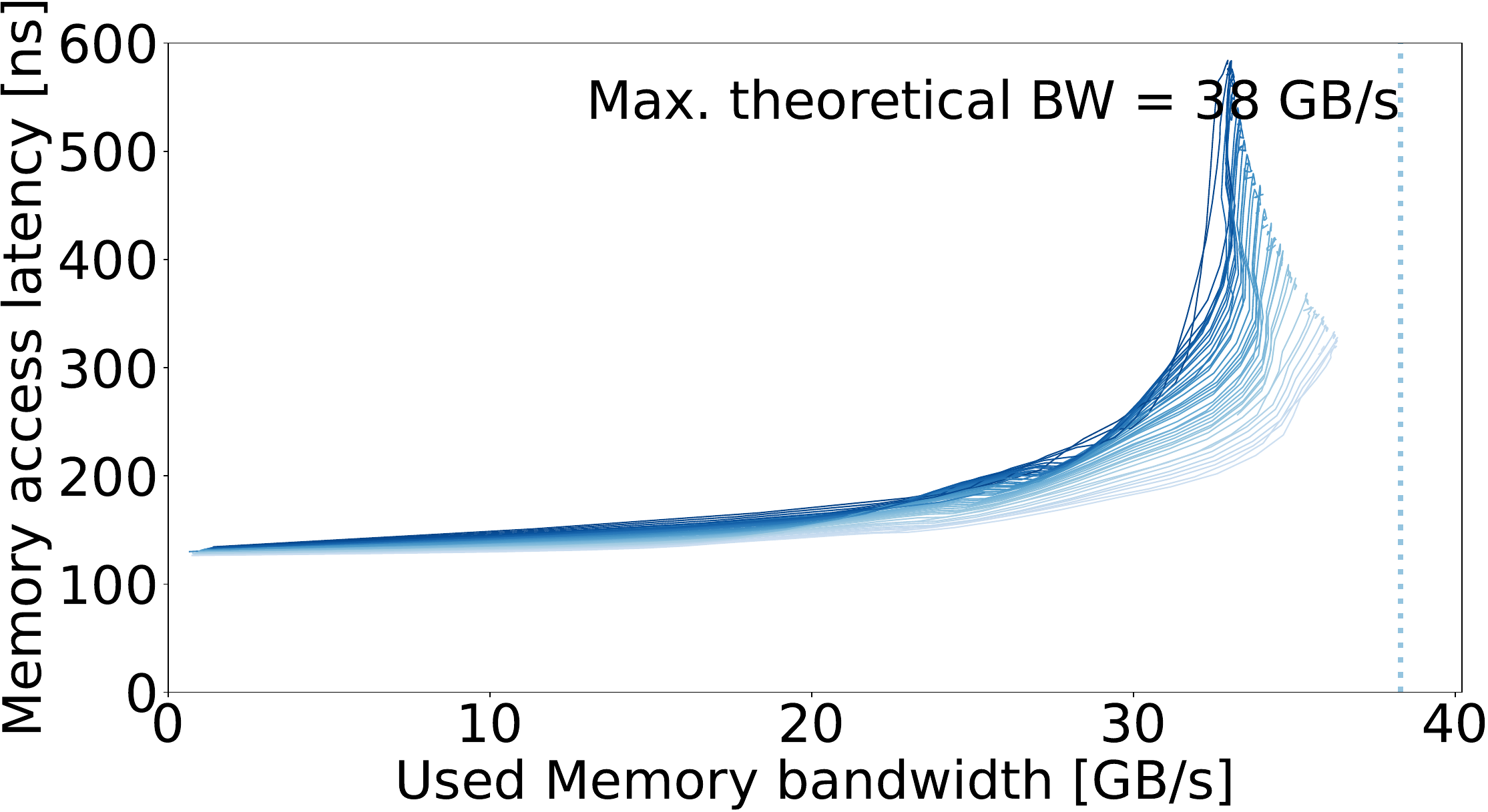}}~%
\subfigure[][\fontsize{7.69}{0}\selectfont gem5:\,16\,cores\,\&\,1$\times$\,HBM2\,channel]{%
\label{fig:}%
\includegraphics[width=.5\linewidth]{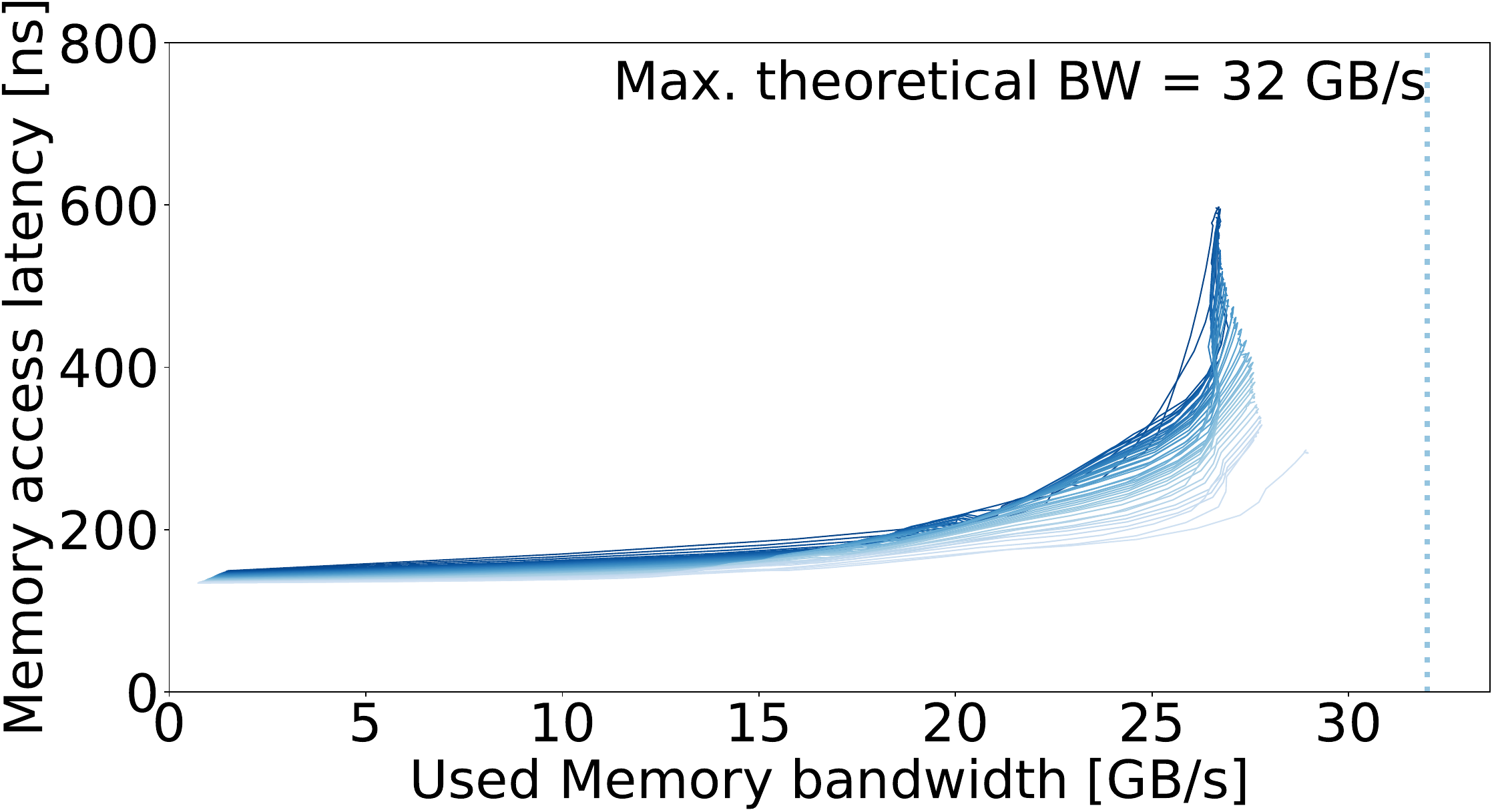}}\\%
\caption[A set of four subfigures.]{gem5 with the Mess simulator, when scaled, closely follows the actual memory system curves (Figures~\ref{fig:char-bw-lat-result-graviton} and ~\ref{fig:char-bw-lat-result-a64fx}).}
\label{fig:gem5-bw-lat-curves-validation}%
\end{figure}

\looseness -1 Figure~\ref{fig:gem5-bw-lat-curves-validation} shows the DDR5 and HBM2 bandwidth--latency curves simulated with the gem5 connected to the Mess memory simulator. %
In all experiments, the gem5 is configured to model Graviton\,3 cores~\cite{Amazon:Graviton3} described in 
Section~\ref{sec:gem5-characterization}.
%
%
To reduce simulation time, we simulate 16 CPU cores connected to a single memory channel.\footnote{Simulation of a whole server comprising 64 Graviton\,3 cores and 8$\times$DDR5-4800 requires more than five hours to obtain a single bandwidth--latency datapoint. The full simulation of all the curves would require more than a year.} 
The simulated bandwidth--latency curves, when scaled to eight DDR5 channels or 32 HBM2 channels, closely resemble actual system behavior (Figures~\ref{fig:char-bw-lat-result-graviton} and~\ref{fig:char-bw-lat-result-a64fx}).

\looseness -1 We also evaluate the Mess memory simulation against the gem5's built-in simple memory model and internal DDR5 model \rev{as well as cycle-accurate Ramulator2}
when running STREAM, LMbench, and Google Multichase benchmarks. 
In these experiments we simulate a whole server comprising 64 Graviton\,3 cores and 8$\times$DDR5-4800,  
and compare the results against the benchmark executions on the actual server. 
\rev{The gem5+Mess simulation time is practically the same as the gem5 internal DDR model, while it provides much better accuracy.}
%
The Mess memory simulator decreases the average error from 30\% (gem5 simple memory model), 15\% (internal DDR5 model), \rev{and 52\% (Ramulator~2)} to only 3\%. Such a low error is unprecedented in any prior validation attempts~\cite{Akram:simulationSurvey,Butko:gem5ARM,Gutierrez:gem5ARM,Akram:haswel}.
%




\subsection{Simulation of novel memory systems: CXL memory expanders}
\label{sec:novel-memory-systems}

\looseness -1 The memory system complexity and the scarcity of publicly-available information often result in a considerable gap between a technology release and the support for its detailed hardware simulation. 
For example, public memory simulators started to support the DDR5 in 2023~\cite{Haocong:Ramulator2}, 
three years after production servers with DDR5 DIMMs hit the market.  
%
%

\looseness -1 The Mess simulator provides a fundamental solution for this gap because it can simulate emerging memory systems 
as soon as their bandwidth--latency curves are available. 
For memory technologies available on the market, the curves can be measured on a real platform.
For emerging memory devices that are not yet available in off-the-shelf servers, the bandwidth–latency curve can be measured on a
developer board with a prototype of the new device, or alternatively it can be provided by the manufacturers, e.g. based on their detailed proprietary RTL models.   




We will demonstrate the Mess simulation of novel memory systems with an example of the Compute Express Link~(CXL) memory expanders. 
CXL is an emerging interconnect standard for processors, accelerators and memory devices. 
The CXL memory expanders enable a straightforward enlargement of the memory system capacity and bandwidth, 
as well as the exploration of unconventional disaggregated memory systems~\cite{CXL2020}. 
One of the main limitations for an academic research in this field, however, is the lack of reliable performance models for these devices. 
%
%

The Mess simulation is performed with the CXL memory expanders bandwidth--latency curves provided by the memory manufacturer 
based on their detailed hardware model. 
%
In particular, we model a CXL memory expander connected to the host via the CXL~2.0 PCIe~5.0 interface with 1$\times$8~Lanes.  
The device comprises one memory controller connected to a DDR5-5600 DIMM with two ranks. 
All the CXL modules, Front end, Central controller and Memory controller, are implemented in SystemC.
The modules communicate by using the manufacturer's proprietary SystemC Transaction Level Modeling~(TLM~\cite{Ghenassia:TLM}) framework, 

 
%

\begin{figure}[!t]%
\centering
\includegraphics[width=\linewidth]{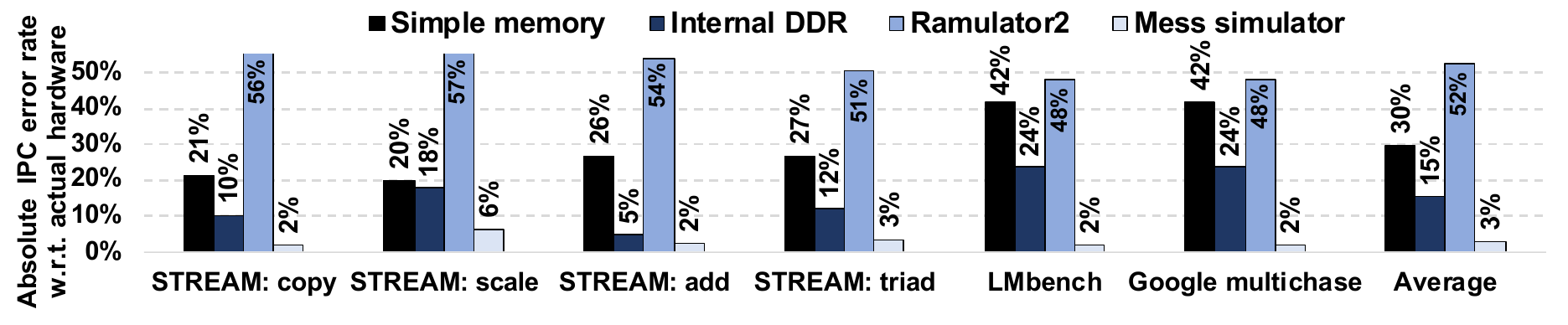}%
\caption[A set of four subfigures.]{The gem5+Mess simulation error for STREAM, LMbench and Google multichase is only 3\%.}
\label{fig:simulation-platform-benchmark-gem5}%
\end{figure}

\begin{figure}[!t]%
\centering
\hspace{-0.1cm}\subfigure[][\fontsize{7.22}{0}\selectfont Manufacturer's SystemC model]{%
\label{fig:CXL-curves}%
\includegraphics[width=.5\linewidth]{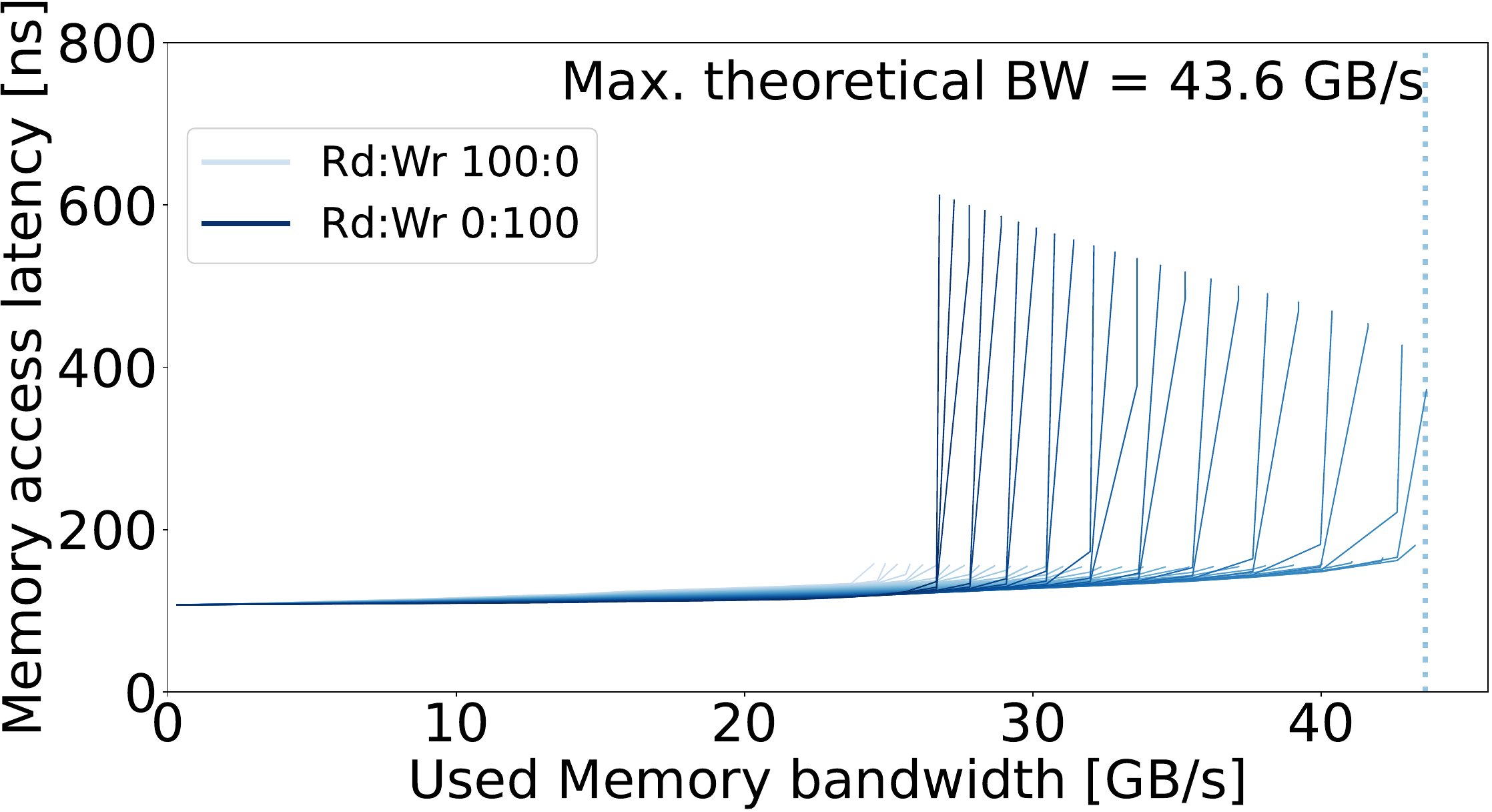}}~%
\subfigure[][\fontsize{7.22}{0}\selectfont OpenPiton: Ariane 64-core CPU]{%
\label{fig:CXL-bw-lat-curves-openPiton}%
\includegraphics[width=.5\linewidth]{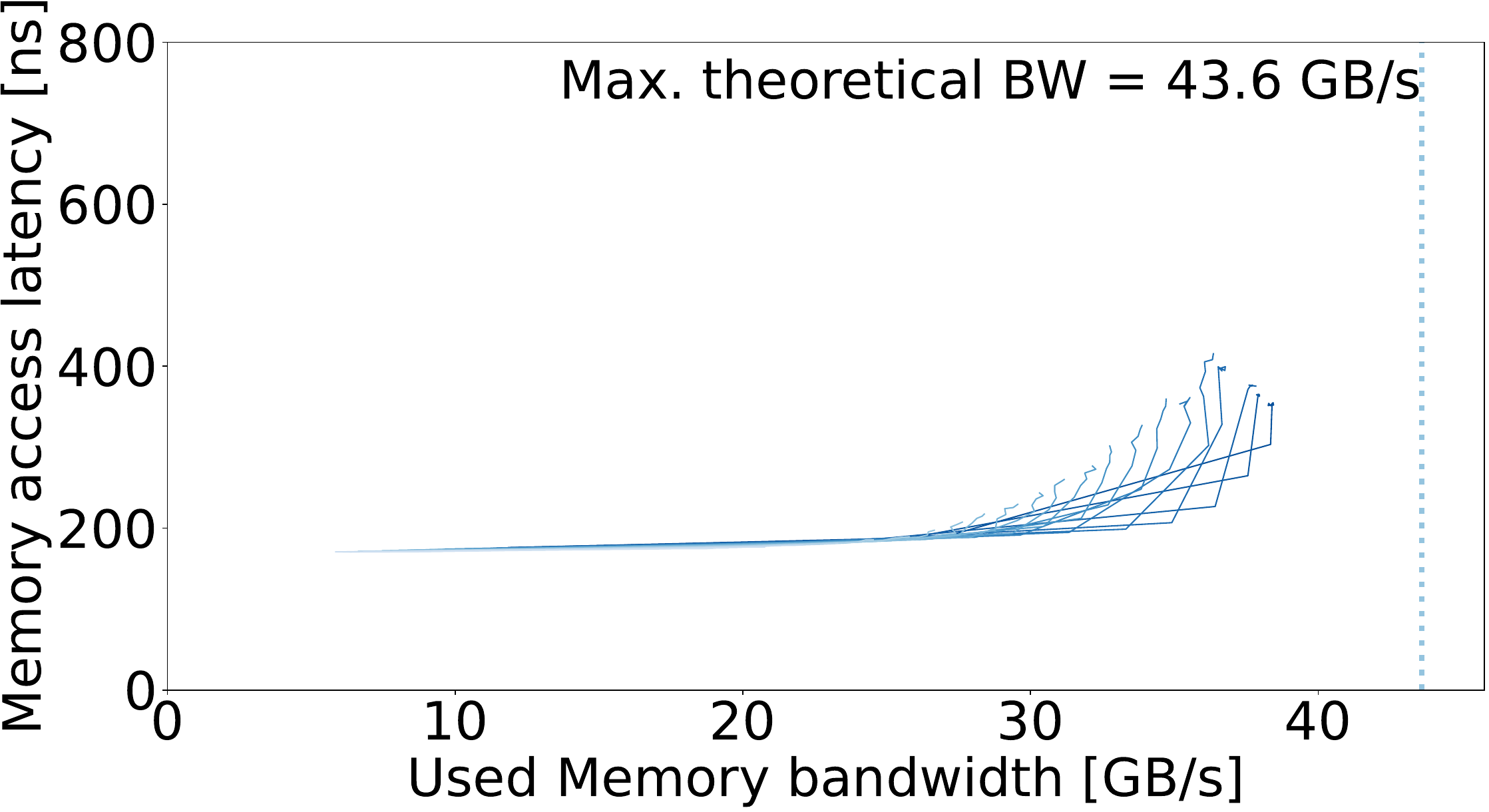}}\\%
\hspace{-0.1cm}\subfigure[][\fontsize{7.22}{0}\selectfont gem5:\,Amazon\,Graviton3\,16-core\,CPU]{%
\label{fig:}%
\includegraphics[width=.5\linewidth]{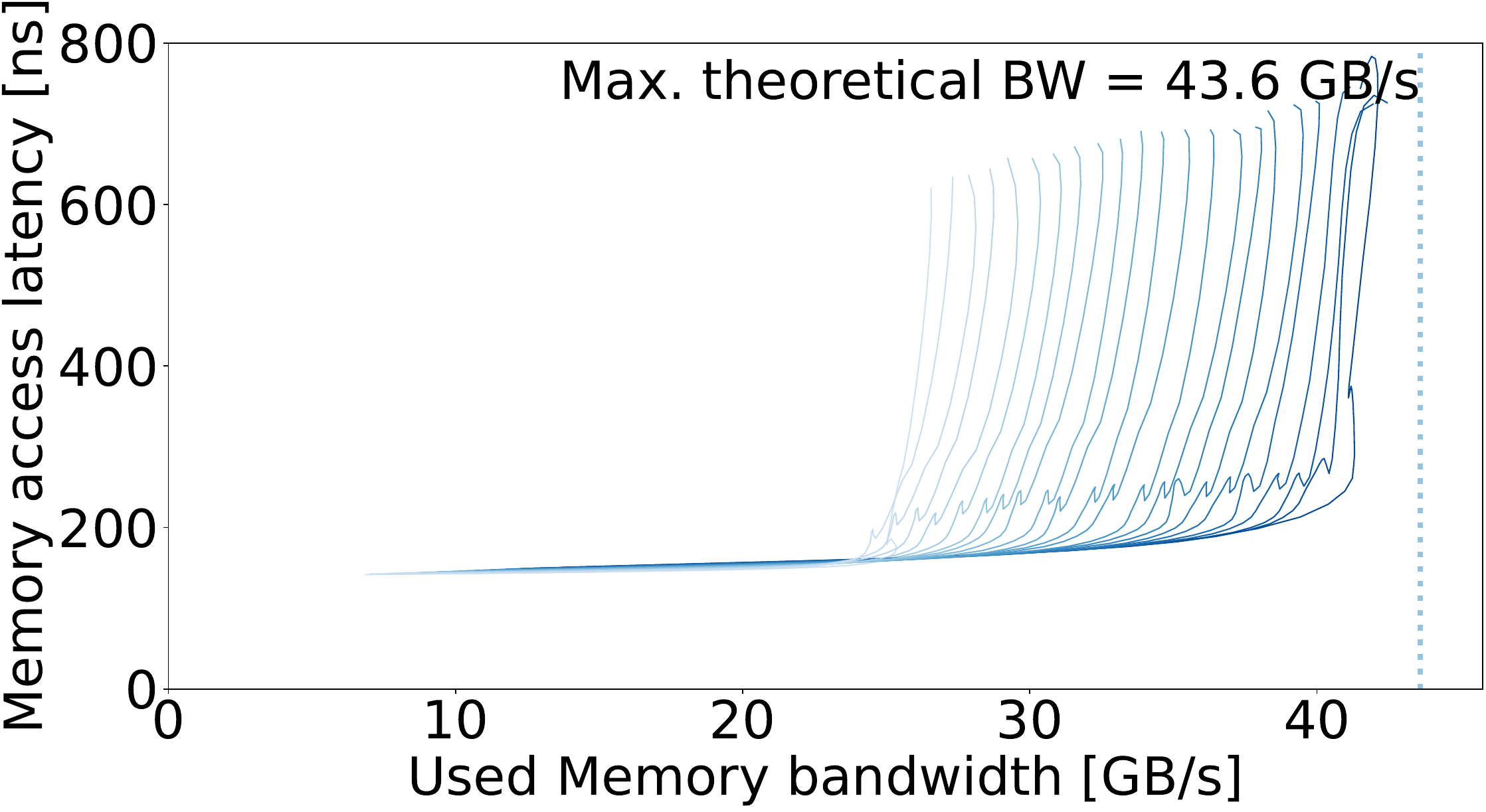}}~%
\subfigure[][\fontsize{7.22}{0}\selectfont ZSim: Intel Skylake 24-core CPU]{%
\label{fig:}%
\includegraphics[width=.5\linewidth]{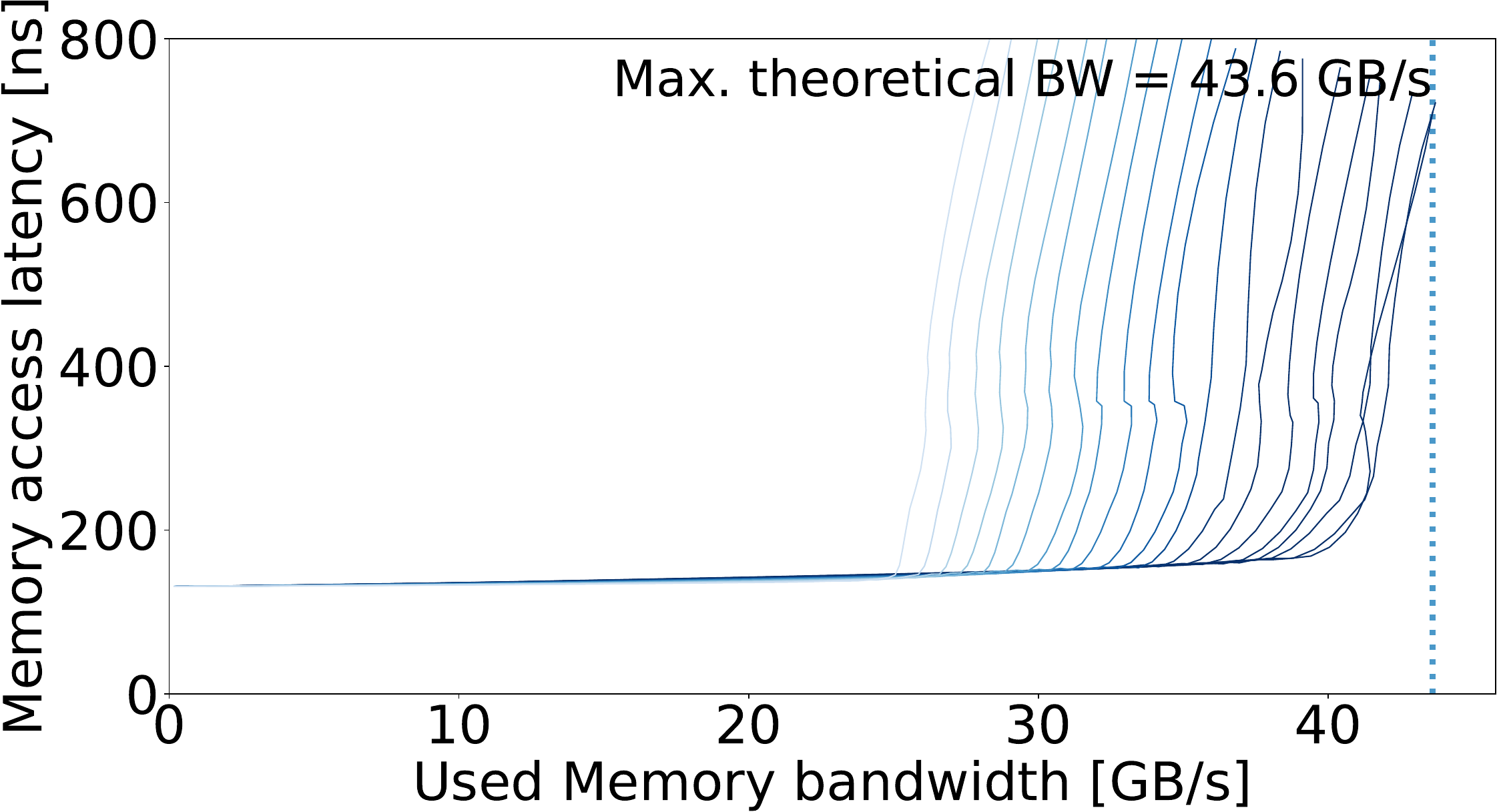}}
\caption[A set of four subfigures.]{Bandwidth--latency curves of the CXL memory expander. Mess simulator closely follows the Manufacturer's SystemC model.}
\label{fig:CXL-bw-lat-curves}%
\end{figure}

\looseness -1 The obtained bandwidth--latency curves are shown in Figure~\ref{fig:CXL-curves}.\footnote{The maximum theoretical bandwidth of the CXL.mem protocol is influenced by the read/write ratio of the workload~\cite{Sharma:cxlMaxBW}. In this figure, we present the highest value among all possible scenarios.} 
The figure plots the round-trip latency from the CXL host input pins. 
To consider a full load-to-use latency, a user should add the round-trip time between the CPU core and the CXL host. 
We measure this latency component with the Intel MLC~\cite{Intel:MLC}. 
%
The CXL memory expanders show a similar performance trends as the DDRx or HBM memory systems: latency that increases with the system load, significant non-linear increase after a saturation point, and the impact of the traffic read/write ratio. 
One major difference is that the CXL interface provides the best performance for a balanced reads and writes traffic, while its performance drops significantly for the 100\%-read or 100\%-write traffic. This is because, unlike the DDRx interfaces, CXL is a full-duplex interconnect with independent read and write links. Therefore, the CXL can transmit simultaneously in both directions, but in the case of the unbalanced traffic one CXL transmission direction could be saturated while other direction is negligibly used. 

We use the obtained CXL memory bandwidth--latency curves in the Mess simulator integrated with 
ZSim, gem5 and OpenPiton Metro-MPI (see Figure~\ref{fig:CXL-bw-lat-curves}). 
In all configurations,  the Mess simulator closely follows the Manufacturer’s SystemC model. 
To reduce long OpenPiton Metro-MPI simulation time, we model only 25 curves with a small number of experimental points in each curve. For this reason some segments of the curves are discrete. 
Nevertheless, the OpenPiton Metro-MPI simulations match the general trend and the saturated bandwidth range of the manufacturer's curves.
The maximum latency range is below the manufacturers CXL curves 
because the simulated small in-order Ariane cores with only 2-entry MSHRs cannot saturate the target memory system. 
This behavior is already detected and discussed in Section~\ref{sec:characterization-openpiton}.  
ZSim and gem5 results practically match the manufacturer's CXL curves. In Appendix~\ref{sec:CXL-emulators}, we compare our CXL simulation platform against prior approach to emulate memory-over-CXL.


\section{Mess application profiling}
\label{sec:paraver-profet}

The Mess framework also enhances the memory-related application profiling. 
We demonstrate this functionality with the Mess extension of Extrae and Paraver, production HPC performance tools for detailed application tracing and analysis~\cite{BSCTools}.
%
The Mess application profiling adds a new layer of information related to the application's memory performance metrics. 
This information can be correlated with other application runtime activities and the source code, leading to a better overall understanding of the application's characteristics and behavior.

\subsection{Background: Extrae and Paraver} 

\looseness -1 Paraver is a flexible data browser for application performance analysis~\cite{bsctools:paraver,bsctools:Pillet1995Paraver}. It can display and analyze application MPI calls, duration of the computing phases, values of the hardware counters, etc. 
Paraver can also summarized application behavior in histograms and link it with the corresponding source code. 
The input data format for Paraver is a timestamped trace of events, states and communications~\cite{bsctools:paraver_format}. 
For parallel applications, the traces are usually generated with the Extrae tool~\cite{bsctools:extrae}. 

Extrae automatically collects entry and exit call points to the
programming model runtime, source code references, hardware counters metrics, dynamic
memory allocation, I/O system calls, and user functions. 
%
It is is compatible with programs written in C, Fortran, Java, Python, and combinations of different languages. It supports a wide range of parallel programming models.  
Extrae is available for most UNIX-based operating systems and it is deployed in all relevant 
HPC architectures, including CPU-based systems and accelerators.


\subsection{Use cases}
\looseness -1  We illustrate the capabilities of the Mess application profiling 
with an example of the memory-intensive HPCG benchmark~\cite{Heroux:HPCG,Dongarra:HPCG} running in a dual-socket Cascade Lake server (Table~\ref{tab:hpc-server-description}). We fully utilize the one CPU socket by executing 16~benchmark copies, one on each core. 
Extrae monitors the application memory behavior with a dedicated profiling process which traces the memory bandwidth counters.  
The sampling frequency is configurable, and it is 10\,ms by default.
Even with this fine-grain profiling, the introduced overhead is negligible, below 1\%. 

The extended Paraver tool correlates the application memory bandwidth measurements with Mess memory curves.  
The application measurements are plotted on the curves as a set of points, 
each of them corresponding to 10\,ms of the application runtime.
The application memory use can be also incorporated into the Paraver trace file,   
so a user can analyze its evolution over time, and correlate it with other application's behavior and the source code.



\begin{figure}[t!]
  \centering
  \includegraphics[width=.8\linewidth]{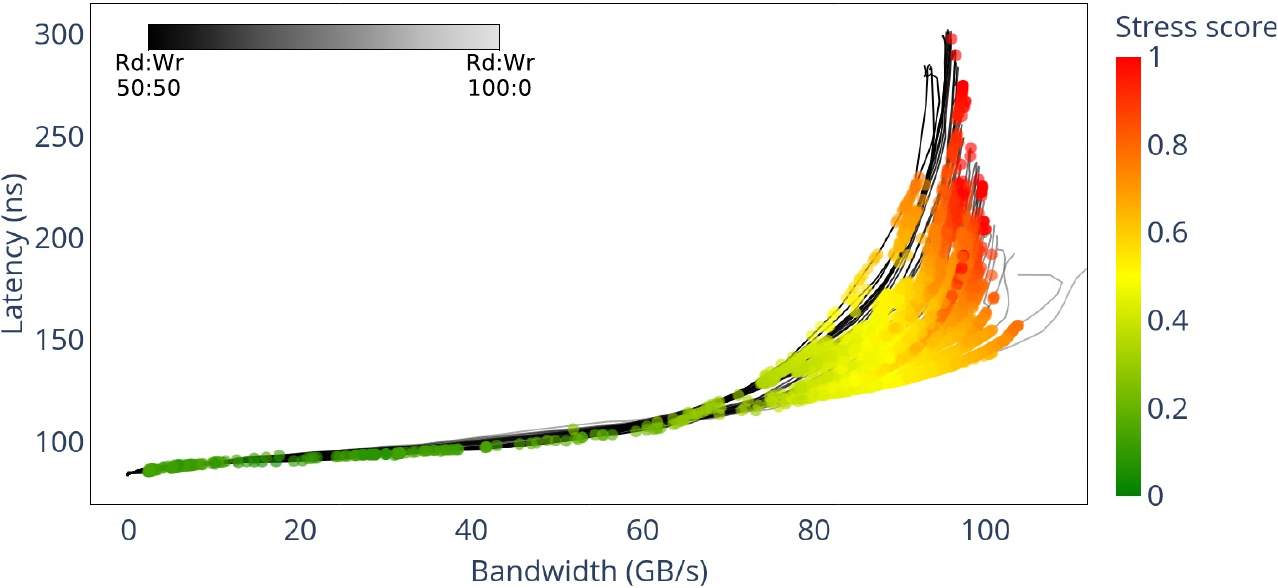}
  \caption{Most of the HPCG execution is located in the saturated bandwidth area,
  above 75\,GB/s. The Mess application-profiling extension of Paraver already includes the memory
  stress score visualization with a green--yellow--red gradient.}
  \label{fig:xhpcg-bw-lat-curve}
\end{figure}

\subsubsection{Bandwidth--latency curves}
\label{sec:bw-lat_curves}

Figure~\ref{fig:xhpcg-bw-lat-curve} depicts the Mess profile of the HPCG benchmark. 
Most of the HPCG execution is located 
in the saturated bandwidth area, above 75\,GB/s. 
Sporadically, the benchmark even reaches the maximum sustained bandwidth with peak memory
latencies in the range of 260--290\,ns.
Also, each HPCG point on the curves is
associated with a memory stress score.  The score value ranges from~0, for the
unloaded memory system, to~1, corresponding to the right-most area of the
bandwidth--latency curves.  Memory stress score in a given point is calculated
as a weighted sum of the memory latency and the curve
inclination. The latency itself is a good proxy of the system stress, 
while the inclination shows the memory system sensitivity to a bandwidth change. Gentle inclination indicates that
a memory bandwidth change would have a minor impact on the memory access latency
and the overall performance.  In the steep curve segments,
e.g. 95--100\,GB/s area in Figure~\ref{fig:xhpcg-bw-lat-curve}, small bandwidth
changes can rapidly saturate the memory system leading to a major latency
increase.
The Mess extension of Paraver already includes the stress score visualization
with a green--yellow--red gradient that can be easily interpreted by application
developers.

\subsubsection{Timeline analysis}
\label{sec:Paraver-TimelineAnalysis}

\begin{figure}[t!]
 \centering
 \includegraphics[width=\linewidth]{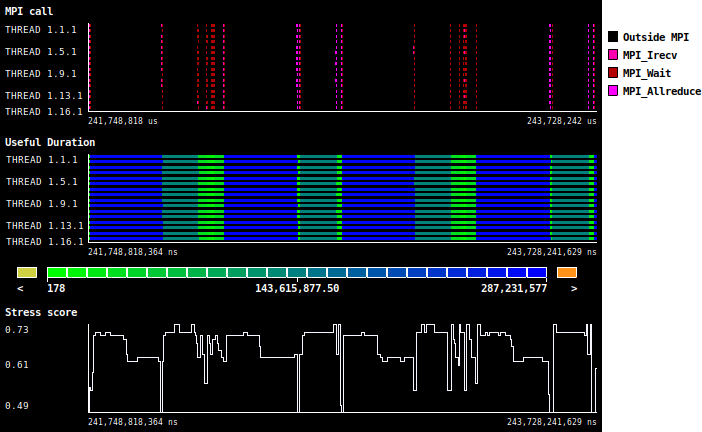}
 \caption{Timeline showing two iterations of the HPCG benchmark with MPI calls
 (top), computations duration (middle) and memory stress score (bottom).}
 \label{fig:analysis-xhpcg-timeline-mpi-useful-stress}
\end{figure}

Once the memory stress score is incorporated into the application's Paraver trace, 
it can be combined with other aspects of the application analysis, 
as illustrated in Figure~\ref{fig:analysis-xhpcg-timeline-mpi-useful-stress}. 
%
The figure analyzes around two seconds of the HPCG runtime, from 241,748,818\,$\mu$$s$ to 243,728,242\,$\mu$$s$
($x$-axis). Guided by the sequence of MPI calls illustrated in
Figure~\ref{fig:analysis-xhpcg-timeline-mpi-useful-stress} (top), we identify
the application's main iterative loop and, using MPI\_Allreduce (pink) as
delimiter, we select two iterations for our analysis. The middle Figure~\ref{fig:analysis-xhpcg-timeline-mpi-useful-stress} analyzes compute applications phases. 
The color
gradient corresponds to the compute phase length: green to blue gradient for short to long phases.
Figure~\ref{fig:analysis-xhpcg-timeline-mpi-useful-stress} (bottom) shows the
memory stress score for this region. The longest compute phases
(blue) exhibits two distinct memory behaviors:  
at the start of the phase, the memory stress score rises to 0.71, and
then halfway through the phase it decreases to 0.64. 
The fine-grain application profiling can detect  
different memory stress score values even within a single compute phase.  

\subsubsection{Links to the source code}
\label{sec:Paraver-SourceCode}

Extrae also collects callstack information of the MPI calls,
referred to as the MPI call-points,\footnote{A callpoint refers to the file and line
number where the program initiates an MPI call at a given level of the
callstack, typically the last one. This point serves as a boundary for a region
that begins with the current MPI call and extends until the next one. In the
tables, we indicate the starting callpoint.} which are used to link the application runtime behavior with the
source code. 
With the Mess application-profiling extension of Paraver, the application source code can be linked to
its memory-related behavior. 
This is fundamental for making data placement decisions in heterogeneous memory systems, e.g. comprising DDRx DIMMs and HBM devices~\cite{intel:max,mccalpin:xeonMax}.



\section{Related work} 
\label{sec:Related-work} 

Mess framework provides a unified view of the memory system performance 
that covers the memory benchmarking, simulation and application profiling. 
Although these three memory performance aspects are inherently
interrelated, they are currently analyzed with distinct and decoupled tools.

\subsection{Memory system benchmarks} 

Memory access latency and utilized bandwidth are commonly treated as independent concepts measured by separate memory benchmarks. 
LMbench~\cite{lmbench} and Google Multichase~\cite{Google:multichase} measure the load-to-use latency in an unloaded memory system, while 
STREAM~\cite{mccalpin:streamLink, mccalpin:streamBenchmark}, STREAM2~\cite{mccalpin:stream2Link}, Hopscotch~\cite{Ahmed:Hopscotch}, CAMP~\cite{Peng:camp}, and HPCG~\cite{Dongarra:HPCG,Heroux:HPCG}  measure the maximum obtainable memory bandwidth or performance of the 
application kernels that are proportional to it. 
Only recently the community started to make the first steps in measuring latencies in loaded memory systems. 
The Intel Memory Latency Checker~(MLC) tool~\cite{Intel:MLC} is used to show 
how memory access latency increases for higher used memory bandwidths~\cite{Helm:perfmemplus, helm:bw-latCurve} 
and to compare systems based on fundamentally different memory technologies, such as the DRAM and Optane~\cite{Yang:optaneCurve}.  
X-Mem benchmark~\cite{Gottscho:X-Mem} reports loaded access latencies for cache subsystem, main memory, and NUMA memory nodes. 
The impact of the read and write traffic to the memory system performance is not measured nor analyzed.  
Overall, current memory systems benchmarks provide a small number of data points in a very large and complex memory-performance space. 

The Mess benchmark is designed for holistic close-to-the-hardware memory
system performance characterization that is easily adaptive to different target
platforms.
It significantly increases the coverage and the level of detail of the previous tools, leading to new findings 
in memory behavior of the hardware platforms and simulators under study.  
%

\subsection{Memory system simulation}
The Mess framework tightly integrates the memory performance characterization into the memory simulation. 
We compare the Mess \rev{analytical memory system simulator} with the state-of-the-art memory
models included in the CPU simulators~\cite{carlson:sniperOriginal,carlson:sniperUpdated,Miller:graphite,Balkind:OpenPitonandAriane2,Leyva:OpenPitonupdateBSC,Balkind:OpenPitonandAriane,Butko:gem5ARM,Gutierrez:gem5ARM,sanchez:zsimEnhance} as well as the external cycle-accurate memory simulators~\cite{Shangli:dramsim3,kim:ramulator,Haocong:Ramulator2}. 
The Mess memory simulator is fast, accurate and easily-integrated with the CPU simulators. 
It can support novel memory systems as soon as their detailed bandwidth--latency curves are measured
on actual production systems or prototypes, or provided by the manufacturers. 
Therefore, Mess can simulate high-end and future memory systems much sooner than 
the standard memory simulators which consider detailed memory device sequences and timings~\cite{Wang:DRAMSIM,kim:ramulator,Shangli:dramsim3,Haocong:Ramulator2,steiner:dramsys4,chatterjee:usimm}. 
Mess is the first memory simulator to support CXL memory expanders.

 Apart from the hardware simulation, system performance can be analyzed with analytical models. 
 The PROFET model~\cite{Radulovic:PROFET} predicts
how an application’s performance and energy consumption change when it is executed on different (future) memory systems. The model is based on the instrumentation of an application execution on actual hardware, so it already takes into account CPU microarchitectural details. PROFET is evaluated on Intel and Huawei servers with DDR3, DDR4, HBM and Optane memory~\cite{profet:urlLink}. The main objective is to provide an alternative to complex and slow CPU simulations. 
 This is orthogonal and complementary to the Mess simulator that targets the main memory simulation.




\subsection{Application profiling}

 ProfileMe~\cite{Dean:profileMe} and PerfMemPlus~\cite{Helm:perfmemplus} 
determine whether the application is memory bound 
based on its memory access latency, measured with the Intel's Event Based Sampling (PEBS)~\cite{Intel:ProgrammingGuide}.   
The Roofline model~\cite{williams:roofline, ilic:roofline-cache-aware} analyzes compute performance and memory bandwidth. 
The application is classified as compute or memory bound based on the comparison with the performance roofs of the target hardware platform.    
The Top-down model~\cite{Yasin:topdown} analyzes the application CPI stack. 
The application is categorized as memory bound if its significant CPI component is caused by the main memory accesses. 
The model also distinguishes between the memory latency and bandwidth stalls depending on the occupancy of the memory controller queues. 
%




\section{Conclusions}
\label{sec:Conclusions}

\looseness -1 The \textbf{Memory stress (Mess) framework} offers a comprehensive and unified approach to memory system benchmarking, simulation, and application profiling. 
The \textbf{Mess benchmark} provides a detailed and holistic view of memory systems performance.  
It covers x86, ARM, Power, RISC-V, and NVIDIA's PTX ISAs, 
and it is already deployed to characterize servers from Intel, AMD, IBM, Fujitsu, Amazon, and NVIDIA with DDR4, DDR5, HBM2, and HBM2E main memory. 
The \rev{\textbf{Mess analytical memory simulator}} is integrated with ZSim, gem5, and OpenPiton Metro-MPI CPU simulators
and it supports a wide range of memory systems, including the CXL memory expanders.
It closely matches the actual memory systems performance and it shows an unprecedented low error of 0.4--6\% when simulating widely-used memory benchmarks. 
Finally, the \textbf{Mess application profiling} is already integrated into a suite of production HPC performance analysis tools, 
adding a new layer of information related to application's memory performance metrics. 
The Mess framework is publicly-released and ready to be used by the community for better understanding of the current and exploration of future memory systems.


\section{Acknowledgment}
\label{sec:Acknowledgment}
\looseness -1 This work was funded by the Collaboration Agreement between Micron Technology, Inc. and BSC. The results have been partially funded by the Spanish Ministry of Science and Innovation MCIN/AEI/10.13039/501100011033 (contracts PID2019-107255GB-C21, PCI2022-132935, PCI2021-121958, PID2023-146511NB-I00 and CEX2021-001148-S), the Generalitat of Catalunya (contract 2021-SGR-00763, 2021-SGR-00807 and 2021-SGR-01264), European Union (Grant Agreement No. 955606), European Processor Initiative (EPI) with grant agreement No. 800928, 826647 and 101036168, and by the Barcelona Zettascale Lab (BZL) project with reference REGAGE22e00058408992 and co-financed by the Ministry for Digital Transformation and Public Services, within the framework of the Resilience and Recovery Fund – and the European Union – NextGenerationEU. The work is also supported by the Arm-BSC Center of Excellence and the European HiPEAC Network of Excellence. A. Armejach is a Serra Hunter Fellow. We would also like to acknowledge valuable comments from the following individuals: Julian Pavon (BSC), Mohammad Bakhshalipour (CMU), Vicenc Beltran (BSC), Arash Yadegari (BSC), Nam Ho (FJZ) and Carlos Falquez (FJZ), Stepan Vanecek (TUM), John McCalpin (TACC).



\appendix

\subsection{Mess benchmark implementation}
\label{app:Mess-implementation}

In this appendix, we discuss Mess benchmark implementation. It includes implementation of pointer-chase and traffic generator.
The scripts and manuals for the experimental setup, execution and measurements is included in the public Mess repository.  
The measurements post-processing removes the outliers, mitigates the noise and plots the results.

\subsubsection{Pointer-chase}
\label{app:ptr-chase-implementation}
The pointer-chase contains a sequence of dependent back-to-back load instructions that access the main memory. 
Since the instructions are dependent, their execution is serialized, 
so the average memory access latency is calculated as the ratio between the total execution time and the number of instructions executed. 
This is guaranteed by the pointer-chase design.  
In the benchmark initialization, we allocate a contiguous section of memory and initialize it in such a way that a given
array element contains the address of the next array element (memory location) that
we want to access. This ensures the dependency between the consecutive load instructions. 
To force accesses to the main memory (and not on-chip caches), 
the pointer-chase traverses the whole array whose size exceeds the last-level cache. 
To avoid any cache-line spatial locality, each array element occupies the whole cache line (64\,Byte). 
 %
Finally, to diminish the impact of data prefetching and temporal locality during traversing, 
the pointer-chase traverses the array in a random pattern. 
The pointer-chase source code is detailed in Listing~\ref{tab:pointer-chase-assembly}: 

\begin{itemize}
\item Lines~0001--0004 initialize the benchmark: set the number of load instructions, initialize the registers, etc. 
\item Lines~0006--1007 are the core of the benchmark. They are a series of back-to-back load instructions for traversing the array. Each load instruction has data dependency on the previous one, 
so their execution is serialized. 
\item Line~1008 decrements the loop counter.
\item Line~1009 checks the loop exit condition.
\end{itemize}

\begin{table}[t]
	\centering
       \begin{lstlisting}[language=C++, caption={Pointer-chase source code. x86 assembly. In our study we also use the benchmark for the ARM, IBM Power, and RISC-V ISA.}, label=tab:pointer-chase-assembly]
	\end{lstlisting}
	\resizebox{\columnwidth}{!}{%
	\begin{tabular}{@{}l|l|l@{}}
		\toprule
		Line & Source code & Explanation \\		
		\midrule
		\midrule
		0001 & \texttt{register struct line *next asm("rax");}    & struct line owns pointer to the next access      \\
		0002 & \texttt{register int i asm("ecx");}                 & ecx is the loop counter   \\
		0003 & \texttt{i = 1000000;}                               & initialization of the loop counter. C format.  \\
		0004 & \texttt{next = ptr-\textgreater{}next}             & first memory access in C format  \\
		0005 & \texttt{start\_loop: }                              & beginning of the loop  \\
		0006 & \hspace*{0.2cm}\texttt{mov (\%rax), \%rax} & \hspace*{0.2cm}load instruction (pointer-chase) \\
		0007 & \hspace*{0.2cm}\texttt{mov (\%rax), \%rax} & \hspace*{0.2cm}load instruction (pointer-chase) \\
		...      & \hspace*{0.2cm}...                            & \hspace*{0.2cm}…  \\
		1007 & \hspace*{0.2cm}\texttt{mov (\%rax), \%rax} & \hspace*{0.2cm}load instruction (pointer-chase) \\
		1008 & \hspace*{0.2cm}\texttt{dec \%ecx}                                  & \hspace*{0.2cm}decrement loop counter    \\
		1009 & \hspace*{0.2cm}\texttt{jnz start\_loop}                            & \hspace*{0.2cm}if counter is not zero jump to start of the loop \\
		
		\bottomrule
	\end{tabular}
	}
\end{table}

\subsubsection{Traffic generator}
\label{app:traffic-generator-implementation}

We write the core of the memory traffic generator directly in assembly. This enables the development of more precise code (changes at the level of the assembly instruction) and prevents any system software optimizations (e.g. compiler optimizations). 
The benchmark generates different read and write memory traffic by executing a different mix of load and store instructions in the source code, each of them in a separate assembly file. 
In the current implementation, the benchmark can execute from 100\% load to 100\% store operations, 
with a step of 2\%.\footnote{The ratio of loads and stores executed by the benchmark does not lead to the same ratio of the reads and writes in the memory traffic. 
This is because most of the state-of-the-art HPC servers deploy a write-allocate cache policy. 
With this policy each store operation targeting data in main memory is first read into the cache (memory read), 
then modified, and finally written to the main memory once the cache line is evicted.   
Since each store operation causes one memory read and one memory write, e.g. 100\% store kernel causes 50\%-read 50\%-write memory traffic.}  
The generated memory bandwidth is adjusted the issue rate of the memory operations. 
This is done by interleaving the load/store operation sequence with a configurable dummy loop of \texttt{nop} operations. %
To reduce the overhead of the loop management, loop counter decrement and jump operations, 
we employ a long loop of around 100 load/store operations. 
Listings~\ref{tab:adjusted-stream}, and \ref{tab:stream-nop-function} show the 
memory traffic generator implementation in the x86 ISA. The benchmark is also available for ARM, IBM Power, and RISC-V ISAs.

\begin{itemize}
\item Lines 001--008 initialize the benchmark: 
    \begin{itemize}
    \item Lines 001--002 loads the addresses of the arrays \texttt{a} and \texttt{c} into the registers.
    \item Lines 003--004 initialize a register as the loop counter.
    \item Lines 005--006 set the size of the array \texttt{a} and \texttt{c}.
    \item Lines 007--008 set the iteration of the dummy \texttt{nop} function. This controls the rate of the memory operations, i.e. used memory bandwidth. 
    \end{itemize}
\item The main loop of the kernel starts at line 009. The loop comprises a series of load and store instructions (Lines 010--015) followed by the call to the dummy \texttt{nop} function (Lines 016--017). 

    \begin{itemize}
	\item The \texttt{nop} function is shown in Listing~\ref{tab:stream-nop-function}. The number of \texttt{nop} operations 
		is configurable: the larger the number of \texttt{nop} operations in each function call, the lower the rate of 
		memory instruction, and therefore the lower used memory bandwidth. This configuration is managed by the value of \texttt{nopCount}
		parameter. When \texttt{nopCount} parameter is set to \texttt{0}, \texttt{i=0} in Listing~\ref{tab:stream-nop-function},
		the function is finalized without entering the \texttt{nop} loop. This causes only a negligible halt in
		the load and store sequence, leading to the maximum pressure to the memory system. As we increase the value of the
		\texttt{nopCount} parameter, we increase the number of the iterations in the dummy \texttt{nop} loop, and therefore
		reduce the rate of the memory operations issued by the memory traffic generator.
    \end{itemize}

\item The sequence of interleaved load and store instructions, and calls to the \texttt{nop} function is repeated until the line 143. This is the core of the benchmark. In the current implementation, each iteration of the main loop comprises 100 load and store instructions. 
\item Lines 144--146 finalize the loop iteration: increment the loop counter, check the exit condition, and conditionally jump to the beginning of the loop. 
\end{itemize}

\begin{table}[!t]
	\centering
       \begin{lstlisting}[language=C++, caption={Memory traffic generator, x86 ISA. Different read and write memory traffic is generated by executing a different mix of load and store instructions in the source code, each of them in a separate assembly file. 
The generated memory bandwidth is adjusted the issue rate of the memory operations. This is done by interleaving the load/store operation sequence with a configurable dummy loop of \texttt{nop} operations.}, label=tab:adjusted-stream]
	\end{lstlisting}
	\resizebox{\columnwidth}{!}{%
	\begin{tabular}{@{}l|l|l@{}}
		\toprule
		Line & Source code & Explanation \\		
		\midrule
		\midrule
		001 & \texttt{register double *a asm("r10");}    & load array \texttt{a} address into the register \texttt{r10}      \\
		002 & \texttt{register double *c asm("r11");}                 & load array \texttt{c} address into the register \texttt{r11}   \\
		003 & \texttt{register ssize\char`_t i asm("rbx");}                               & register \texttt{rbx} holds the loop counter  \\
		004 & \texttt{i = 0;}             & initialize the counter (\texttt{i}) by 0  \\
		005 & \texttt{register ssize\char`_t n asm("r12");}                              & \texttt{r12} holds the size of the arrays  \\
		006 & \texttt{n = *array\char`_size;}                              & load \texttt{r12} with size of the arrays  \\
		007 & \texttt{register int *p asm("r15");}                              & register \texttt{r15} holds the nopCount value  \\
		008 & \texttt{p = nopCount;}                              & load \texttt{r15} with value of nopCount  \\
		009 & \texttt{..L\char`_50:}                              & the beginning of the kernel loop  \\
		010 & \hspace*{0.2cm}\texttt{vmovupd   (\%r10,\%rbx,8), \%ymm0;} & \hspace*{0.2cm}load instruction  \\
		011 & \hspace*{0.2cm}\texttt{vmovupd  \%ymm1, (\%r11,\%rbx,8);} & \hspace*{0.2cm}store instruction  \\
		012 & \hspace*{0.2cm}\texttt{vmovupd   32(\%r10,\%rbx,8), \%ymm0;} & \hspace*{0.2cm}load instruction  \\
		013 & \hspace*{0.2cm}\texttt{vmovupd  \%ymm1, 32(\%r11,\%rbx,8);} & \hspace*{0.2cm}store instruction  \\
		014 & \hspace*{0.2cm}\texttt{vmovupd   64(\%r10,\%rbx,8), \%ymm0;} & \hspace*{0.2cm}load instruction  \\
		015 & \hspace*{0.2cm}\texttt{vmovupd  \%ymm1, 64(\%r11,\%rbx,8);} & \hspace*{0.2cm}store instruction  \\
		016 & \hspace*{0.2cm}\texttt{movq      \%r15, \%rdi;} & \hspace*{0.2cm}send number of nopCount as an input of \texttt{nop} function \\
		017 & \hspace*{0.2cm}\texttt{call      nop;} & \hspace*{0.2cm}call \texttt{nop} function   \\
		...      & \hspace*{0.2cm}...                            & \hspace*{0.2cm}…  \\
		...      & \hspace*{0.2cm}...                            & \hspace*{0.2cm}…  \\
		138 & \hspace*{0.2cm}\texttt{vmovupd   1536(\%r10,\%rbx,8), \%ymm0;} & \hspace*{0.2cm}load instruction  \\
		139 & \hspace*{0.2cm}\texttt{vmovupd  \%ymm1, 1536(\%r11,\%rbx,8);} & \hspace*{0.2cm}store instruction  \\
		140 & \hspace*{0.2cm}\texttt{vmovupd   1568(\%r10,\%rbx,8), \%ymm0;} & \hspace*{0.2cm}load instruction  \\
		141 & \hspace*{0.2cm}\texttt{vmovupd  \%ymm1, 1568(\%r11,\%rbx,8);} & \hspace*{0.2cm}store instruction  \\
		142 & \hspace*{0.2cm}\texttt{movq      \%r15, \%rdi;} & \hspace*{0.2cm}send number of nopCount as an input of \texttt{nop} function  \\
		143 & \hspace*{0.2cm}\texttt{call      nop;} & \hspace*{0.2cm}call \texttt{nop} function  \\
		144 & \hspace*{0.2cm}\texttt{add       \$200, \%rbx;} & \hspace*{0.2cm}increment the loop counter  \\
		145 & \hspace*{0.2cm}\texttt{cmp       \%r12, \%rbx;} & \hspace*{0.2cm}check the exit condition  \\
		146 & \hspace*{0.2cm}\texttt{jb        ..L\char`_50;} & \hspace*{0.2cm}conditional jump to beginning of the kernel loop  \\
		
		\bottomrule
	\end{tabular}
	}
\end{table}

\begin{table}[!t]
	\centering
	\begin{lstlisting}[language=C++, caption={Dummy \texttt{nop} function is used to generate a pause in the sequence of the load and store instructions in the Memory traffic generator (see Listing~\ref{tab:adjusted-stream}). The input parameter \texttt{nopCount} is the number of the dummy loop iterations. The larger the number of iterations, the lower the rate of memory instruction and the lower the generated memory bandwidth.}, label=tab:stream-nop-function]
	\end{lstlisting}
	\resizebox{\columnwidth}{!}{%
	\begin{tabular}{@{}l|l|l@{}}
		\toprule
		Line & Source code & Explanation \\		
		\midrule
		\midrule
		001 & \texttt{int nop(int *nopCount) \{}    & definition of \texttt{nop} function       \\
		002 & \hspace*{0.2cm}\texttt{unsigned int i  = *nopCount;}                 & \texttt{i} is the loop counter; number of loops for \texttt{nop} operations    \\
		003 & \hspace*{0.2cm}\texttt{if ( !i ) \{}                 &  if \texttt{i} is zero then return immediately    \\
		004 & \hspace*{.4cm}\texttt{return 0;}                 &      \\
		005 & \hspace*{0.2cm}\texttt{\} else \{}                 &  assembly section of \texttt{nop} function    \\
		006 & \hspace*{0.4cm}\texttt{asm(}                 &  it is written in extended assembly format    \\
		007 & \hspace*{.8cm}\texttt{mov \%0, \%\%ecx;}                 &  move the value of \texttt{\%0} register  into \texttt{ecx }register    \\
		008 & \hspace*{.8cm}\texttt{the\char`_loop\%=:}                 &  the beginning of the loop    \\
		009 & \hspace*{.8cm}\texttt{nop;}                 &  \hspace*{0.5cm}\texttt{nop} instruction    \\
		010 & \hspace*{.8cm}\texttt{dec \%\%ecx}                 &  \hspace*{0.5cm}decrementing the counter    \\
		011 & \hspace*{.8cm}\texttt{jnz the\char`_loop\%=}                 & \hspace*{0.5cm}conditional jump to beginning of the loop    \\
		012 & \hspace*{.8cm}\texttt{: "r" (i) : ecx }                 &  put the value of i into register \texttt{\%0} (extended assembly)   \\
		013 & \hspace*{.4cm}\texttt{);}                 &      \\
		014 & \hspace*{.2cm}\texttt{\}}                 &      \\
		014 & \hspace*{.2cm}\texttt{return 0;}                 &      \\
		015 & \texttt{\}}                 &      \\
		
		\bottomrule
	\end{tabular}
	}
\end{table}


\subsection{Case study: Remote-socket memory emulation of CXL memory expanders}
\label{sec:CXL-emulators}

Recent industrial studies emulate CXL memory expansion in cloud and datacenter servers with conventional dual-socket systems: 
one socket is used as a host CPU and the other socket as a CPU-less memory expander~\cite{maruf:cxl-tieredmem, Li:Pond}. 
To evaluate this approach, we perform a detailed simulation of both systems.  
The CPU host is simulated with the ZSim configured to model the Intel 24-core Skylake processor (Section~\ref{sec:ZSim-characterization}).  
We use Micron Technology's bandwidth--latency curves to model the CXL memory expander, 
and measurements from the actual dual-socket server to model the remote-socket memory. 
In each system configuration, we simulate 500\,billion instructions of the multiprogramed SPEC~CPU2006 workloads~\cite{Henning:SPEC}.

\begin{figure}[t]%
	\centering
	\includegraphics[width=\linewidth]{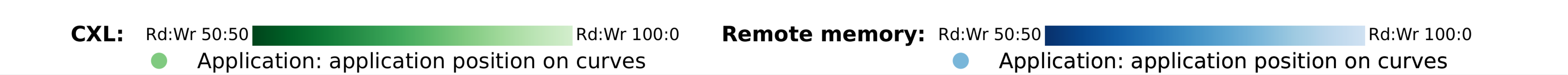}\vspace{-.14cm}\\%
	\subfigure[][The perlbench experiences higher remote-memory latency which leads to 5\% lower performance.]{%
		\label{fig:cxl-remote-app-profile-latency-bound}%
		\includegraphics[width=.5\linewidth]{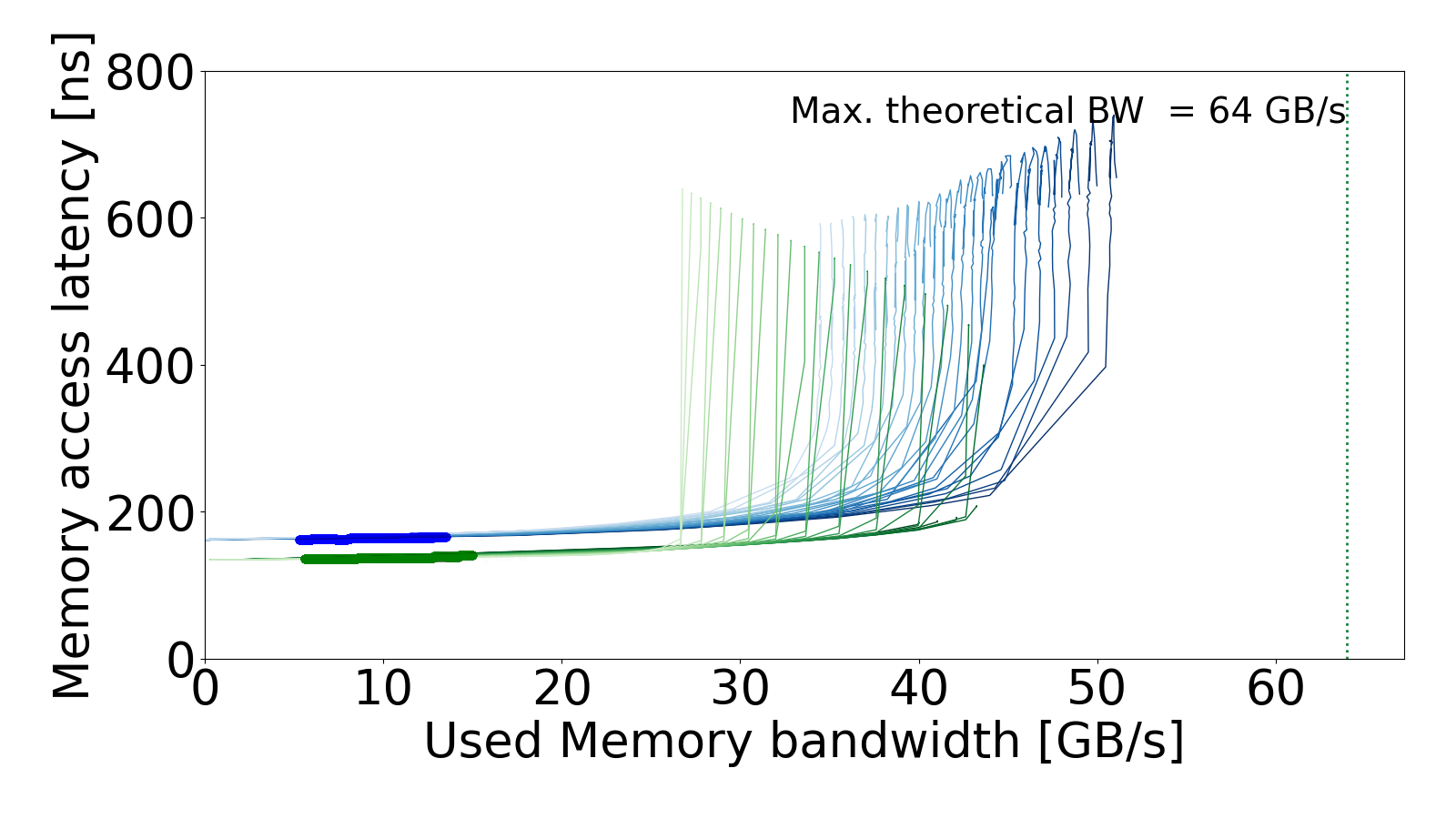}}~%
	\subfigure[][The lbm reaches higher remote-memory bandwidth which leads to 11\% higher performance.]{%
		\label{fig:cxl-remote-app-profile-bw-bound}%
		\includegraphics[width=.5\linewidth]{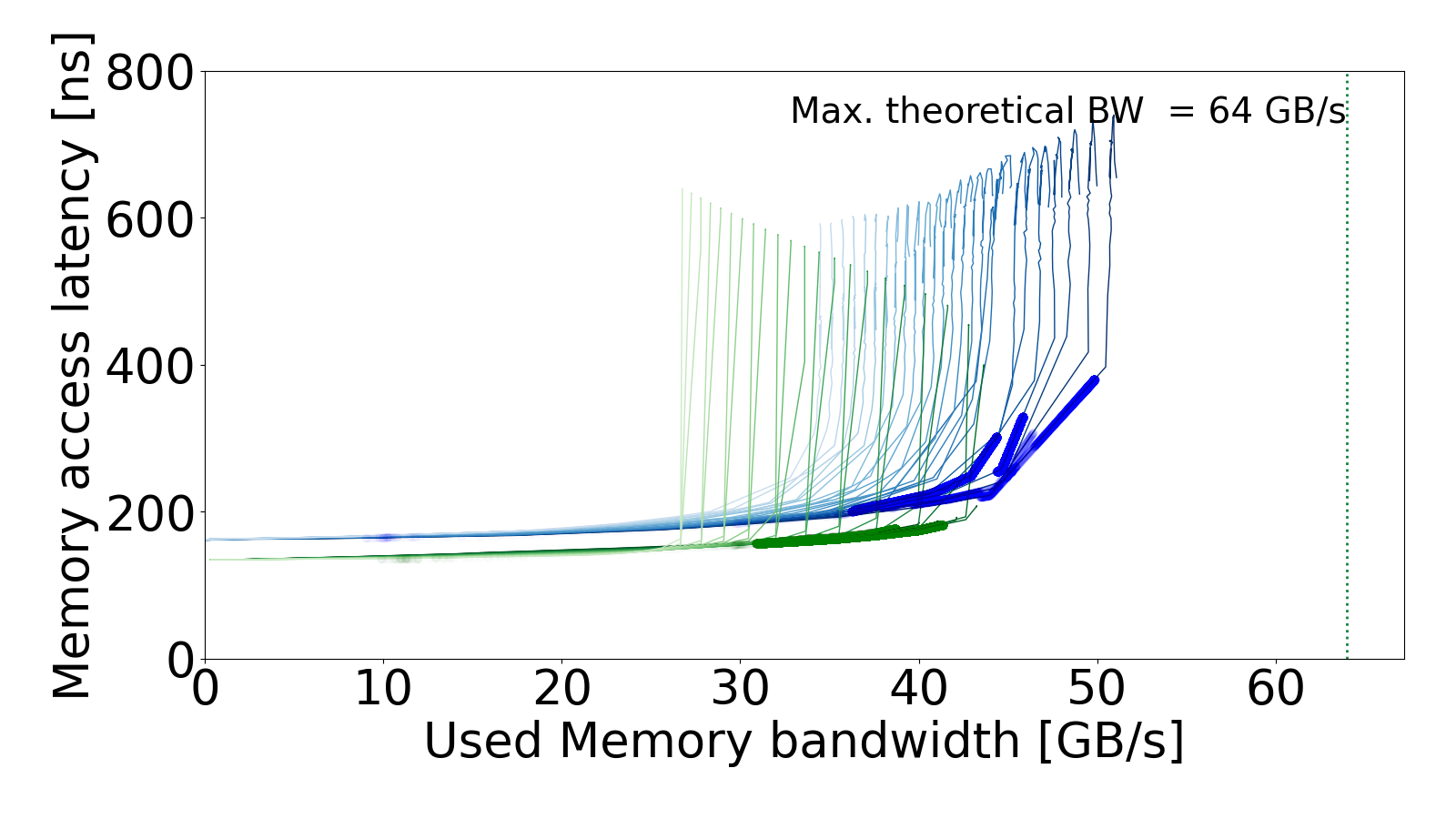}}%
	\caption[A set of two subfigures.]{Remote-socket emulation of CXL memory expanders: Bandwidth--latency curves and implications}
	\label{fig:numa-vs-cxm-input-sim}%
\end{figure}

\begin{figure}[t!]
	\centering
	\includegraphics[width=\linewidth]{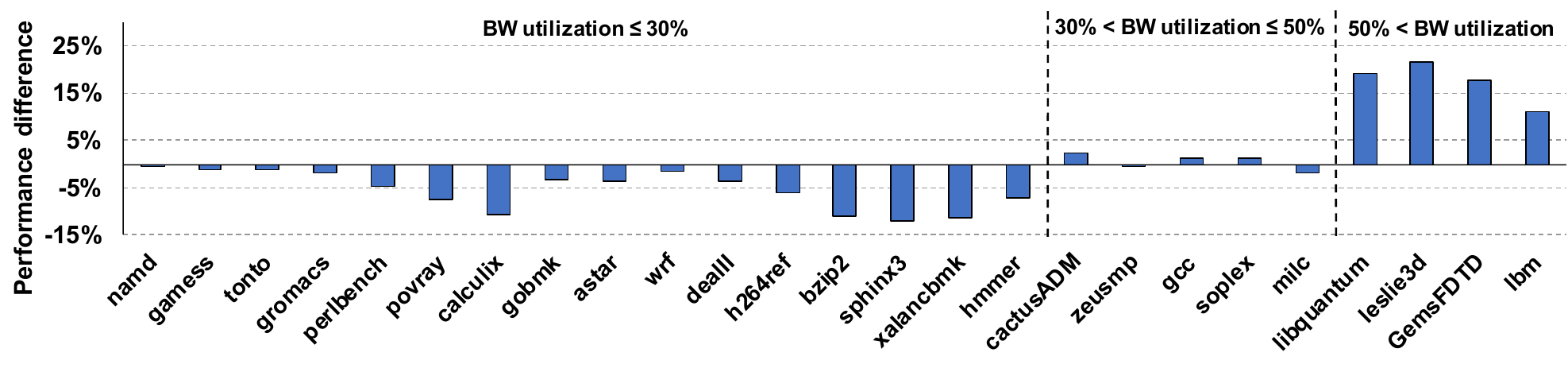}
	\caption{Remote-socket memory emulation of CXL memory expanders: Application performance difference is correlated with its bandwidth utilization. SPEC2006 benchmarks, 24-core Intel Skylake CPU.}
	\label{fig:cxl-vs-numa}
\end{figure}

Figure~\ref{fig:numa-vs-cxm-input-sim} plots the CXL (green) and remote-socket (blue) memory curves, and the benchmark behavior. 
Each 20.000 simulated instructions we monitor the benchmark read and write memory bandwidth, 
and then plot each observation as a point on the corresponding bandwidth--latency curve. 
The figure shows two characteristic use-cases. 
The perlbench benchmark (Figure~\ref{fig:cxl-remote-app-profile-latency-bound}) has a low bandwidth utilization. 
In this area remote-socket memory curves shows approximately 28\,ns higher latency, which leads to somewhat lower performance 
w.r.t. target CXL system. 
We detect the opposite behavior for high-bandwidth benchmarks.  
The remote-socket curves have a higher bandwidth saturation area.  
This leads to higher achieved bandwidths and the overall performance of the bandwidth-intensive benchmarks, such as the  
lbm illustrated in Figure~\ref{fig:cxl-remote-app-profile-bw-bound}. 
The overall performance trend is illustrated in Figure~\ref{fig:cxl-vs-numa}. 
The figure plots the results for all SPEC~CPU2006 benchmarks, sorted from the lowest to the highest bandwidth utilization. 
For low-bandwidth benchmarks, remote-socket emulation provides up to 12\% lower performance w.r.t. to the target CXL system. 
Both system provide a similar performance for the benchmarks with medium bandwidth utilization between 30\% and 50\% of CXL max theoretical bandwidth. 
For the high-bandwidth benchmarks, the remote-socket memory provides 11\%--22\% higher performance.

To the best of our knowledge, this is the first study that enables a detailed simulation of CXL memory expanders 
based on the manufacturer's SystemC model. The model is already integrated and publicly released with ZSim, gem5 and OpenPiton Metro-MPI simulators.  
We also provide the performance analysis and correction factors that can be useful for future studies who want to keep useing remote-socket memory emulation, e.g. due to its much faster execution time.  

\subsection{{Artifact Appendix}}
\label{sec:Artifact}

\subsubsection{Abstract}

{

This artifact comprises the implementation of the Mess benchmark for various platforms, including actual hardware platforms (e.g., Intel, IBM, NVIDIA), system simulators (ZSim, gem5, and OpenPiton), and memory simulators (DRAMsim3, Ramulator, Ramulator2). It also includes the Mess memory simulator integrated with ZSim, gem5, and OpenPiton simulators.

This artifact also contains all the scripts and guidelines necessary to reproduce the major figures presented in this paper. In addition to the scripts, it includes raw hardware measurements, processed measurements, and the final bandwidth-latency curves.

This study involves more than eight different hardware platforms with a diverse set of runtime environments, including various compilers, ISAs, and tools. Therefore, in this artifact, we mention the packages, tools, and applications without specifying the exact versions. In the Git repository, we provide detailed dependencies and version information for each experiment. \\

\subsubsection{Artifact check-list (meta-information)}

{\small
\begin{itemize}
  \item {\bf Program:} Pointer-chase and workload generator implemented in C, C++ with its kernel implemented in inline assembly (included in the benchmark). Mess simulator integrated with ZSim, gem5, and OpenPiton.
  \item {\bf Compilation:} GCC, G++, ICX, MPI++, and Python\,3.
  \item {\bf Data set:} All the raw values measured from hardware counters and simulation tracecs as well as final curves for all the figures are included.
  \item {\bf Run-time environment:} For Intel Cascade Lake curves, we use small server with single node. For Fujitsu A64FX, we use PJM batch processing support. For Graviton server, we use Amazon AWS. For the rest of the systems, we use production servers with Slurm Workload Manager environment.
  \item {\bf Hardware:} Servers or Supercomputers with the following CPUs and GPUs: Intel Sandy Bridge, Skylake, Cascade Lake and Sapphire Rapids. AMD Zen\,2, IBM Power\,9, Amazon Graviton\,3, Fujitsu A64FX, NVIDIA Hopper H100.
  \item {\bf Metrics:} For latency, we use nanoseconds and cycles. For bandwidth, we use GB/s.
  \item {\bf Output:} We plot bandwidth--latency curves. Moreover, we print detailed data points in a .csv file.
  \item {\bf Experiments:} Generate experiments using supplied scripts.
  \item {\bf How much disk space required (approximately)?:} 10s of GB.
  \item {\bf How much time is needed to prepare workflow (approximately)?:} For each experiment, approximately one hour.
  \item {\bf How much time is needed to complete experiments (approximately)?:} To generate bandwidth--latency curves for actual hardware, ZSim, gem5, and OpenPiton, we need approximately 3-6 days, 1-2 weeks, 2-3 weeks, and 1-2 weeks, respectively.
  \item {\bf Publicly available:} Yes.
  \item {\bf Code licenses:} MIT License.
  \item {\bf Data licenses:} MIT License.
  \item {\bf Archived (provide DOI)?:} 10.5281/zenodo.13748673 
\end{itemize}
}

\subsubsection{Description}

Figure~\ref{fig:benchmark-repo} and~\ref{fig:simulator-repo}} show where to replicate each major result presented in the paper. A detailed explanation of dependencies, system configurations, experimental setups, and result validation is available in the \texttt{README.md} file in each folder. To fit within the page limit of this artifact, this guideline introduces the general approach for replicating each result, along with an example to reproduce Figure~6.b. To replicate other figures, a similar approach should be followed (detailed guidelines are available in the Git repositories).

\paragraph{How to access}

The Mess Benchmark artifact can be cloned from GitHub at \url{https://github.com/bsc-mem/Mess-benchmark.git}.
The structure of the repository is depicted in Figure~\ref{fig:benchmark-repo}. The Mess simulator artifact can also be cloned from \url{https://github.com/bsc-mem/Mess-simulator}. 
The structure of the repository is depicted in Figure~\ref{fig:simulator-repo}. Each folder in the repositories replicates one or more figures presented in the main manuscript (figures are indicated in blue text). This artifact can also be downloaded as a \texttt{.zip} file from \url{https://zenodo.org/records/13748674}.

\paragraph{Hardware dependencies}
To run the Mess benchmark on actual hardware, access to a full node is required. The CPU/GPUs must support hardware counters to measure memory bandwidth (preferably \texttt{uncore} counters). For simulation experiments, a single core is sufficient. However, ZSim and OpenPiton can benefit from multicore or multinode parallelism. 

\paragraph{Software dependencies}
The benchmark and simulations run on Linux OS. To measure \texttt{uncore} counters, we primarily use the Linux perf tool, which is supported by all major Linux versions. In some cases, we also use Intel VTune and LIKWID.

\paragraph{Data sets}
All the data sets are included in the repositories.


\subsubsection{Installation}
To install, first clone the repository. Then, navigate to the directory corresponding to the figure from the main manuscript that you want to replicate (see Figures~\ref{fig:benchmark-repo} and~\ref{fig:simulator-repo}). For Figure~6.b of the main manuscript: 

\begin{lstlisting}[language=sh, frame=single, basicstyle=\scriptsize]
git clone https://github.com/bsc-mem/Mess-benchmark.git
cd ./CPU/Simulators/Trace-driven/DRAMsim3
\end{lstlisting}

\begin{figure}[!t]
\footnotesize
\hspace{-.5cm}\begin{minipage}{.29\textwidth}
\dirtree{%
.1 Mess-benchmark.
.2 Actual-hardware.
.3 CPU. 
.4 x86 \textcolor{blue}{(Fig.2.a,b,c,f)}.
.4 ARM \textcolor{blue}{(Fig.2.e,g)}. 
.4 IBM power \textcolor{blue}{(Fig.2.d)}.
.4 RISC-V.
.3 GPU.
.4 NVIDIA PTX\,\textcolor{blue}{(Fig.2.h)}.
.2 Simulators.
.3 Execution-driven.
.4 gem5 \textcolor{blue}{(Fig.4.b,c,d)}.
.4 ZSim \textcolor{blue}{(Fig.5.b--f)}.
.3 Trace-driven.
.4 Ramulator2~\textcolor{blue}{(Fig.6.a)}.
.4 DRAMsim3 \textcolor{blue}{(Fig.6.b)}.
.4 Ramulator \textcolor{blue}{(Fig.6.c)}.
}
\captionof{figure}{Mess benchmark repo}
  \label{fig:benchmark-repo}
\end{minipage}%
\begin{minipage}{.22\textwidth}
\dirtree{%
.1 Mess-simulator.
.2 Standalone.
.2 Integrated.
.3 ZSim \textcolor{blue}{\\(Fig.10.a--c) \\(Fig.14.d)}.
.3 gem5 \textcolor{blue}{\\(Fig.11.a--b) \\(Fig.14.c)}.
.3 OpenPiton \textcolor{blue}{\\ (Fig.14.b)}.
}
\vspace{2.55cm}
\captionof{figure}{Mess simulator repo}
  \label{fig:simulator-repo}
\end{minipage}%

\end{figure}


\begin{figure}[h]

\hspace{-.5cm}\begin{minipage}{.51\textwidth}
\footnotesize
\dirtree{%
.1 DRAMsim3. 
.2 measurement\_rdRatio\_Pause \textcolor{blue}{$\rightarrow$raw simulation result}.
.3 dramsim3.json \textcolor{blue}{$\rightarrow$results in json format}.
.3 dramsim3.txt \textcolor{blue}{$\rightarrow$results in txt format}.
.3 dramsim3epoch.json \textcolor{blue}{$\rightarrow$result per time epoch}.
.3 output\_jobID.err \textcolor{blue}{$\rightarrow$simulation error print}.
.3 output\_jobID.out \textcolor{blue}{$\rightarrow$simulation output print}.
.3 submit.batch \textcolor{blue}{$\rightarrow$run a single experiment}.
.2 DRAMsim3\_mn5 \textcolor{blue}{$\rightarrow$DRAMsim3 simulator directory}.
.2 traceInput \textcolor{blue}{$\rightarrow$input trace used in our experiments}.
.2 main.py \textcolor{blue}{$\rightarrow$parser of raw simulation results}.
.2 results.csv \textcolor{blue}{$\rightarrow$final processed outputs}.
.2 results\_original.csv \textcolor{blue}{$\rightarrow$original data to validate against}.
.2 runner.sh \textcolor{blue}{$\rightarrow$the workflow to generate all raw simulation results}. 
.2 submit.batch \textcolor{blue}{$\rightarrow$template to generate a single simulation result}.
.2 replicate.sh \textcolor{blue}{$\rightarrow$the main bash file to replicate Figure 6.b}.
.2 convert.py \textcolor{blue}{$\rightarrow$generate bandwidth--latency curves.}.
.2 output.pdf \textcolor{blue}{$\rightarrow$final bandwidth--latency curves.}.
}
\end{minipage}%
\caption{Directory structure to replicate DRAMsim3 experiments.}
\label{fig:dramsim}
\end{figure}


\subsubsection{Experiment workflow}

For each experiment (each folder in Figures~\ref{fig:benchmark-repo} or~\ref{fig:simulator-repo}), the workflow for running the Mess benchmark/simulation is provided in the \texttt{runner.sh} script. For Figure~6.b of the main manuscript (trace-driven DRAMsim3 simulation), the full workflow takes less than one day.
 
\subsubsection{Evaluation and expected results}

To replicate each experiment, we run \texttt{replicate.sh} script. This script compiles necessary codes, executes the workflow  (i.e., \texttt{runner.sh}), and processes the output raw data. The artifact also includes all the raw measurements and final processed \texttt{.csv} data. For the DRAMsim3 example, the \texttt{replicate.sh} script inside DRAMsim3 directory (Figure~\ref{fig:dramsim}) executes the following commands:

\begin{lstlisting}[language=sh, frame=single, basicstyle=\scriptsize]
# unzip trace files
cd traceInput
for file in *.zip; do
    unzip "$file"
done
cd ..

# compile DRAMsim3
cd DRAMsim3_mn5
mkdir build
cd build
cmake ..
make 
cd ../..

# run the experiment 
./runner.sh

###################
# Post-processing #
###################

# generate results.csv file
python3 main.py .

# generate output.pdf (bandwidth--latency curves)
python3 convert.py 


\end{lstlisting}

The easiest way to validate the result is visually by examining the generated curves (e.g., output.pdf in our example). However, if one wants to evaluate the results in more detail, the \texttt{results.csv} file can be compared to \texttt{results\_original.csv}; rows with the same rw\_ratio and pause values should have a very close latency and bandwidth. 



\bibliographystyle{unsrt}
\bibliography{refs}

\end{document}